\documentclass{emulateapj}
\usepackage{lscape}
\newcommand{\mycolspace}{}
\newcommand{\ssp}{}
\newcommand{\erg}{\mbox{$\rm\,erg$}}
\newcommand{\kev}{\mbox{$\rm\,keV$}}
\newcommand{\cm}{\mbox{$\rm\,cm$}}

\newcommand{\yr}{\mbox{$\rm\,yr$}}
\newcommand{\s}{\mbox{$\rm\,s$}}

\newcommand{\mpc}{\mbox{$\rm\,Mpc$}}
\newcommand{\kpc}{\mbox{$\rm\,kpc$}}

\newcommand{\beq}{\begin{equation}}
\newcommand{\eeq}{\end{equation}}

\newcommand{\RJ}{\ensuremath{R_{\rm J}}}
\newcommand{\LNVSS}{\ensuremath{L_{\rm NVSS}}}
\newcommand{\Lxgas}{\ensuremath{L_{\rm X,gas}}}
\newcommand{\lxgas}{\ensuremath{L_{\rm X,gas}}}
\newcommand{\aout}{\ensuremath{\alpha_{\rm 24}}}
\newcommand{\ain}{\ensuremath{\alpha_{\rm 02}}}
\newcommand{\rhomass}{\ensuremath{\rho_{\rm 2MASS}}}
\newcommand{\rhotully}{\ensuremath{\rho_{\rm Tully}}}
\newcommand{\MK}{\ensuremath{M_{\rm K}}}
\newcommand{\Lsix}{\ensuremath{L_{\rm 6cm}}}
\newcommand{\TX}{\ensuremath{T_{\rm X}}}

\shorttitle{Thermal Structure of Hot ISM in Normal Ellipticals}
\shortauthors{Diehl \& Statler}

\begin{document}

\title{The Hot Interstellar Medium in Normal Elliptical Galaxies\\ III:
The Thermal Structure of the Gas}

\author{Steven Diehl\altaffilmark{1,2} and Thomas S. Statler\altaffilmark{2}}
\altaffiltext{1}{Theoretical Astrophysics Group (T-6)/Computational Methods 
Group (CCS-2), Mailstop B227, Los Alamos National Laboratory, P.O. Box 1663, 
Los Alamos, NM 87545, USA (present address)}
\altaffiltext{2}{Astrophysical Institute, Department of Physics and Astronomy, 
251B Clippinger Research Laboratories, Ohio University, Athens, OH 
45701, USA}
\email{diehl@lanl.gov, statler@ohio.edu}

\begin{abstract}
This is the third paper in a series analyzing X-ray emission from the
hot interstellar medium in a sample of 54 normal elliptical galaxies
observed by Chandra. We focus on a subset of 36 galaxies with
sufficient signal to compute radial temperature profiles.  We
distinguish four qualitatively different types of profile: positive
gradient (outwardly rising), negative gradients (falling),
quasi-isothermal (flat) and hybrid (falling at small radii and rising
at larger radii). We measure the mean logarithmic temperature
gradients in two radial regions: from 0--2 $J$-band effective radii
$R_J$ (excluding the central point source), and from 2--$4R_J$. We
find the outer gradient to be uncorrelated with intrinsic host galaxy
properties, but strongly influenced by the environment: galaxies in
low-density environments tend to show negative outer gradients, while
those in high-density environments show positive outer gradients,
suggesting the influence of circumgalactic hot gas. The inner
temperature gradient, however, is unaffected by the environment but
strongly correlated with intrinsic host galaxy characteristics:
negative inner gradients are more common for smaller, optically faint,
low radio-luminosity galaxies, whereas positive gradients are found in
bright galaxies with stronger radio sources. There is no evidence for
bimodality in the distribution of inner or outer gradients. We propose
three scenarios to explain the inner temperature gradients: (1) Weak
AGN heat the ISM locally, while higher-luminosity AGN heat the system
globally through jets inflating cavities at larger radii; (2) The
onset of negative inner gradients indicates a declining importance of
AGN heating relative to other sources, such as compressional heating
or supernovae; (3) The variety of temperature profiles are snapshots
of different stages of a time-dependent flow, cyclically reversing the
temperature gradient over time.
\end{abstract}

\keywords{galaxies: cooling flows---galaxies: elliptical and lenticular,
cD---galaxies: ISM---X-rays: galaxies---X-rays: ISM}


\section{Introduction}

In the first two papers of this series
\citep[][hereafter Paper I and Paper II]{DiehlGallery,DiehlAGN},
we conducted a comprehensive morphological analysis of the hot gas in normal
elliptical galaxies. In Paper I, we introduced a technique to separate the
hot gas emission from the contamination of
unresolved point sources. We applied this technique to a {\it Chandra}
archive sample of 54 elliptical galaxies and presented a gallery of
adaptively binned gas-only images, which were photon-flux calibrated
and background corrected. We used these gas maps to derive isophotal
ellipticity profiles and conducted a systematic morphological
analysis. A comparison between optical and X-ray ellipticities measured
in the inner, stellar mass dominated regions shows no correlation, contrary
to what would be expected if the gas were in perfect hydrostatic equilibrium.
We modeled the expected correlation under various assumptions, and
concluded that these systems, in general, are at best only approximately
hydrostatic. Moreover, the gas morphologies almost always look disturbed.
In Paper II, we introduced a quantitative measure of morphological asymmetry,
and found it to be tightly correlated with radio continuum power and nuclear
X-ray luminosity. We also found the gas morphology to be influenced, to a
comparable degree, by the ambient intergalactic medium. But the AGN--morphology
correlation forms a continuous trend down to the lowest detectable AGN
luminosities, indicating the importance of AGN feedback, even in rather
X-ray faint elliptical galaxies.

In this third paper, we address the question of whether the central AGN is
merely redistributing the gas, or heating it as well. We produce
radial temperature profiles and find that they fall into a variety of
distinct types. In particular, we confirm that negative (outwardly falling)
temperature gradients \citep{HumphreyDarkmatter,FukazawaMassprofiles,
RandallXMMNGC4649, PonmanNGC6482} are present, and relatively common, in
low-luminosity systems.

Outwardly {\em rising\/} (positive-gradient) temperature profiles,
nearly ubiquitous in galaxy clusters, X-ray groups, and massive
ellipticals, are usually understood as being the result of efficient
radiative cooling in the dense central regions. Accreting gas at large
radii can additionally shock-heat itself, amplifying the positive
gradient. This interpretation is supported by the short central
cooling times observed for galaxies, groups, and clusters, which can
drop well below $100\,{\rm Myr}$. However, cooling times
are equally short in galaxies with negative temperature gradients,
i.e. with a warm center \citep{HumphreyDarkmatter}.  
Several solutions have been proposed to
explain these counter-intuitive objects.

\citet{FukazawaMassprofiles} suggest that the gradients are a function
of environment, with outwardly rising (positive) gradients caused by
the hotter ambient intracluster or intragroup gas surrounding the
galaxy. Galaxies with negative gradients should instead be isolated.

\citet{HumphreyDarkmatter}, on the other hand, propose a bimodal
distribution of temperature gradients, and suggest that the total mass
of the system is the decisive factor for the sign of the gradient. The
division between their two groups happens at a virial mass of $\sim
10^{13}M_\sun$, implying a distinction between normal galaxies and
groups. They hypothesize further that the temperature gradients could
be related to a significant difference in the galaxies' evolutionary
histories.

\citet{HumphreyDarkmatter} also discuss the role of
compressive (gravitational) heating, noting that during a
slow inflow of relatively cool gas ($<1-2\kev$) the energy gain
would exceed the radiative losses. For
the inflow of hotter baryons, radiative cooling would dominate and one
would observe a positive temperature gradient instead. However, they
find no reason for hot baryons to be specific to systems above their
break mass $\sim 10^{13}M_\sun$, and suspect the environment as a fuel
source instead.

\citet{PonmanNGC6482} observe a negative temperature gradient in the
fossil group candidate NGC~6482 and argue along the same lines. They
model NGC~6482's temperature profile successfully with a steady-state
cooling flow with a reasonable cooling rate of $\dot M =
2M_\sun\,\yr^{-1}$, and adopt it as their preferred solution. They
also estimate that type Ia supernovae (SN) may be responsible for
balancing about $1/3$ of the radiative losses in this galaxy. They
find the contribution from type II SN to be insignificant and argue
against AGN feedback on grounds of the very relaxed appearance of
NGC~6482.

In this paper, we show that the distribution of temperature gradients
is {\em not\/} bimodal. We further show that the temperature gradients within
the inner 2 optical effective radii are {\em not\/} strongly influenced by the
environment. Instead, we find evidence that these inner gradients owe their
origins either to the specific nature of low-luminosity AGN feedback or to a
declining importance of AGN relative to compressive heating or supernovae.

In \S\ref{s5.dataanalysis}, we summarize our analyses and results from
Papers I and II, and describe the methodology to derive radial
temperature profiles. We then discuss the
various types of temperature profiles seen in our sample in
\S\ref{s5.temperature}. For a quantitative analysis, we split the
radial range into two regions: the inner region extending out to 2
effective radii and an outer region between $2-4$ effective radii. We
fit and analyze the gradients in these two regions separately,
and demonstrate that the
inner gradient is determined by galaxy properties, while the outer
gradient is strongly influenced by the presence of neighboring galaxies and/or
a hot ambient medium. In
\S\ref{s5.discussion}, we discuss the implications of our findings for
cooling flows, SN heating, and AGN feedback, before we briefly
summarize in \S\ref{s5.conclusions}.

\section{Data Analysis}\label{s5.dataanalysis}

\subsection{Summary of Results from Papers I and II}
\label{s5.preliminary}

We make use of several parameters from Papers I and II. We list those
essential to our analysis in Tables \ref{t.tempprop} and
\ref{t.tempprop2} for completeness, along with some additional
quantities. We extract absolute $K$ magnitudes $M_{\rm K}$ and
$J$-band effective radii $\RJ$ from the 2MASS extended source catalog
\citep{2MASS}. We adopt $20\cm$ radio continuum radio luminosities
$L_{\rm NVSS}$ from the NRAO VLA Sky Survey \citep[NVSS,][]{NVSS}
within $3\RJ$ (see Paper II). In addition, we extract $6\cm$ radio
continuum luminosities $L_{\rm 6cm}$ from the GB6 catalog of radio
sources \citep{GB6cm}, the Parkes-MIT-NRAO 4.85GHz Surveys
\citep{PMN6cm}, and a $6\cm$ radio catalog by \citet{Becker6cm} in the
same region. Central velocity dispersion values are taken from the
Lyon--Meudon Extragalactic Database \citep[LEDA;][]{LEDA}. We also
adopt the projected galaxy density parameter $\rho_{2MASS}$ from Paper
II, which is based on the number of neighbors in the 2MASS extended
source catalog, and corrected for incompleteness. As it is one of the
few accessible parameters to describe the galaxy environment, we also
list the \citet{TullyRho} galaxy density $\rho_{\rm Tully}$.


\begin{deluxetable*}{lrrrrrrrrrrrrrrrr}
\ssp \tablewidth{0pt} \tablecaption{{\it Chandra} X-ray luminosity and
temperature profile parameters. \label{t.tempprop}} \tablehead{
\colhead{Name} & \colhead{\mycolspace} & \colhead{$L_{\rm
X,Gas}$\tablenotemark{a}} & \colhead{$T_{\rm X}$\tablenotemark{b}} &
\colhead{$\alpha_{02}$\tablenotemark{c}} &
\colhead{$\alpha_{24}$\tablenotemark{c}} }

\startdata
IC1262 & & $ 2.0\pm 1.7\times 10^{43} $ & $ 1.30 \pm 0.01 $ & $ 0.29 \pm 0.02 $ & $ 0.21 \pm 0.07 $ \\
IC1459 & & $ 4.3\pm 3.2\times 10^{40} $ & $ 0.48 \pm 0.02 $ & $ -0.00 \pm 0.03 $ & $ -0.27 \pm 0.04 $ \\
IC4296 & & $ 1.1\pm 0.4\times 10^{41} $ & $ 0.88 \pm 0.02 $ & $ 0.23 \pm 0.04 $ & $ 0.08 \pm 0.07 $ \\
NGC0193 & & $ 2.5\pm 0.8\times 10^{41} $ & $ 0.77 \pm 0.01 $ & \nodata & \nodata \\
NGC0315 & & $ 9.4\pm 3.4\times 10^{40} $ & $ 0.64 \pm 0.01 $ & $ 0.02 \pm 0.02 $ & \nodata \\
NGC0383 & & $ < 7.5\times 10^{41} $ & $ 0.98 \pm 0.04 $ & $ 0.42 \pm 0.06 $ & $ 0.50 \pm 0.16 $ \\
NGC0404 & & $ < 2.1\times 10^{38} $ & $ 0.28 \pm 0.07 $ & \nodata & \nodata \\
NGC0507 & & $ > 5.7\times 10^{42} $ & $ 1.03 \pm 0.01 $ & $ 0.01 \pm 0.01 $ & $ 0.16 \pm 0.10 $ \\
NGC0533 & & $ 9.6\pm 3.5\times 10^{41} $ & $ 0.98 \pm 0.01 $ & $ 0.18 \pm 0.04 $ & \nodata \\
NGC0720 & & $ 9.3\pm 2.7\times 10^{40} $ & $ 0.57 \pm 0.01 $ & $ -0.05 \pm 0.03 $ & $ -0.04 \pm 0.15 $ \\
NGC0741 & & $ 3.2\pm 1.3\times 10^{41} $ & $ 0.96 \pm 0.02 $ & $ 0.31 \pm 0.04 $ & $ -0.10 \pm 0.12 $ \\
NGC0821 & & $ < 3.3\times 10^{40} $ & \nodata & \nodata & \nodata \\
NGC1132 & & $ > 9.1\times 10^{42} $ & $ 1.02 \pm 0.01 $ & $ 0.24 \pm 0.07 $ & $ -0.22 \pm 0.11 $ \\
NGC1265 & & $ < 1.1\times 10^{42} $ & $ 0.86 \pm 0.08 $ & \nodata & \nodata \\
NGC1316 & & $ 5.7\pm 2.1\times 10^{40} $ & $ 0.62 \pm 0.01 $ & $ -0.04 \pm 0.04 $ & $ 0.12 \pm 0.26 $ \\
NGC1399 & & $ > 7.9\times 10^{41} $ & $ 1.13 \pm 0.01 $ & $ 0.17 \pm 0.02 $ & $ 0.10 \pm 0.03 $ \\
NGC1404 & & $ 1.7\pm 0.4\times 10^{41} $ & $ 0.58 \pm 0.01 $ & $ -0.10 \pm 0.01 $ & $ 0.15 \pm 0.05 $ \\
NGC1407 & & $ 1.0\pm 0.3\times 10^{41} $ & $ 0.87 \pm 0.01 $ & $ 0.11 \pm 0.04 $ & $ 0.46 \pm 0.23 $ \\
NGC1549 & & $ > 2.0\times 10^{40} $ & $ 0.34 \pm 0.03 $ & \nodata & \nodata \\
NGC1553 & & $ 2.8\pm 2.6\times 10^{40} $ & $ 0.41 \pm 0.01 $ & $ -0.21 \pm 0.13 $ & $ -0.28 \pm 0.12 $ \\
NGC1600 & & $ > 1.2\times 10^{42} $ & $ 1.18 \pm 0.04 $ & \nodata & \nodata \\
NGC1700 & & $ > 3.2\times 10^{41} $ & $ 0.43 \pm 0.01 $ & $ -0.06 \pm 0.14 $ & \nodata \\
NGC2434 & & $ 2.6\pm 2.0\times 10^{40} $ & $ 0.53 \pm 0.03 $ & \nodata & \nodata \\
NGC2865 & & $ < 9.9\times 10^{40} $ & $ 0.66 \pm 0.09 $ & \nodata & \nodata \\
NGC3115 & & $ < 8.7\times 10^{39} $ & $ 0.50 \pm 0.04 $ & \nodata & \nodata \\
NGC3377 & & $ < 6.1\times 10^{39} $ & \nodata & \nodata & \nodata \\
NGC3379 & & $ < 6.3\times 10^{39} $ & $ 0.33 \pm 0.03 $ & $ -0.27 \pm 0.06 $ & \nodata \\
NGC3585 & & $ > 4.2\times 10^{39} $ & $ 0.33 \pm 0.01 $ & \nodata & \nodata \\
NGC3923 & & $ 4.3\pm 1.3\times 10^{40} $ & $ 0.48 \pm 0.02 $ & $ -0.07 \pm 0.01 $ & $ -0.37 \pm 0.11 $ \\
NGC4125 & & $ 7.2\pm 2.7\times 10^{40} $ & $ 0.44 \pm 0.01 $ & $ -0.06 \pm 0.02 $ & $ -0.15 \pm 0.16 $ \\
NGC4261 & & $ 4.8\pm 1.1\times 10^{40} $ & $ 0.78 \pm 0.01 $ & $ 0.25 \pm 0.05 $ & \nodata \\
NGC4365 & & $ > 3.8\times 10^{40} $ & $ 0.64 \pm 0.02 $ & $ 0.14 \pm 0.04 $ & \nodata \\
NGC4374 & & $ 5.9\pm 1.3\times 10^{40} $ & $ 0.71 \pm 0.01 $ & $ 0.16 \pm 0.02 $ & $ 0.44 \pm 0.08 $ \\
NGC4406 & & $ > 1.0\times 10^{42} $ & $ 0.78 \pm 0.01 $ & $ 0.06 \pm 0.01 $ & $ -0.05 \pm 0.04 $ \\
NGC4472 & & $ > 8.5\times 10^{41} $ & $ 0.97 \pm 0.01 $ & $ 0.14 \pm 0.01 $ & \nodata \\
NGC4494 & & $ < 2.1\times 10^{40} $ & \nodata & \nodata & \nodata \\
NGC4526 & & $ 8.8\pm 7.5\times 10^{39} $ & $ 0.35 \pm 0.03 $ & \nodata & \nodata \\
NGC4552 & & $ 2.1\pm 1.2\times 10^{40} $ & $ 0.57 \pm 0.01 $ & $ -0.21 \pm 0.04 $ & $ 0.42 \pm 0.16 $ \\
NGC4555 & & $ > 2.3\times 10^{41} $ & $ 0.97 \pm 0.03 $ & \nodata & $ -0.02 \pm 0.03 $ \\
NGC4564 & & $ > 2.0\times 10^{39} $ & \nodata & \nodata & \nodata \\
NGC4621 & & $ 1.1\pm 0.9\times 10^{40} $ & $ 0.23 \pm 0.03 $ & \nodata & \nodata \\
NGC4636 & & $ 2.7\pm 2.0\times 10^{41} $ & $ 0.69 \pm 0.01 $ & $ 0.11 \pm 0.01 $ & $ 0.04 \pm 0.02 $ \\
NGC4649 & & $ 1.3\pm 0.3\times 10^{41} $ & $ 0.80 \pm 0.01 $ & $ 0.02 \pm 0.01 $ & $ -0.01 \pm 0.01 $ \\
NGC4697 & & $ > 3.5\times 10^{40} $ & $ 0.32 \pm 0.01 $ & \nodata & \nodata \\
NGC5018 & & $ < 1.9\times 10^{41} $ & $ 0.45 \pm 0.09 $ & \nodata & \nodata \\
NGC5044 & & $ 2.6\pm 0.8\times 10^{42} $ & $ 0.91 \pm 0.01 $ & $ 0.09 \pm 0.01 $ & $ 0.12 \pm 0.02 $ \\
NGC5102 & & $ < 1.6\times 10^{39} $ & $ 0.38 \pm 0.08 $ & \nodata & \nodata \\
NGC5171 & & $ > 2.7\times 10^{42} $ & $ 0.80 \pm 0.05 $ & \nodata & \nodata \\
NGC5532 & & $ < 8.7\times 10^{41} $ & $ 0.61 \pm 0.02 $ & \nodata & $ -0.38 \pm 0.07 $ \\
NGC5845 & & $ < 5.2\times 10^{40} $ & $ 0.32 \pm 0.05 $ & \nodata & \nodata \\
NGC5846 & & $ 3.9\pm 0.9\times 10^{41} $ & $ 0.71 \pm 0.01 $ & $ 0.02 \pm 0.02 $ & $ 0.03 \pm 0.17 $ \\
NGC6482 & & $ 1.7\pm 1.3\times 10^{42} $ & $ 0.74 \pm 0.01 $ & $ -0.09 \pm 0.01 $ & $ -0.34 \pm 0.02 $ \\
NGC7052 & & $ > 1.1\times 10^{41} $ & $ 0.53 \pm 0.03 $ & \nodata & \nodata \\
NGC7618 & & $ 2.3\pm 0.9\times 10^{42} $ & $ 0.80 \pm 0.01 $ & $ -0.14 \pm 0.06 $ & $ -0.12 \pm 0.08 $ \\
\enddata
\tablenotetext{a}{Total X-ray gas luminosity in ${\rm ergs\,s^{-1}}$ for the $0.3-5\kev$ band, see Paper I for more details}.
\tablenotetext{b}{Luminosity weighted temperature within 3 optical radii.}
\tablenotetext{c}{Temperature gradients, measured in $\log r/\RJ - \log T/\kev$ space, between $0-2\RJ$ (\ain) and $2-4\RJ$ (\aout).}
\end{deluxetable*}

\begin{deluxetable*}{lrrrrrrrrrrrrrrr}
\ssp
\tablewidth{0pt}
\tablecaption{Optical, radio, and environmental parameters\label{t.tempprop2}}
\tablehead{
\colhead{} & \colhead{} &
\multicolumn{2}{c}{2MASS\tablenotemark{a}} & \colhead{} & \colhead{LEDA\tablenotemark{b}} & \colhead{} &
\multicolumn{2}{c}{Radio\tablenotemark{c}} & \colhead{} & \multicolumn{2}{c}{Environment\tablenotemark{d}} \\
\cline{3-4} \cline{6-6} \cline{8-9} \cline{11-12}\\
\colhead{Name} & \colhead{\mycolspace} & \colhead{$M_{\rm K}$} & \colhead{$\RJ$} & \colhead{\mycolspace} & \colhead{$\sigma$} & \colhead{\mycolspace} & \colhead{$L_{\rm NVSS}$} & \colhead{$L_{\rm GB6}$} & \colhead{\mycolspace} & \colhead{$\log \rho_{\rm 2MASS}$} & \colhead{$\rho_{\rm Tully}$}
}
\startdata
IC1262 & & $ -25.43 \pm 0.33 $  & $ 14.4 $ & & $ 266 \pm 36 $ & & $ 1.8\pm 0.5\times 10^{30} $ & $ < 4.5\times 10^{29} $      & & $ 3.47 \pm 0.22 $ & \nodata \\
IC1459 & & $ -25.53 \pm 0.28 $  & $ 29.1 $ & & $ 308 \pm 6 $  & & $ 1.3\pm 0.3\times 10^{30} $ & $ 1.2\pm 0.1\times 10^{30} $ & & $ 2.41 \pm 0.31 $ & $ 0.28 $ \\
IC4296 & & $ -26.06 \pm 0.33 $  & $ 25.5 $ & & $ 333 \pm 6 $  & & $ 6.0\pm 1.8\times 10^{30} $ & $ 6.0\pm 0.5\times 10^{30} $ & & $ 2.77 \pm 0.25 $ & \nodata \\
NGC0193 & & $ -24.71 \pm 0.33 $ & $ 14.5 $ & & \nodata        & & $ 5.6\pm 1.7\times 10^{30} $ & $ 3.8\pm 0.3\times 10^{30} $ & & $ 2.66 \pm 0.31 $ & \nodata \\
NGC0315 & & $ -26.33 \pm 0.33 $ & $ 22.9 $ & & $ 296 \pm 21 $ & & $ 9.8\pm 2.9\times 10^{30} $ & $ 5.7\pm 0.5\times 10^{30} $ & & $ 2.43 \pm 0.43 $ & \nodata \\
NGC0383 & & $ -25.84 \pm 0.33 $ & $ 17.8 $ & & $ 277 \pm 6 $  & & $ 1.3\pm 0.4\times 10^{31} $ & $ 6.8\pm 0.6\times 10^{30} $ & & $ 3.52 \pm 0.13 $ & \nodata \\
NGC0404 & & \nodata             & \nodata  & & $ 38 \pm 3 $   & & $ 4.3\pm 0.6\times 10^{25} $ & $ < 2.3\times 10^{26} $      & & \nodata           & $ 0.20 $ \\
NGC0507 & & $ -25.98 \pm 0.33 $ & $ 26.1 $ & & $ 315 \pm 9 $  & & $ 6.1\pm 1.8\times 10^{29} $ & $ < 1.1\times 10^{29} $      & & $ 3.03 \pm 0.22 $ & \nodata \\
NGC0533 & & $ -26.01 \pm 0.33 $ & $ 25.2 $ & & $ 275 \pm 6 $  & & $ 2.1\pm 0.6\times 10^{29} $ & $ < 1.3\times 10^{29} $      & & $ 3.17 \pm 0.19 $ & \nodata \\
NGC0720 & & $ -24.94 \pm 0.17 $ & $ 27.4 $ & & $ 242 \pm 5 $  & & $ < 2.3\times 10^{27} $      & $ < 3.8\times 10^{28} $      & & $ 2.57 \pm 0.25 $ & $ 0.25 $ \\
NGC0741 & & $ -26.19 \pm 0.33 $ & $ 25.9 $ & & $ 290 \pm 8 $  & & $ 6.4\pm 1.9\times 10^{30} $ & $ 2.0\pm 0.2\times 10^{30} $ & & $ 2.95 \pm 0.25 $ & $ 0.05 $ \\
NGC0821 & & $ -24.01 \pm 0.17 $ & $ 23.9 $ & & $ 199 \pm 2 $  & & $ < 1.7\times 10^{27} $      & $ < 1.2\times 10^{28} $      & & \nodata           & $ 0.08 $ \\
NGC1132 & & $ -25.70 \pm 0.33 $ & $ 19.8 $ & & $ 247 \pm 13 $ & & $ 1.1\pm 0.3\times 10^{29} $ & $ < 4.6\times 10^{29} $      & & $ 3.07 \pm 0.25 $ & \nodata \\
NGC1265 & & \nodata             & \nodata  & & \nodata        & & $ 3.3\pm 1.0\times 10^{31} $ & $ < 2.6\times 10^{29} $      & & \nodata           & \nodata \\
NGC1316 & & $ -26.07 \pm 0.17 $ & $ 49.8 $ & & $ 227 \pm 4 $  & & $ 1.6\pm 0.3\times 10^{29} $ & $ < 2.2\times 10^{28} $      & & $ 3.13 \pm 0.13 $ & $ 1.15 $ \\
NGC1399 & & $ -25.19 \pm 0.16 $ & $ 36.9 $ & & $ 337 \pm 5 $  & & $ 3.0\pm 0.4\times 10^{29} $ & $ < 3.4\times 10^{28} $      & & $ 3.28 \pm 0.12 $ & $ 1.59 $ \\
NGC1404 & & $ -24.79 \pm 0.19 $ & $ 19.3 $ & & $ 233 \pm 3 $  & & $ 2.1\pm 0.5\times 10^{27} $ & $ < 3.8\times 10^{28} $      & & $ 3.27 \pm 0.11 $ & $ 1.59 $ \\
NGC1407 & & $ -25.60 \pm 0.26 $ & $ 36.4 $ & & $ 272 \pm 5 $  & & $ 9.7\pm 2.3\times 10^{28} $ & $ < 4.2\times 10^{28} $      & & $ 3.10 \pm 0.14 $ & $ 0.42 $ \\
NGC1549 & & $ -24.69 \pm 0.18 $ & $ 29.0 $ & & $ 203 \pm 3 $  & & \nodata                      & $ < 1.9\times 10^{28} $      & & $ 2.74 \pm 0.22 $ & $ 0.97 $ \\
NGC1553 & & $ -25.06 \pm 0.17 $ & $ 33.9 $ & & $ 177 \pm 4 $  & & \nodata                      & $ < 1.6\times 10^{28} $      & & $ 2.96 \pm 0.18 $ & $ 0.97 $ \\
NGC1600 & & $ -26.06 \pm 0.33 $ & $ 24.8 $ & & $ 335 \pm 6 $  & & $ 3.6\pm 1.1\times 10^{29} $ & $ < 2.1\times 10^{29} $      & & $ 3.09 \pm 0.19 $ & \nodata \\
NGC1700 & & $ -25.59 \pm 0.33 $ & $ 15.9 $ & & $ 235 \pm 3 $  & & $ < 8.8\times 10^{27} $      & $ < 1.4\times 10^{29} $      & & $ 2.31 \pm 0.43 $ & \nodata \\
NGC2434 & & $ -23.78 \pm 0.29 $ & $ 19.3 $ & & $ 188 \pm 5 $  & & \nodata                      & $ < 2.2\times 10^{28} $      & & $ 2.56 \pm 0.25 $ & $ 0.19 $ \\
NGC2865 & & $ -24.43 \pm 0.20 $ & $ 14.8 $ & & $ 170 \pm 2 $  & & $ < 4.3\times 10^{27} $      & $ < 7.2\times 10^{28} $      & & \nodata           & $ 0.11 $ \\
NGC3115 & & $ -24.05 \pm 0.09 $ & $ 36.4 $ & & $ 257 \pm 5 $  & & $ < 2.8\times 10^{26} $      & $ < 4.5\times 10^{27} $      & & \nodata           & $ 0.08 $ \\
NGC3377 & & $ -22.81 \pm 0.09 $ & $ 27.7 $ & & $ 139 \pm 2 $  & & $ < 3.8\times 10^{26} $      & $ < 2.7\times 10^{27} $      & & \nodata           & $ 0.49 $ \\
NGC3379 & & $ -23.85 \pm 0.11 $ & $ 29.9 $ & & $ 205 \pm 2 $  & & $ 3.2\pm 0.7\times 10^{26} $ & $ < 2.4\times 10^{27} $      & & $ 3.82 \pm 0.10 $ & $ 0.52 $ \\
NGC3585 & & $ -24.81 \pm 0.18 $ & $ 32.3 $ & & $ 207 \pm 4 $  & & $ < 1.2\times 10^{27} $      & $ < 2.0\times 10^{28} $      & & $ 2.43 \pm 0.31 $ & $ 0.12 $ \\
NGC3923 & & $ -25.30 \pm 0.28 $ & $ 43.8 $ & & $ 247 \pm 6 $  & & $ < 1.6\times 10^{27} $      & $ < 2.6\times 10^{28} $      & & $ 2.89 \pm 0.16 $ & $ 0.40 $ \\
NGC4125 & & $ -25.03 \pm 0.25 $ & $ 33.0 $ & & $ 226 \pm 6 $  & & $ 1.7\pm 0.4\times 10^{28} $ & $ < 1.2\times 10^{28} $      & & $ 3.13 \pm 0.13 $ & $ 0.34 $ \\
NGC4261 & & $ -25.24 \pm 0.19 $ & $ 25.5 $ & & $ 320 \pm 8 $  & & $ 1.0\pm 0.2\times 10^{31} $ & $ 4.8\pm 0.1\times 10^{30} $ & & $ 3.09 \pm 0.14 $ & $ 0.84 $ \\
NGC4365 & & $ -24.91 \pm 0.17 $ & $ 40.7 $ & & $ 255 \pm 2 $  & & $ < 1.2\times 10^{27} $      & $ < 9.0\times 10^{27} $      & & $ 3.16 \pm 0.13 $ & $ 2.93 $ \\
NGC4374 & & $ -25.10 \pm 0.11 $ & $ 34.8 $ & & $ 280 \pm 2 $  & & $ 2.5\pm 0.3\times 10^{30} $ & $ 1.3\pm 0.0\times 10^{30} $ & & $ 3.18 \pm 0.14 $ & $ 3.99 $ \\
NGC4406 & & $ -25.07 \pm 0.14 $ & $ 59.7 $ & & $ 235 \pm 2 $  & & $ 1.1\pm 0.2\times 10^{27} $ & $ < 6.3\times 10^{27} $      & & $ 3.31 \pm 0.13 $ & $ 1.41 $ \\
NGC4472 & & $ -25.66 \pm 0.10 $ & $ 59.2 $ & & $ 288 \pm 2 $  & & $ 8.1\pm 0.8\times 10^{28} $ & $ 2.7\pm 0.0\times 10^{28} $ & & $ 3.41 \pm 0.12 $ & $ 3.31 $ \\
NGC4494 & & $ -24.16 \pm 0.11 $ & $ 30.8 $ & & $ 149 \pm 3 $  & & $ < 8.7\times 10^{26} $      & $ < 6.3\times 10^{27} $      & & \nodata           & $ 1.04 $ \\
NGC4526 & & $ -24.67 \pm 0.20 $ & $ 43.8 $ & & $ 263 \pm 18 $ & & $ 6.6\pm 1.2\times 10^{27} $ & $ < 6.1\times 10^{27} $      & & $ 2.54 \pm 0.31 $ & $ 2.45 $ \\
NGC4552 & & $ -24.20 \pm 0.14 $ & $ 25.4 $ & & $ 253 \pm 2 $  & & $ 2.9\pm 0.4\times 10^{28} $ & $ 1.9\pm 0.0\times 10^{28} $ & & $ 3.62 \pm 0.09 $ & $ 2.97 $ \\
NGC4555 & & $ -25.78 \pm 0.33 $ & $ 10.9 $ & & \nodata        & & $ < 2.8\times 10^{28} $      & $ < 2.0\times 10^{29} $      & & $ 3.19 \pm 0.22 $ & \nodata \\
NGC4564 & & $ -22.94 \pm 0.17 $ & $ 19.9 $ & & $ 158 \pm 2 $  & & $ < 6.7\times 10^{26} $      & $ < 4.8\times 10^{27} $      & & \nodata           & $ 4.09 $ \\
NGC4621 & & $ -24.56 \pm 0.20 $ & $ 32.9 $ & & $ 225 \pm 3 $  & & $ < 10.0\times 10^{26} $     & $ < 7.2\times 10^{27} $      & & $ 2.66 \pm 0.25 $ & $ 2.60 $ \\
NGC4636 & & $ -24.41 \pm 0.13 $ & $ 59.3 $ & & $ 202 \pm 3 $  & & $ 2.7\pm 0.3\times 10^{28} $ & $ 1.8\pm 0.0\times 10^{28} $ & & $ 3.47 \pm 0.12 $ & $ 1.33 $ \\
NGC4649 & & $ -25.39 \pm 0.15 $ & $ 45.2 $ & & $ 334 \pm 3 $  & & $ 9.8\pm 1.4\times 10^{27} $ & $ 1.3\pm 0.0\times 10^{28} $ & & $ 3.39 \pm 0.12 $ & $ 3.49 $ \\
NGC4697 & & $ -23.98 \pm 0.14 $ & $ 42.4 $ & & $ 173 \pm 2 $  & & $ < 4.1\times 10^{26} $      & $ < 6.6\times 10^{27} $      & & $ 3.65 \pm 0.11 $ & $ 0.60 $ \\
NGC5018 & & $ -25.27 \pm 0.33 $ & $ 15.6 $ & & $ 214 \pm 8 $  & & $ < 4.8\times 10^{27} $      & $ < 8.0\times 10^{28} $      & & $ 2.50 \pm 0.31 $ & $ 0.29 $ \\
NGC5044 & & $ -24.76 \pm 0.28 $ & $ 25.3 $ & & $ 238 \pm 8 $  & & $ 4.0\pm 1.0\times 10^{28} $ & $ < 4.9\times 10^{28} $      & & $ 3.17 \pm 0.13 $ & $ 0.38 $ \\
NGC5102 & & $ -21.09 \pm 0.14 $ & $ 79.3 $ & & $ 66 \pm 4 $   & & $ 6.1\pm 1.5\times 10^{25} $ & $ < 1.4\times 10^{27} $      & & \nodata           & $ 0.17 $ \\
NGC5171 & & $ -24.95 \pm 0.33 $ & $ 10.8 $ & & \nodata        & & $ < 2.9\times 10^{28} $      & $ < 2.1\times 10^{29} $      & & \nodata           & \nodata \\
NGC5532 & & $ -26.33 \pm 0.33 $ & $ 16.0 $ & & $ 293 \pm 18 $ & & $ 5.8\pm 1.7\times 10^{31} $ & $ 1.6\pm 0.1\times 10^{31} $ & & $ 3.11 \pm 0.25 $ & \nodata \\
NGC5845 & & $ -22.96 \pm 0.21 $ & $ 4.9 $  & & $ 234 \pm 8 $  & & $ < 2.0\times 10^{27} $      & $ < 1.4\times 10^{28} $      & & \nodata           & $ 0.84 $ \\
NGC5846 & & $ -25.04 \pm 0.20 $ & $ 34.5 $ & & $ 239 \pm 3 $  & & $ 1.6\pm 0.3\times 10^{28} $ & $ < 1.3\times 10^{28} $      & & $ 2.97 \pm 0.15 $ & $ 0.84 $ \\
NGC6482 & & $ -25.48 \pm 0.33 $ & $ 12.6 $ & & $ 303 \pm 9 $  & & $ < 1.0\times 10^{28} $      & $ < 7.5\times 10^{28} $      & & $ 2.64 \pm 0.31 $ & \nodata \\
NGC7052 & & $ -25.66 \pm 0.33 $ & $ 21.8 $ & & $ 271 \pm 9 $  & & $ 1.2\pm 0.4\times 10^{30} $ & $ 6.7\pm 0.6\times 10^{29} $ & & $ 2.72 \pm 0.31 $ & \nodata \\
NGC7618 & & $ -25.40 \pm 0.33 $ & $ 11.6 $ & & \nodata        & & $ 2.7\pm 0.8\times 10^{29} $ & $ < 1.3\times 10^{29} $      & & $ 2.76 \pm 0.31 $ & \nodata \\
\enddata
\tablenotetext{a}{2MASS data from the extended source catalog \citep{2MASS}. $K$-band absolute magnitude $M_{\rm K}$ and $J$-band effective radius \RJ\ in $\kpc$.}
\tablenotetext{b}{Velocity dispersion (in ${\rm km}\, \s^{-1}$) from LEDA \citep{LEDA}.}
\tablenotetext{c}{\LNVSS (in $\erg \, \s^{-1}$): $20\cm$ radio continuum luminosity from NVSS \citep{NVSS}; \Lsix\ (in $\erg \, \s^{-1}$): Combination of $6\cm$ radio continuum luminosities from \citet{GB6cm}, \citet{PMN6cm}, and \citet{Becker6cm}. }
\tablenotetext{d}{Projected galaxy density \rhomass\ is scaled logarithmically and in units of $\mpc^{-2}$, Tully galaxy density \rhotully\ is scaled linearly and in units of $\mpc^{-3}$.}
\end{deluxetable*}


\subsection{Temperature Profiles}\label{s5.tempprofile}

To produce radial temperature profiles, we divide the X-ray counts
image of each galaxy into elliptical annuli, according to the X-ray
ellipticity profiles computed in Paper I. For those galaxies with
insufficient signal to fit ellipses, we revert to circular annuli. We
find no evidence that this choice affects our results in any way. We
adapt the width of our annuli to contain a minimum of 900 counts above
the background level, which we determine by the appropriately rescaled
Markevitch blank-sky background
files\footnote{http://cxc.harvard.edu/cal/Acis/Cal\_prods/bkgrnd/acisbg/COOKBOOK}.

We then extract a source and background spectrum for each annulus and
fit them with a two-component model using the CIAO analysis package
Sherpa. The first component consists of an APEC
\footnote{Astrophysical Plasma Emission Code} plasma model to
represent the hot gas emission. A quantitative comparison with its
better-known predecessor, the Mekal model, shows nearly identical
fitting results. We fix the gas metallicity at the solar abundance
value. Unresolved point sources are represented by a power-law model
with the power law index fixed at 1.6. This ``universal'' spectral
model is an adequate representation for the emission of low-luminosity
low-mass X-ray binaries, as demonstrated in Paper I and determined
independently by \citet{Irwin03}. We also add a multiplicative
absorption component, for which we fix the hydrogen column density to
the Galactic value, evaluated at the target position with the CIAO
tool {\it Colden}\footnote{http://cxc.harvard.edu/toolkit/colden.jsp}.

We repeat our spectral analysis for a few objects with the gas
abundance as a free parameter, and find that our choice to fix them to
the solar value does not affect the fitted temperature. Since the
metallicity is poorly constrained by the fits in low signal-to-noise
systems, we fix the metallicity for all of our galaxies, in order not
to introduce systematic differences in the analysis.

\section{Results}
\label{s5.temperature}

\subsection{Radial Temperature Profile Types}

\begin{figure}
\begin{center}
\includegraphics[width=0.45\textwidth]{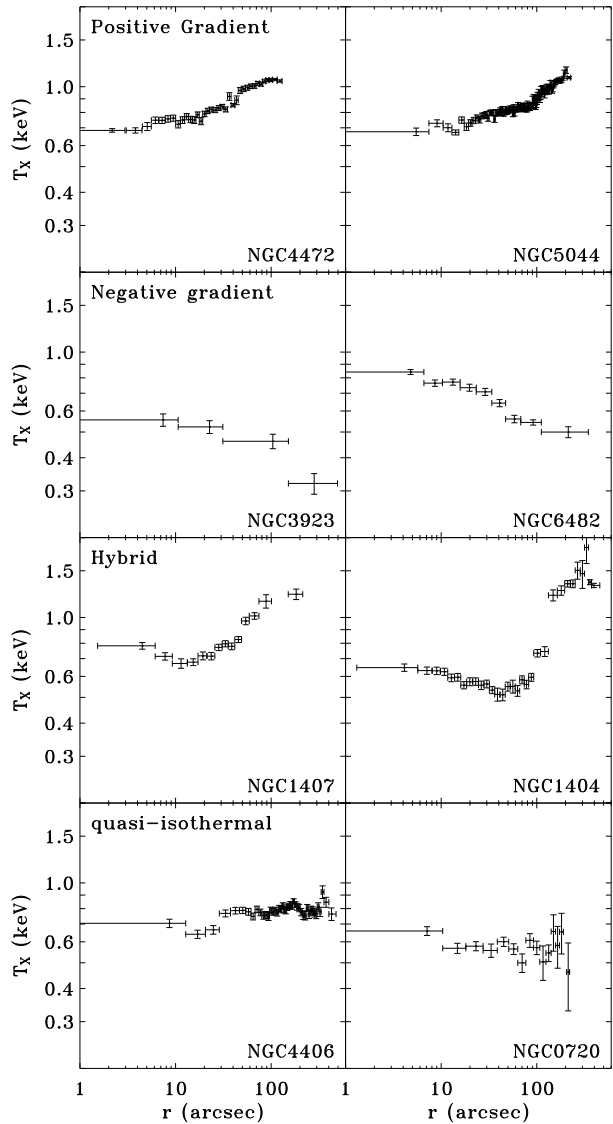}
\end{center}
\caption{Examples of different projected temperature profiles as a
function of radius. Temperature profile types can be divided into 4
major groups (top to bottom rows): (1) Positive gradient (outwardly
rising) at all radii; (2) Negative gradient (outwardly falling) at all
radii; (3) Hybrid, negative gradient in the core and positive gradient
at larger radii; (4) Quasi-isothermal, no apparent temperature change
with radius. The complete set of temperature profiles is available
online for all 36 galaxies with two or more valid temperature profile
points. \label{f.temptypes}}
\end{figure}

We categorize the observed temperature profiles into four major
groups, described below. Two examples from each are shown in Figure
\ref{f.temptypes}. (Note that the distinctions between the groups are
not always clear cut.)

\paragraph{Positive Gradients.} 
These temperature profiles show a positive gradient at all radii,
i.e. the temperature continuously rises outwardly. These profiles
resemble those found in clusters of galaxies, which generally harbor
cool cores.

\paragraph{Negative Gradients.} 
The temperature profiles show a negative gradient at all radii,
i.e. temperatures monotonically decline outward. This phenomenon is
less well-known, and has been reported only recently
\citep{RandallXMMNGC4649,FukazawaMassprofiles,HumphreyDarkmatter,
PonmanNGC6482}.

\paragraph{Hybrid.} 
These peculiar cases exhibit a dramatic change in the temperature
gradients. The gradient changes its sign from negative to positive at
some intermediate radius, generally between $1-3R_{\rm J}$. These
galaxies have warm centers, outside of which their temperatures drop
to a minimum level and rise back up again. These profiles have first
been noted by \citet{HumphreyDarkmatter}.

\paragraph{Quasi-Isothermal.} 
The radial temperature profiles in this category are consistent with
being almost flat at all radii. These galaxies form the transition
point between galaxies with positive and negative gradients.

We observe only hybrid temperature profiles that change their gradient
from negative to positive. Some cooling flow clusters have been found
to exhibit the opposite behavior \citep{PiffarettiClusters}. Their
temperature profiles show a cool center, then rise to a peak
temperature and fall back down on the outskirts. This ``break''
usually happens at around 10\% of the virial radius, which is larger
than the radii that we are probing in normal galaxies. A {\it ROSAT}
study by \citet{OSullivan} exhibits similar trends for elliptical
galaxies at larger radii.

We split the profiles into two radial regions and analyze the inner
and outer temperature gradients separately. As most hybrid profiles
exhibit their turnover in slope somewhere around $2R_{\rm J}$, we use
this radius as the boundary between our two regions. Accordingly, we
define the inner region from outside the central point source
extending out to $2R_{\rm J}$ and the outer region between $2-4R_{\rm
J}$. We then fit each part of the profile with a power law to derive
effective temperature gradients for each region. We will refer to the
logarithmic gradients ${\rm d}\ln T /{\rm d} \ln R$ evaluated within
2\RJ\ and from $2-4\,\RJ$ as $\alpha_{02}$ and $\alpha_{24}$,
respectively.

The best-fit values for \ain\ and \aout\ are listed in Table
\ref{t.tempprop}. The reported errors are the formal $1\sigma$
statistical errors obtained from the fitting procedure. For cases with
only 2 valid data points within the fitting range, we use the
difference between these two points to derive a gradient; the errors
are derived by propagating the statistical errors of the individual
temperature measurements.

\begin{figure}
\begin{center}
\includegraphics[width=0.45\textwidth]{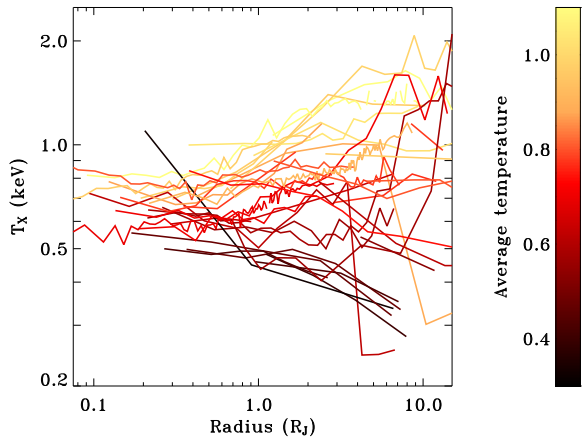}
\end{center}
\caption{Combined plot of all projected temperature profiles, as a
function of radius scaled by the $J$-band effective radius $R_{\rm
J}$. Error bars are omitted for clarity; for typical error estimates,
refer to Figure \ref{f.temptypes}. The profiles are colored according
to the luminosity weighted average temperature of the galaxy within 3
optical radii \TX, as indicated by the color scale bar. The
temperature gradient changes continuously from positive gradients at
the top to isothermal and hybrid profiles to negative gradient
profiles at the bottom, along with the average
temperature. \label{f.tempprofiles}}
\end{figure}

\subsection{The Inner Temperature Gradient $\alpha_{02}$}

\begin{figure}
\begin{center}
\includegraphics[width=0.45\textwidth]{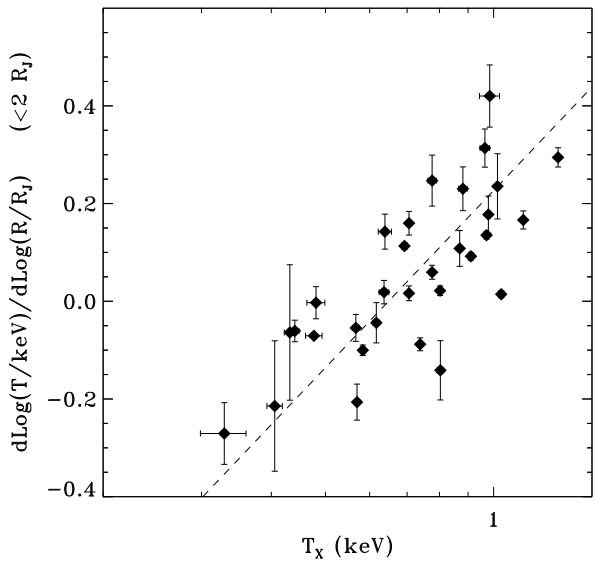}
\end{center}
\caption{Inner temperature gradient within $2\,R_{\rm J}$ as a
function of the average luminosity weighted temperature within 3
optical radii. The dashed line indicates the best
fit. \label{f.txtslope02}}
\end{figure}

Figure \ref{f.tempprofiles} shows the compilation of
all profiles overlaid in one plot, with the
radial axes scaled by their $J$-band effective radius $\RJ$. This plot
already clearly indicates the absence of any real bimodality in the
temperature profiles.  
Each profile is colored according to its luminosity weighted
temperature $T_{\rm X}$ within 3 optical radii. The coloring changes
smoothly from top to bottom, indicating overall positive gradients for
intrinsically hotter galaxies and negative gradients for cooler
galaxies. We plot $\alpha_{02}$ as a function of $T_{\rm X}$ in Figure
\ref{f.txtslope02}. Again, there is no sign of bimodality in $\alpha_{02}$.
The plot shows a tight correlation, significant at the $9.1\sigma$ level,
reflecting the fact that we observe only a
very small range in central temperatures between $0.6-0.7\kev$ (Figure
\ref{f.tempprofiles}). Thus, any average temperature will obviously be
strongly correlated with the gradient as well. Consistently, the fit
suggests that the transition from positive to negative inner
temperatures gradient occurs at a mean temperature around $0.64\kev$.

To establish the underlying cause for the negative inner temperature
gradients, we perform a correlation analysis with various galaxy
properties, the most interesting of which are listed in the upper half
of Table \ref{t.gradcorrelations}: X-ray gas luminosities $L_{\rm
X,gas}$, absolute $K$ magnitudes $M_{\rm K}$, central velocity
dispersions $\sigma$, radio luminosities at 20cm ($L_{\rm NVSS}$) and
6cm ($L_{\rm6cm}$), and environmental measures of local galaxy
density, \rhomass\ and \rhotully.

We assess the correlation of $\alpha_{02}$ with each of these properties
using the linear fitting algorithm \texttt{bandfit} (see Appendix of
Paper II). This algorithm models the distribution of $(x,y)$ points as
a linear band with a Gaussian intrinsic width. The model is fitted by
maximizing the likelihood of the data; this is similar to fitting a
straight line by minimizing the error-weighted perpendicular residuals.
The standard error $\sigma$ in the best-fit slope is obtained from the
covariance matrix, and the significance of the correlation is the
number of $\sigma$ by which the slope differs from zero.
In these fits, the quantities assigned as abscissae are
scaled logarithmically (except for $M_{\rm K}$, which is
already intrinsically logarithmic), while the ordinate \ain\ is
scaled linearly. The top half of Table \ref{t.gradcorrelations} lists
the best-fit parameters and the statistical significance of the correlation.
The fitted parameters are omitted where the significance is $<2\sigma$.
The tabulated fits are obtained from the full data set including upper
limits, but we also include in the last column the significance of the
correlation obtained with the upper limits omitted.\footnote{The exception
is $L_{\rm GB6}$, for which we give the parameters for the fit without upper
limits. This is because the number of galaxies with upper limits only
(49) greatly outweighs that of galaxies with actual detections (15).}
The $x_0$ column indicates the transition point from negative to positive
temperature gradients in each fit.

\begin{figure*}
\begin{center}
\includegraphics[width=0.4\textwidth]{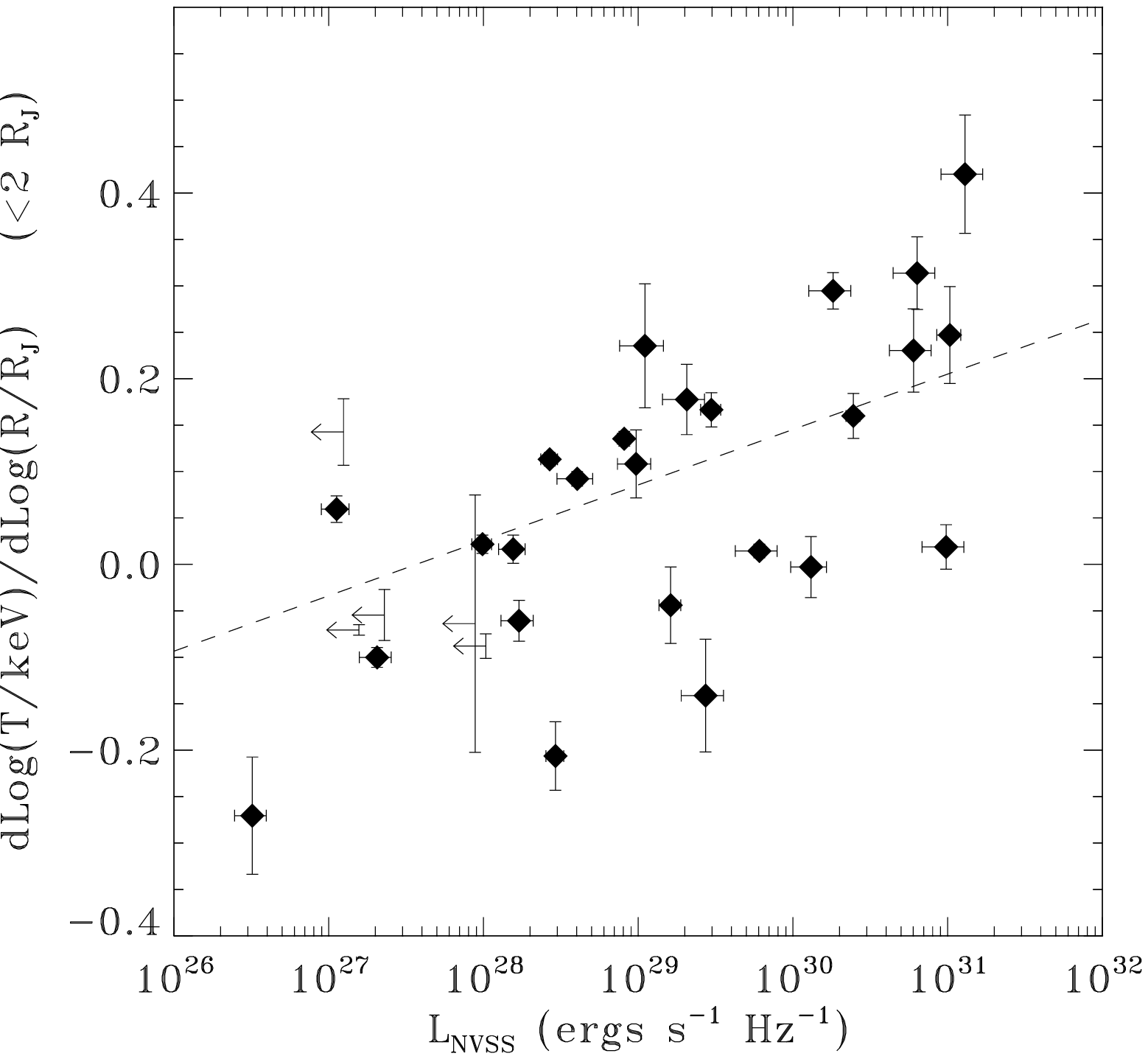}
\includegraphics[width=0.4\textwidth]{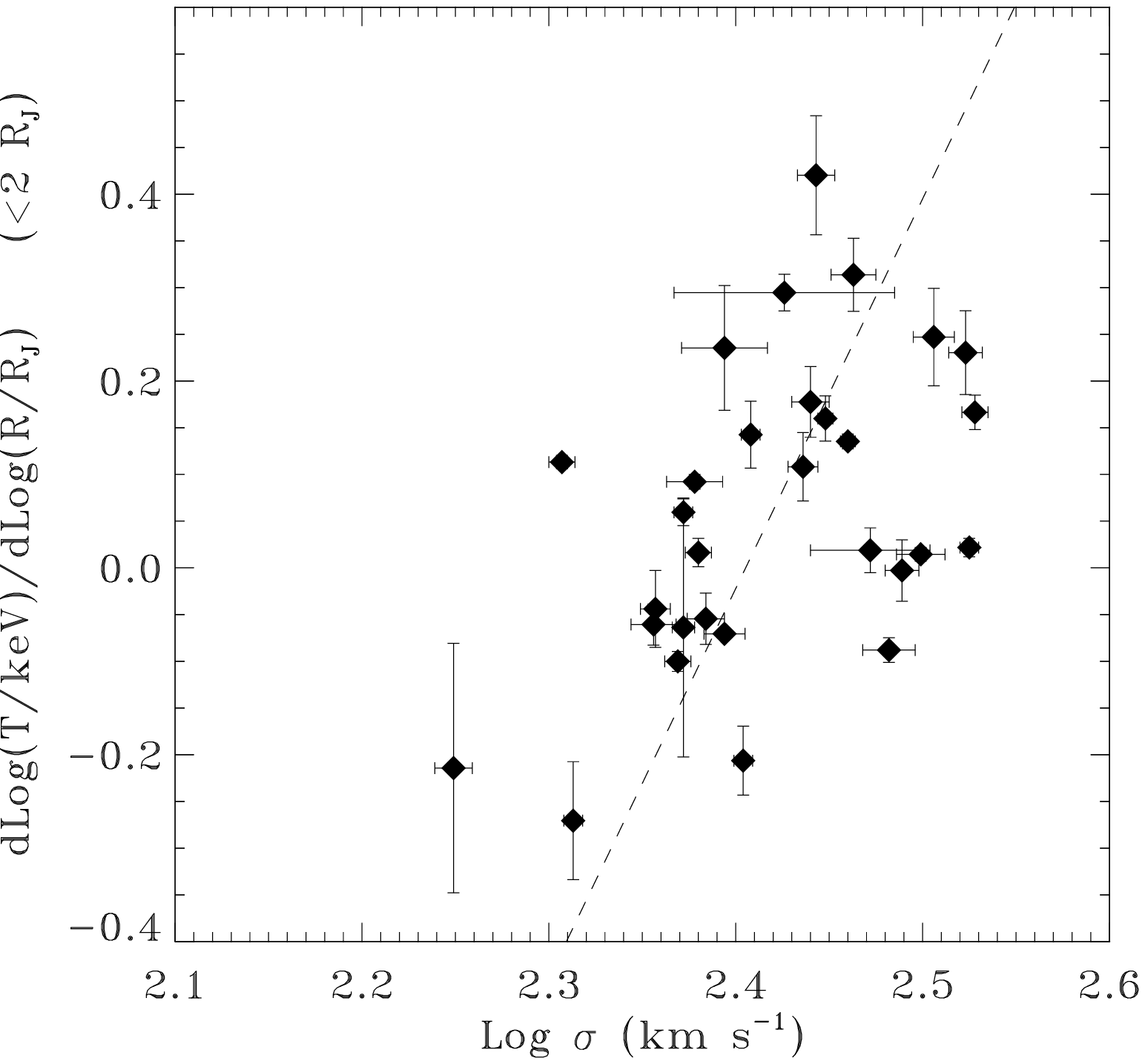}
\includegraphics[width=0.4\textwidth]{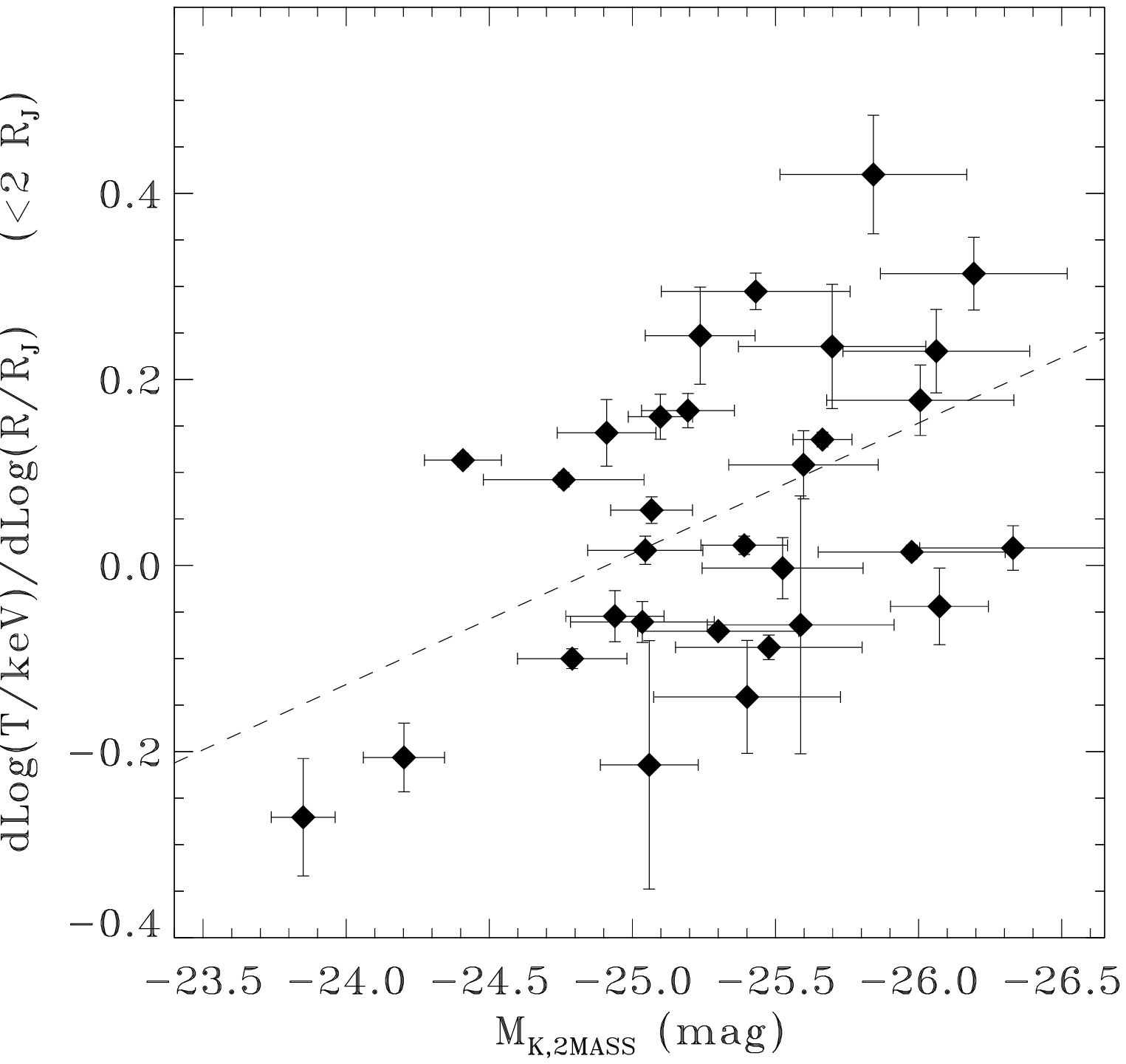}
\includegraphics[width=0.4\textwidth]{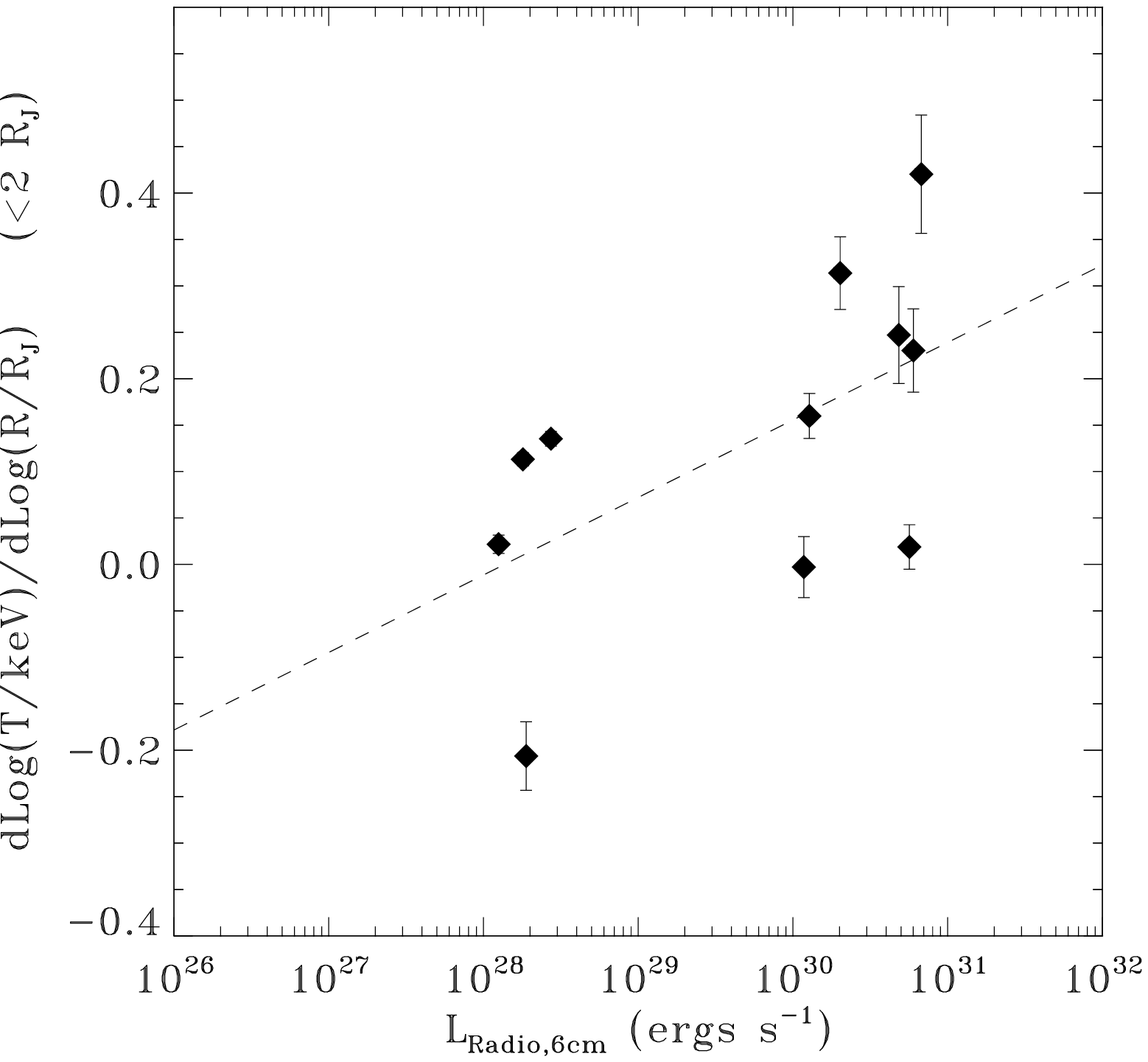}
\end{center}
\caption{Inner temperature gradient within $2\,R_{\rm J}$ as a
function of $20\cm$ NVSS radio luminosity (top left), central velocity
dispersion (top right), absolute $K$ magnitude (or stellar mass,
bottom left) and $6\cm$ radio luminosity (bottom right). The dashed
lines indicate the best-fit correlations from bandfit, as given in
Table \ref{t.gradcorrelations}. \label{f.kmaglradiotslope02} }
\end{figure*}

As the table and Figure \ref{f.kmaglradiotslope02} show, the strongest
correlations with \ain\ are found with the 20 cm NVSS radio luminosity
\LNVSS, the velocity dispersion $\sigma$, and the absolute K-magnitude
\MK. All of these three correlations are of roughly the same
significance, with comparable intrinsic widths, as Figure
\ref{f.kmaglradiotslope02} shows. The correlation with the $6\cm$
radio luminosity $L_{\rm GB6}$ seems equally tight, but has a lower
significance, due to the smaller sample size with $6\cm$ radio
luminosity measurements.

\begin{figure}
\begin{center}
\includegraphics[width=0.45\textwidth]{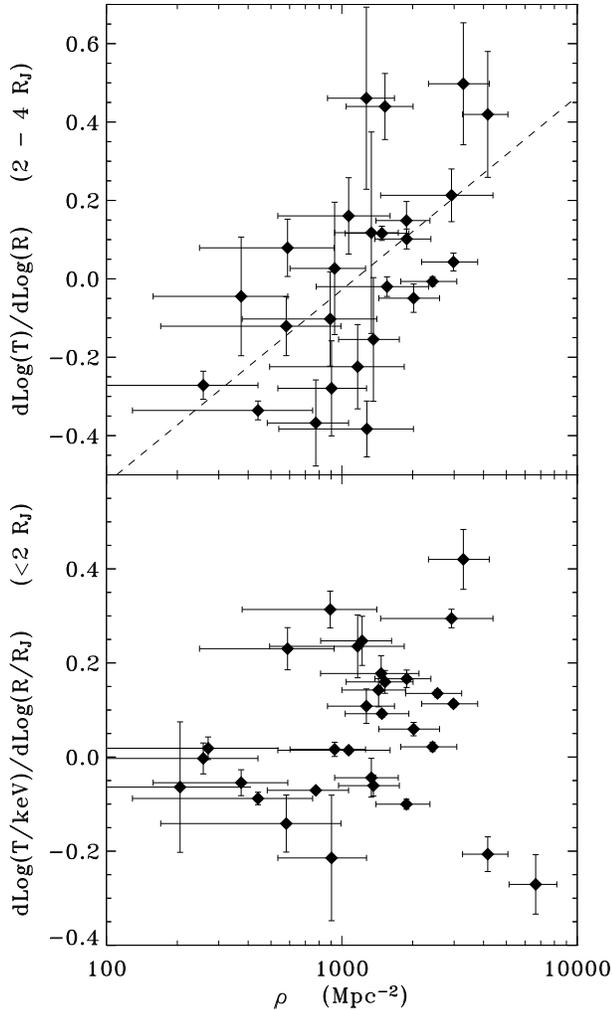}
\end{center}
\caption{Projected galaxy number density \rhomass\ vs. outer (\aout,
top panel) and inner (\ain, bottom panel) temperature gradients. Note
that \ain\ is evidently unaffected by environment, whereas \aout\
depends strongly on the density of nearby systems, suggesting the
influence of hot ambient
gas.\label{f.tempenvironment}}
\end{figure}

We do not find {\it any} evidence that \ain\ is correlated with the
environmental galaxy densities \rhomass\ or \rhotully. The
\rhomass--\ain\ plot is shown in the bottom panel of Figure
\ref{f.tempenvironment}. We conclude that the inner temperature
gradients are neither the result of interactions with neighbor galaxies, nor
with ambient intragroup
or intracluster gas. Instead, we find that they are connected to
intrinsic galaxy properties. We can generally characterize galaxies
with negative inner temperature gradients as being smaller, optically
fainter galaxies with lower velocity dispersions, lower X-ray gas
luminosities, lower average temperatures, and lower radio luminosities
than their positive gradient counterparts.

Unfortunately, {\it all} of these galaxy properties are intimately
connected with each other through well-known correlations such as the
\TX--$\sigma$ relation \citep[e.g.][]{OSullivan}, the Faber-Jackson
relation \citep{FaberJackson}, the \Lxgas--\TX\ relation
\citep[e.g.][]{OSullivan} and the $L_{\rm Radio}$--$\sigma$ relation
\citep[e.g.][]{SnellenLsigma}. This makes it difficult to distinguish
between fundamental correlations that are really responsible for
determining the inner temperature structure and others that are simply
``riding along'' via other correlations. 

\begin{figure}
\begin{center}
\includegraphics[width=0.45\textwidth]{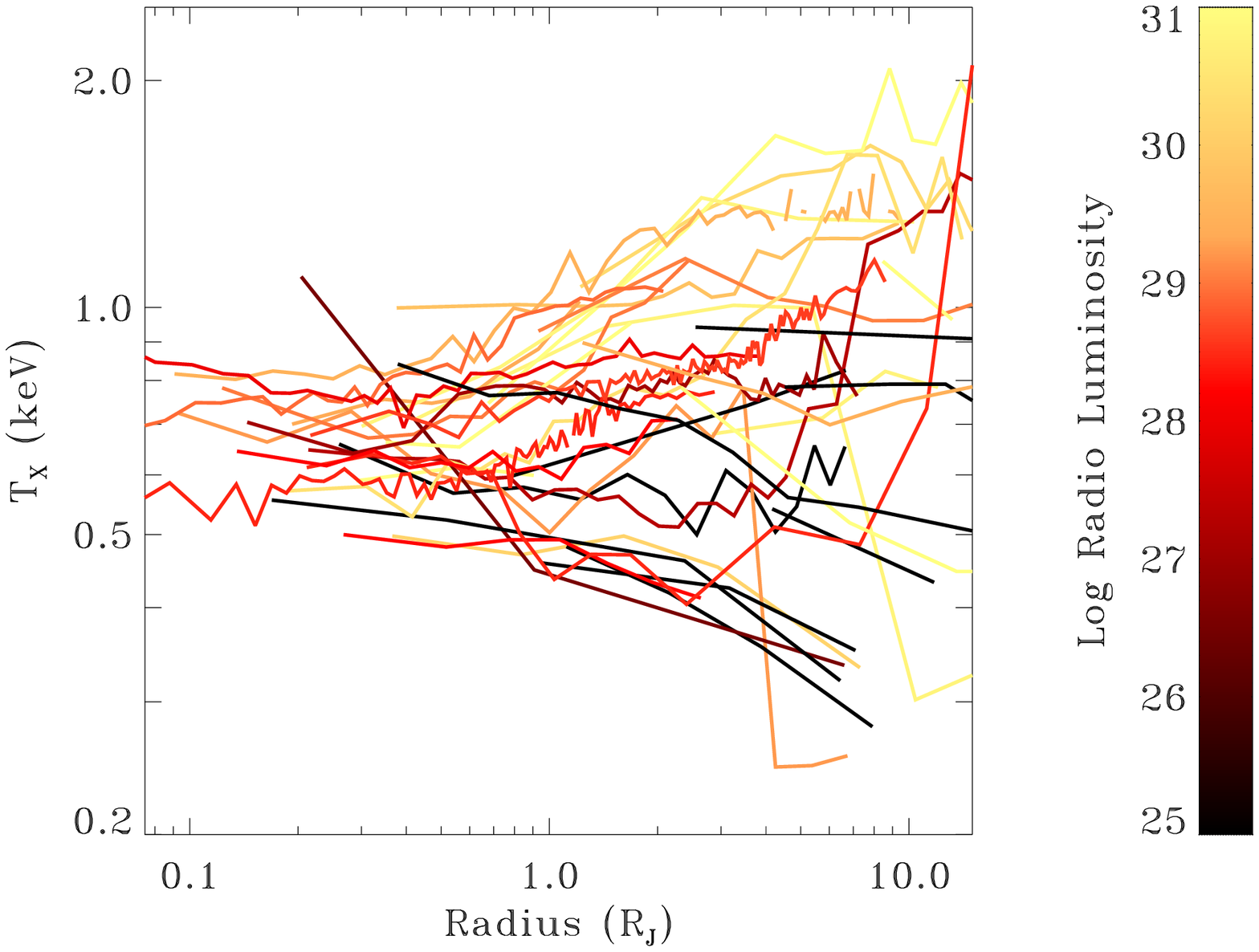}
\end{center}
\caption{Temperature profiles, as in Figure \ref{f.tempprofiles}, but
with colors indicating NVSS $20\cm$ continuum radio luminosity within
$3 R_{\rm J}$,
as shown by the color bar. The radio luminosity changes
continuously as profiles change from positive to negative
temperature gradients. \label{f.tempprofilesradio}}
\end{figure}

We check the robustness of our results by deriving inner temperature
gradients for different cutoff radii and find that all of the observed
trends are confirmed, as long as the cutoff-radius does not exceed
$\sim 3\RJ$. In particular, we find that for smaller cutoff-radii
(e.g. 1\RJ), the significance of the correlations with $\sigma$, \MK\
and \Lxgas\ slightly decreases, while the correlations with the radio
luminosities \LNVSS\ and \Lsix\ strengthen even further. This may suggest
that the correlations with radio luminosities are intrinsically the
strongest. Figure \ref{f.tempprofilesradio} shows a combined plot of
all temperature profiles, similar to Figure \ref{f.tempprofiles}, but
this time colored according to the NVSS radio luminosities. A trend
with radio luminosity is clearly evident. We will discuss the
implications of our results in \S\ref{s5.discussion}.

\begin{deluxetable*}{llrrrcc}
\ssp
\tablewidth{0pt}
\tablecaption{Correlations involving inner and outer temperature gradients. \label{t.gradcorrelations}}
\tablehead{
\colhead{$y$} & \colhead{$x$} & \colhead{$a$} & \colhead{$b$} & \colhead{$x_0$} & \colhead{Significance}  & \colhead{Significance} \\
 &  &  &  &  & \colhead{(with limits)} & \colhead{(without limits)} 
}
\startdata
$\alpha_{02}$ & $\log L_{\rm NVSS}$                 & $ 0.060\pm0.019$ & $-1.65\pm0.02$ & $ 27.4$ & $3.2\sigma$  & $4.3\sigma$  \\
$\alpha_{02}$ & $\log \sigma$                       & $ 4.171\pm1.192$ & $-10.0\pm0.01$ & $ 2.40$ & $3.6\sigma$  & $3.6\sigma$  \\
$\alpha_{02}$ & $M_{\rm K}$                         & $-0.141\pm0.044$ & $-3.50\pm0.02$ & $-24.8$ & $3.2\sigma$  & $3.2\sigma$  \\
$\alpha_{02}$ & $\log L_{\rm GB6}$\tablenotemark{*} & $ 0.083\pm0.036$ & $-2.35\pm0.04$ & $ 28.3$ & $0.4\sigma$  & $2.3\sigma$  \\
$\alpha_{02}$ & $\log L_{\rm X,gas}$                & \nodata          & \nodata        & \nodata & $1.0\sigma$  & $1.0\sigma$  \\
$\alpha_{02}$ & $\log \rho$                         & \nodata          & \nodata        & \nodata & $0.7\sigma$  & $0.7\sigma$  \\
$\alpha_{02}$ & $\log \rho_{\rm Tully}$             & \nodata          & \nodata        & \nodata & $0.4\sigma$  & $0.4\sigma$  \\
\\
$\alpha_{24}$ & $\log \rho$                         & $ 0.735\pm0.223$ & $-2.28\pm0.03$ & $ 3.10$ & $4.4\sigma$  & $4.4\sigma$  \\
$\alpha_{24}$ & $\log \rho_{\rm Tully}$             & $ 0.272\pm0.102$ & $ 0.04\pm0.04$ & $ 0.15$ & $2.8\sigma$  & $2.8\sigma$  \\
$\alpha_{24}$ & $M_{\rm K}$                         & \nodata          & \nodata        & \nodata & $1.6\sigma$  & $1.6\sigma$  \\
$\alpha_{24}$ & $\log \sigma$                       & \nodata          & \nodata        & \nodata & $0.5\sigma$  & $0.5\sigma$  \\
$\alpha_{24}$ & $\log L_{\rm GB6}$                  & \nodata          & \nodata        & \nodata & $<0.1\sigma$ & $0.6\sigma$  \\
$\alpha_{24}$ & $\log L_{\rm NVSS}$                 & \nodata          & \nodata        & \nodata & $<0.1\sigma$ & $0.3\sigma $ \\
$\alpha_{24}$ & $\log L_{\rm X,gas}$                & \nodata          & \nodata        & \nodata & $<0.1\sigma$ & $<0.1\sigma$ \\

\enddata

\tablecomments{Results are listed in order of decreasing correlation
significance. Parameters refer to linear fits of the form $y=ax+b$
(for correlations of $>2\sigma$ significance). $x_0$ denotes the
point where the fit yields 0, i.e. where the temperature gradients
change sign. The last two columns quote the correlation significances
with and without upper and lower limits included, respectively.}

\tablenotetext{*}{The fitted parameters for the $L_{\rm GB6}$ --
$\alpha_{02}$ correlation are quoted for the fit without upper and
lower limits included.}
\end{deluxetable*}

\subsection{The Outer Temperature Gradient $\alpha_{24}$}\label{s5.touter}

\begin{figure}
\begin{center}
\includegraphics[width=0.45\textwidth]{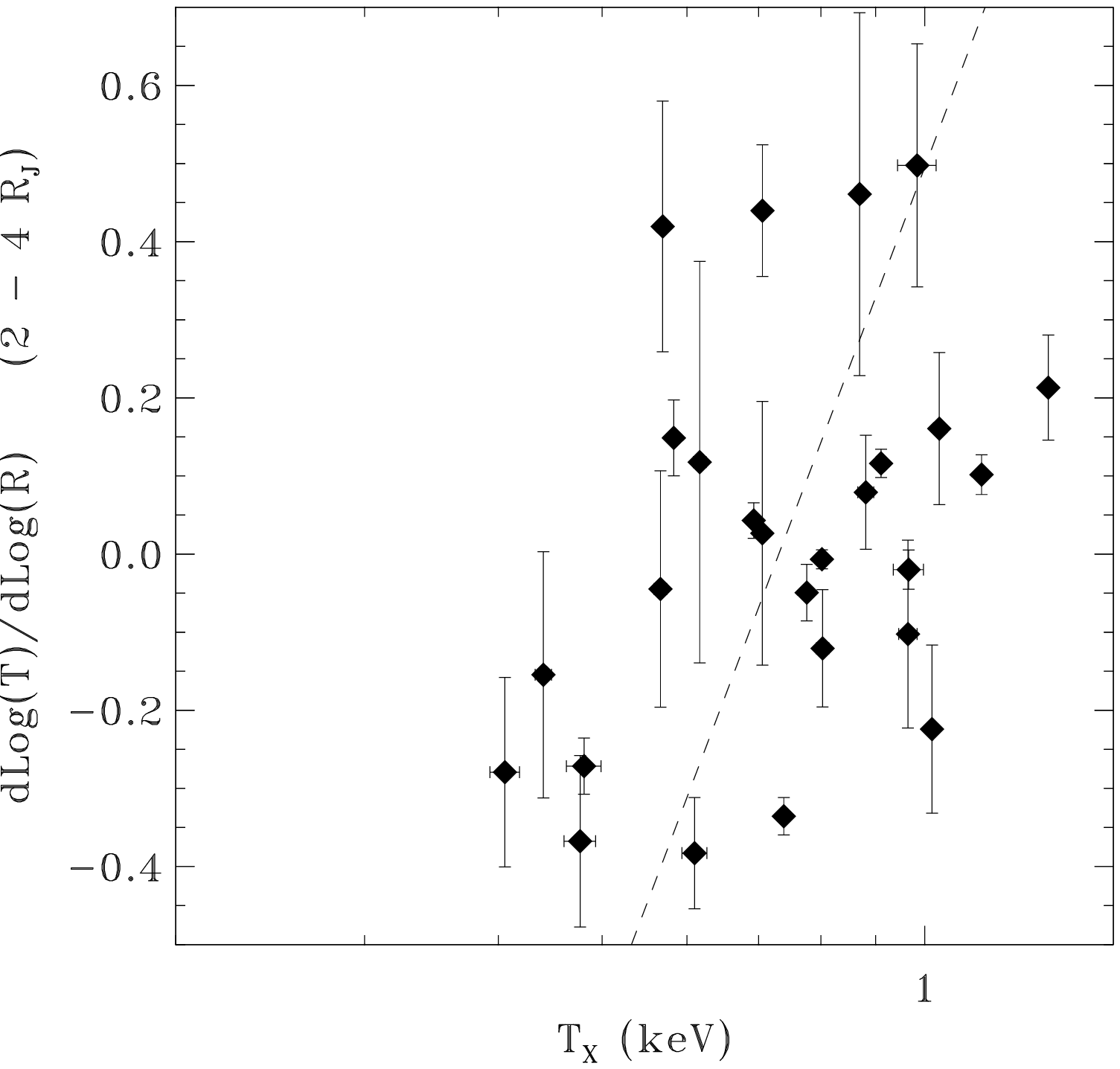}
\end{center}
\caption{Outer temperature gradient within $2-4\,R_{\rm J}$ as a
function of the average luminosity weighted temperature within
$3\,R_{\rm J}$. The dashed line indicates the best
fit. \label{f.txtslope24}
}
\end{figure}

We now look at the outer temperature gradient between $2$ and
$4\,R_{\rm J}$, and with what it is correlated. Like $\alpha_{02}$,
$\alpha_{24}$ is correlated with the average temperature within
$3\,R_{\rm J}$ (Fig. \ref{f.txtslope24}), though less strongly
($2.5\sigma$, compared with $9.1\sigma$). This trend is such that
galaxies with a hotter average temperature have stronger positive
temperature gradients. Some correlation is expected because these
quantities are not truly independent. As a consequence, the outer
temperature gradient is also weakly correlated ($2.4\sigma$) with the
inner temperature gradient. 

We now repeat the same analysis for the outer temperature gradient
\aout . The results of the correlation analysis are listed in the
bottom half of Table \ref{t.gradcorrelations}. Unlike \ain,
\aout\ does {\it not} depend on the intrinsic galaxy properties
\LNVSS, \Lsix, $\sigma$, or \MK. Instead, we find strong evidence that
\aout\ depends only on the environmental density parameters \rhomass\
and \rhotully. This trend with environment is statistically even
stronger than the one with the luminosity weighted temperature \TX,
even though those parameters are not independent measurements.

To check the robustness of these results, we repeat our analysis using
larger outer radial boundaries and confirm all trends. The
significance of the environmental dependence gets even stronger when
extending the analysis to larger radii. These relations are
strongest, when one fits temperature gradients to all radii beyond
2\RJ without imposing an outer radial limit. However, since
the gradients tend to get stronger with radius, and our galaxies have
very different cutoff radii owing to different surface brightness
profiles, we do not report the functional form of the fit, as it is
driven by the brightest galaxies. Nevertheless, this strengthens the
confidence in the observed correlation.

We conclude that the inner and outer temperature gradients are
essentially decoupled. While the inner gradient depends only on
intrinsic galaxy properties, the outer gradient shows no correlations
{\it but} with the environment.

\section{Discussion}\label{s5.discussion}

\subsection{Implications for Cooling Flows}

Steady-state cooling flow models have gone out of fashion recently due to
extensive work on galaxy clusters, which show insufficient amounts of
cooling gas at the center \citep[][]{PetersonCoolingspec}. These simple
models are unlikely to apply to X-ray bright
elliptical galaxies either. We showed in Paper II that the hot gas in
these systems is almost always disturbed, and we see evidence linking the
origin of these disturbances to the central AGN. 

However, it is far from proven that the same is true for
low-luminosity galaxies, in which we find negative inner
temperature gradients. Compressive heating during a gradually cooling
inflow of relatively cool gas may be able to offset radiative losses
for low-temperature gas in steep gravitational potentials
\citep[e.g.][]{MathewsReview}. This counter-intuitively results in a
cooling flow that gets heated during inflow and may even produce a hot
center, i.e. a negative gradient. We also find in Paper II
that these systems are generally less disturbed, which could be
consistent with a steady state cooling flow solution.

\citet{PonmanNGC6482} observe a falling temperature profile for the
fossil group candidate NGC~6482 and successfully fit a steady-state
cooling flow model with a reasonable cooling rate of $\dot M =
2M_\sun\,\yr^{-1}$. However, they derive the inner gravitational
potential from the X-ray profiles themselves assuming hydrostatic
equilibrium, which yields a steep inner potential gradient. They then
use this potential to fit the cooling flow model to the negative
temperature gradient. This could be circular reasoning: a central
temperature peak together with the assumption of hydrostatic
equilibrium implies a steep gravitational potential, which leads to
increased compressional heating \citep[e.g.][]{MathewsReview} and a
central temperature peak. In addition, we have argued in Paper I
that an assumption of hydrostatic equilibrium is generally not secure.
It would be much safer to derive the gravitational
potential from independent stellar dynamics, as the inner region is
most likely stellar-mass dominated, at least within two effective
radii \citep[e.g.][]{MamonDarkmatterI}. In any event, NGC~6482 is the
only negative-gradient object that has been successfully fitted with a
cooling flow model so far. Only modeling a more complete sample will
show if this idea can generally hold.

\subsection{Implications for the Existence of Circumgalactic Gas}

Negative temperature gradients have been recognized only recently
\citep{HumphreyDarkmatter,FukazawaMassprofiles,RandallXMMNGC4649,
PonmanNGC6482}. Because earlier observations revealed only positive
gradients, many theoretical flow models have been dismissed on the
grounds that they produce negative gradients
\citep[e.g.][]{MathewsReview}. Instead, theoretical effort has focused
on finding an explanation for the prevalence of positive
gradients. \citet{BrighentiCircumgas} argue that a hot circumgalactic
gas reservoir is able to reverse a negative temperature gradient. This
explanation is consistent with our observations that the outer
temperature gradient is correlated with the environment. Whether
models with circumgalactic gas can quantitatively account for the more
complex hybrid temperature profiles remains to be seen.

\subsection{Implications for Supernova Feedback}

Another means of producing negative temperature gradients involves
supernova (SN) feedback. Since star formation should be negligible in
elliptical galaxies, this mechanism would involve only contributions
from type Ia SN. Early proposed wind models involving SN feedback
\citep{BinneyCoolevolution} were later dismissed, since a main feature
was a negative temperature gradient throughout their evolution, which
had not been observed at that time. However, \citet{MathewsReview}
point out that these models are able to reproduce observed gas
profiles only for a very short period of time ($\sim 10^8\yr$) just
before a cooling catastrophe sets in. Furthermore, these models are
sensitive to the assumed SN rate, resulting in abrupt transitions to
SN driven winds, and thus require fine-tuning. This fine-tuning
problem can be circumvented by the presence of circumgalactic gas
\citep{MathewsReview}. However, our observations of purely negative
gradients without an outer rise in temperature may present a challenge
to this model, though we cannot exclude the possibility of an external
gas reservoir below our detection limit.

In addition, the predicted metallicities in galactic wind models are
generally super-solar, significantly exceeding the historically
observed extremely low abundances in the hot gas
\citep[][]{ArimotoIrondiscrepancy}. A {\it Chandra} spectral analysis
of abundance gradients in a sample of 28 elliptical galaxies by
\citet{Humphrey}, on the other hand, no longer shows strongly
sub-solar abundances. They attribute this difference to previously
imperfect modeling of the spectra, mainly caused by the neglect of the
unresolved point source component and attempting to fit
multi-temperature gas with a single-temperature model, the so-called
iron-bias \citep[e.g.][]{BuoteIronbias}. Nevertheless, they conclude
that their abundances are still far too low to be consistent with
galactic wind models, and favor the circulation flow model of
\citet{Circflow} instead.

However, if we consider only the energy input by SNIa feedback, we
find that SN feedback could play a role in heating the hot gas. The
average SNIa rate for elliptical and S0 galaxies is $r_{\rm
SN}=0.18\pm 0.06\, (100\yr)^{-1}\, (10^{10} L_{B\sun})^{-1}$
\citep{CappellaroSNrates}. With an average energy injection of $\sim
10^{51}$ ergs per SN, this results in a SN heating rate of $L_{\rm
SN}=5.7\times 10^{30} (L_{\rm B}/L_{\rm B, \sun})\, \erg \s^{-1} $. An
inspection of the \Lxgas--$L_{\rm B}$ diagram in Paper I shows that
$L_{\rm SN}>\Lxgas$ for galaxies below the blue luminosity where the
inner temperature gradients turn from negative to positive ($\sim
4\times 10^{10} L_{\rm B\sun}$). Although this is somewhat suggestive,
the large scatter in X-ray luminosity of almost 2 orders of magnitude
at a given blue luminosity, combined with the also rather large
scatter in the $L_{\rm B}$--\ain\ relation, make it impossible to tell
if this is simply coincidence.

However, if SN heating were the main cause, we would expect a
correlation between the inner temperature gradient and the SN heating
to X-ray cooling ratio ($L_{\rm SN}/\lxgas$), in the sense that
negative gradients would correspond to high ratios of heating to
cooling. We do not find such a correlation. $L_{\rm SN}/\lxgas$
correlates with the inner temperature gradient very weakly on the less
than $0.5\sigma$ level. We conclude that supernova feedback may be
important for balancing part of the radiative losses in X-ray faint
galaxies, but our analysis suggests that it is most likely not the
dominant factor.

\subsection{Implications for AGN feedback}

In Paper II we have measured the amount of asymmetry in the hot gas,
and find a strong correlation between asymmetry and AGN power. This
correlation persists all the way down to the weakest AGN luminosities
at the detection limit of the NVSS $20\cm$ survey.
We now find that the temperature structure is also strongly correlated
with the AGN luminosities, another hint toward the importance of AGN
feedback throughout the elliptical galaxy population.
However, we cannot completely rule out compressionally
heated cooling flows or SN feedback to explain the prevalence of
negative temperature gradients. Thus, we propose three possible
scenarios involving AGN feedback to explain our results:

\begin{enumerate}
\item
    Weak AGN with smaller black holes heat the ISM
    locally, while higher-luminosity sources feed powerful jets
    that distribute the heat globally by blowing large cavities into
    the ISM. This is consistent with the observation that smaller
    elliptical galaxies have rather weak AGN and generally less
    extended radio emission, and also in agreement with our findings
    from Papers I and II that the amount of asymmetry correlates with
    AGN luminosity. In this scenario, weak AGN would still be
    disturbing the gas, but on a scale and surface brightness level
    that is simply less detectable, resulting in a lower
    asymmetry. Negative temperature gradients could then be a sign of
    very localized heating by the central AGN.
\item
    AGN are responsible for globally heating the hot gas only in X-ray
    bright galaxies with positive temperature gradients. The onset of
    negative inner temperature gradients marks the point where AGN
    heating becomes unimportant, relative to other sources. These
    other sources could include compressional heating or supernovae.
\item
    The observed temperature
    gradients are snapshots of different stages of a time-dependent
    flow, which cyclically reverses the temperature gradient over
    time. If such solutions exist, it will be challenging for
    theoretical models to explain the fact that none of our galaxies
    exhibit central temperatures below $\sim 0.6\kev$. Thus, any
    cyclic solution has to keep the central temperature rather
    constant, while reversing the temperature gradient by heating or
    cooling only at large radii. The best chance to achieve this may
    be for galaxies to cycle through wind and inflow phases, possibly
    intimately correlated with the time-dependent AGN activity of the
    central black hole.
\end{enumerate}

The possible importance of AGN heating for elliptical galaxies has
also recently been pointed out by \citet{BestAGNcooling}. They combine
two empirical results to derive an estimate of time-averaged heating
by radio sources in galaxies. They use a result by
\citet{BirzanCavities} for galaxy clusters that empirically links the
$p{\rm d}V$ work associated with inflating X-ray cavities into the
intracluster medium with the observed $20\cm$ radio continuum power of
the associated radio source. Although this correlation exhibits
significant scatter, Best et al. derive a linear fit and use it to
convert their radio powers to mechanical energy. In an earlier study,
\citet{BestRadioloudAGN} find that the fraction of elliptical galaxies
hosting radio-loud AGN correlates with black hole mass and radio
luminosity. Assuming that all elliptical galaxies have AGN at their
centers, Best et al. interpret the fraction of galaxies with active
AGN as the fraction of time that they are turned on. By combining the
computed mechanical work per unit radio luminosity derived from
\citet{BirzanCavities} with the fraction of time the radio source is
turned on, Best et al. calculate the time-averaged mechanical heat
input of the AGN. A comparison with the \Lxgas--$L_{\rm B}$ relation
for normal ellipticals shows a remarkable agreement between the
time-average AGN heat input and the averaged radiative losses of
elliptical galaxies \citep[Figure 2]{BestAGNcooling}. This good
agreement is actually surprising, since the conversion factor from
radio power to mechanical energy has a rather large scatter and is
only based on observations of cluster cavities.

Further support for AGN heating has been provided by
\citet{FabianAccretionJet}, who measure the mechanical energy
associated with X-ray cavities in 9 X-ray luminous elliptical
galaxies. They compare this value to the Bondi accretion rate, which
they derive from deprojected density and temperature profiles,
evaluated at the accretion radius. Allen et al. find a tight
correlation between the Bondi accretion rate and the mechanical energy
injected into the ISM, and find that this energy input may be
sufficient to prevent the gas from cooling.

\subsection{What is so special about $\sim 0.6\kev$?}

A close inspection of Figure \ref{f.tempprofiles} shows a remarkably
small range in central temperature, which falls between $0.6$ to
$0.7\kev$. The upper limit owes its origin to our explicit exclusion
of brightest cluster galaxies, with higher temperatures, from our
sample. Including cluster cDs in our sample would add the missing
profiles, adding positive temperature gradients with higher central
temperatures.

However, the lower limit is quite mysterious. We find it unlikely that
this is simply a {\it Chandra} sensitivity effect. We know that our
temperature fits are sensitive to lower temperatures, as we can see
them in fits to the outer regions of the same objects. Conceivably,
this could represent a selection effect imposed on the {\it
Chandra} archive through the proposal process, which disfavors
observations of systems with lower temperature due to the
drop in instrument sensitivity at lower energy. We find this
explanation also difficult to believe, as galaxies with negative
gradients would have been characterized simply as having a lower mean
temperature, since {\it ROSAT} would not have been able to detect the
rise in temperature toward the center. However, we do see that the
faintest galaxies in our sample exclusively build the lower envelope
in the temperature profiles, with luminosity weighted temperatures of
$\sim 0.4\kev$. Thus, fainter galaxies could lower this envelope even
further, and with it the central temperature. The lower envelope may
also mark the transition to a galactic wind, which would render the
temperature gradient for these galaxies unobservable due to low gas
densities.

Nevertheless, something {\it is} special about $\sim 0.6\kev$. First,
we do not observe any central temperature below this value. Second,
all hybrid temperature profiles drop below $0.6\kev$ at some
intermediate radius and then rise back up again. And third, the best
fit for the \TX--\ain\ relation puts the transition between negative
and positive gradients at $0.64\kev$. Any flow model on the galaxy
scale has to be able to reproduce these properties.


\section{Conclusions}\label{s5.conclusions}

We have reported on the shape of temperature profiles in 36 normal elliptical
galaxies. 
These profiles show a variety of different profile types:
purely positive gradients, purely negative gradients, quasi-isothermal
and even hybrid profiles. To understand this complexity, we derive
mean temperature gradients for an inner region within $2\RJ$,
excluding the central point source, and an outer region between
$2-4\RJ$. We find that the outer temperature gradient is independent
of intrinsic galaxy properties, but a strong function of environment,
such that positive outer temperature gradients are restricted to
cluster and group environments. This suggests that the outer gradients
are caused by interaction with hotter ambient gas, whereas galaxies
with negative outer gradients are in less dense environments and lack
this intergalactic gas reservoir.

The inner temperature gradient, on the other hand, is completely
independent of the environmental
influence. Instead, we find that it is correlated with a number of
intrinsic galaxy properties; in decreasing order of significance, the
$20\cm$ radio luminosity, the central velocity dispersion, the
absolute $K$ magnitude, and the $6\cm$ radio luminosity.

The data cannot rule out the idea that negative gradients can be
produced by compressional heating in low-temperature systems, during a
slow cooling inflow in a steep gravitational potential.
SN feedback may also provide sufficient energy to offset
cooling in X-ray faint galaxies, but we find no direct evidence that
SN heating dominates.

Our preferred feedback model involves the central AGN. The 
inner temperature gradient is most strongly correlated with radio
luminositiy and central velocity dispersion, which may be
interpreted as a surrogate for black hole mass
\citep{TremaineBHsigma}. The nature of these correlations is such that
weak AGN hosts show negative temperature gradients, whereas more
luminous AGN exclusively live in positive gradient systems. Thus, we
propose three scenarios, to explain the observed features. (1) Weak
AGN distribute their heat locally, whereas luminous AGN heat the gas
more globally with their extended jets. (2) The onset of negative
gradients marks the point where AGN heating becomes unimportant, and
compressional heating or SN feedback becomes dominant. (3) A cyclic
model in which the AGN drives an outflow, which shuts the AGN
activity off until the flow reverses itself, fuels the black
hole and starts another cycle.

These findings are in agreement with the results from Paper I, which
showed that precise hydrostatic equilibrium does not hold for the hot gas in
elliptical galaxies, and established the prevalence of disturbances
in the X-ray gas morphology. The results of Paper II indicate that the
central AGN probably causes these disturbances. Combining these
results with the connection between the temperature structure and the radio
luminosity of the system produces a strong argument for the general
importance of AGN feedback in nearly all normal elliptical galaxies.


\acknowledgments We have made use of the HyperLEDA database
(http://leda.univ-lyon1.fr).  Support for this work was provided by
the National Aeronautics and Space Administration (NASA) through
Chandra Awards G01-2094X and AR3-4011X, issued by the {\em Chandra
X-Ray Observatory Center}, which is operated by the Smithsonian
Astrophysical Observatory for and on behalf of NASA under contract
NAS8-39073, and by National Science Foundation grant AST0407152.


\bibliographystyle{apj}
\bibliography{../bibtex/allreferences}

\begin{thebibliography}{33}
\expandafter\ifx\csname natexlab\endcsname\relax\def\natexlab#1{#1}\fi

\bibitem[{{Allen} {et~al.}(2006){Allen}, {Dunn}, {Fabian}, {Taylor}, \&
  {Reynolds}}]{FabianAccretionJet}
{Allen}, S.~W., {Dunn}, R.~J.~H., {Fabian}, A.~C., {Taylor}, G.~B., \&
  {Reynolds}, C.~S. 2006, \mnras, 372, 21

\bibitem[{{Arimoto} {et~al.}(1997){Arimoto}, {Matsushita}, {Ishimaru},
  {Ohashi}, \& {Renzini}}]{ArimotoIrondiscrepancy}
{Arimoto}, N., {Matsushita}, K., {Ishimaru}, Y., {Ohashi}, T., \& {Renzini}, A.
  1997, \apj, 477, 128+

\bibitem[{{Becker} {et~al.}(1991){Becker}, {White}, \& {Edwards}}]{Becker6cm}
{Becker}, R.~H., {White}, R.~L., \& {Edwards}, A.~L. 1991, \apjs, 75, 1

\bibitem[{{Best} {et~al.}(2006){Best}, {Kaiser}, {Heckman}, \&
  {Kauffmann}}]{BestAGNcooling}
{Best}, P.~N., {Kaiser}, C.~R., {Heckman}, T.~M., \& {Kauffmann}, G. 2006,
  \mnras, 368, L67

\bibitem[{{Best} {et~al.}(2005){Best}, {Kauffmann}, {Heckman}, {Brinchmann},
  {Charlot}, {Ivezi{\'c}}, \& {White}}]{BestRadioloudAGN}
{Best}, P.~N., {Kauffmann}, G., {Heckman}, T.~M., {Brinchmann}, J., {Charlot},
  S., {Ivezi{\'c}}, {\v Z}., \& {White}, S.~D.~M. 2005, \mnras, 362, 25

\bibitem[{{Binney} \& {Tabor}(1995)}]{BinneyCoolevolution}
{Binney}, J., \& {Tabor}, G. 1995, \mnras, 276, 663+

\bibitem[{{B{\^\i}rzan} {et~al.}(2004){B{\^\i}rzan}, {Rafferty}, {McNamara},
  {Wise}, \& {Nulsen}}]{BirzanCavities}
{B{\^\i}rzan}, L., {Rafferty}, D.~A., {McNamara}, B.~R., {Wise}, M.~W., \&
  {Nulsen}, P.~E.~J. 2004, \apj, 607, 800

\bibitem[{{Brighenti} \& {Mathews}(1998)}]{BrighentiCircumgas}
{Brighenti}, F., \& {Mathews}, W.~G. 1998, \apj, 495, 239+

\bibitem[{{Buote}(2000)}]{BuoteIronbias}
{Buote}, D.~A. 2000, \apj, 539, 172

\bibitem[{{Cappellaro} {et~al.}(1999){Cappellaro}, {Evans}, \&
  {Turatto}}]{CappellaroSNrates}
{Cappellaro}, E., {Evans}, R., \& {Turatto}, M. 1999, \aap, 351, 459

\bibitem[{{Condon} {et~al.}(1998){Condon}, {Cotton}, {Greisen}, {Yin},
  {Perley}, {Taylor}, \& {Broderick}}]{NVSS}
{Condon}, J.~J., {Cotton}, W.~D., {Greisen}, E.~W., {Yin}, Q.~F., {Perley},
  R.~A., {Taylor}, G.~B., \& {Broderick}, J.~J. 1998, \aj, 115, 1693

\bibitem[{{Diehl} \& {Statler}(2007)}]{DiehlGallery}
{Diehl}, S., \& {Statler}, T.~S. 2007, \apj, 668, 150

\bibitem[{{Diehl} \& {Statler}(2008)}]{DiehlAGN}
---. 2008, \apj, in press

\bibitem[{{Faber} \& {Jackson}(1976)}]{FaberJackson}
{Faber}, S.~M., \& {Jackson}, R.~E. 1976, \apj, 204, 668

\bibitem[{{Fukazawa} {et~al.}(2006){Fukazawa}, {Botoya-Nonesa}, {Pu}, {Ohto},
  \& {Kawano}}]{FukazawaMassprofiles}
{Fukazawa}, Y., {Botoya-Nonesa}, J.~G., {Pu}, J., {Ohto}, A., \& {Kawano}, N.
  2006, \apj, 636, 698

\bibitem[{{Gregory} {et~al.}(1996){Gregory}, {Scott}, {Douglas}, \&
  {Condon}}]{GB6cm}
{Gregory}, P.~C., {Scott}, W.~K., {Douglas}, K., \& {Condon}, J.~J. 1996,
  \apjs, 103, 427+

\bibitem[{{Humphrey} \& {Buote}(2006)}]{Humphrey}
{Humphrey}, P.~J., \& {Buote}, D.~A. 2006, \apj, 639, 136

\bibitem[{{Humphrey} {et~al.}(2006){Humphrey}, {Buote}, {Gastaldello},
  {Zappacosta}, {Bullock}, {Brighenti}, \& {Mathews}}]{HumphreyDarkmatter}
{Humphrey}, P.~J., {Buote}, D.~A., {Gastaldello}, F., {Zappacosta}, L.,
  {Bullock}, J.~S., {Brighenti}, F., \& {Mathews}, W.~G. 2006, \apj, 646, 899

\bibitem[{{Irwin} {et~al.}(2003){Irwin}, {Athey}, \& {Bregman}}]{Irwin03}
{Irwin}, J.~A., {Athey}, A.~E., \& {Bregman}, J.~N. 2003, \apj, 587, 356

\bibitem[{{Jarrett} {et~al.}(2000){Jarrett}, {Chester}, {Cutri}, {Schneider},
  {Skrutskie}, \& {Huchra}}]{2MASS}
{Jarrett}, T.~H., {Chester}, T., {Cutri}, R., {Schneider}, S., {Skrutskie}, M.,
  \& {Huchra}, J.~P. 2000, \aj, 119, 2498

\bibitem[{{Khosroshahi} {et~al.}(2004){Khosroshahi}, {Jones}, \&
  {Ponman}}]{PonmanNGC6482}
{Khosroshahi}, H.~G., {Jones}, L.~R., \& {Ponman}, T.~J. 2004, \mnras, 349,
  1240

\bibitem[{{Mamon} \& {{\L}okas}(2005)}]{MamonDarkmatterI}
{Mamon}, G.~A., \& {{\L}okas}, E.~L. 2005, \mnras, 362, 95

\bibitem[{{Mathews} \& {Brighenti}(2003)}]{MathewsReview}
{Mathews}, W.~G., \& {Brighenti}, F. 2003, \araa, 41, 191

\bibitem[{{Mathews} {et~al.}(2004){Mathews}, {Brighenti}, \&
  {Buote}}]{Circflow}
{Mathews}, W.~G., {Brighenti}, F., \& {Buote}, D.~A. 2004, \apj, 615, 662

\bibitem[{{O'Sullivan} {et~al.}(2003){O'Sullivan}, {Ponman}, \&
  {Collins}}]{OSullivan}
{O'Sullivan}, E., {Ponman}, T.~J., \& {Collins}, R.~S. 2003, \mnras, 340, 1375

\bibitem[{{Paturel} {et~al.}(1997){Paturel}, {Andernach}, {Bottinelli}, {di
  Nella}, {Durand}, {Garnier}, {Gouguenheim}, {Lanoix}, {Marthinet}, {Petit},
  {Rousseau}, {Theureau}, \& {Vauglin}}]{LEDA}
{Paturel}, G., {Andernach}, H., {Bottinelli}, L., {di Nella}, H., {Durand}, N.,
  {Garnier}, R., {Gouguenheim}, L., {Lanoix}, P., {Marthinet}, M.~C., {Petit},
  C., {Rousseau}, J., {Theureau}, G., \& {Vauglin}, I. 1997, \aaps, 124, 109

\bibitem[{{Peterson} \& {Fabian}(2006)}]{PetersonCoolingspec}
{Peterson}, J.~R., \& {Fabian}, A.~C. 2006, \physrep, 427, 1

\bibitem[{{Piffaretti} {et~al.}(2005){Piffaretti}, {Jetzer}, {Kaastra}, \&
  {Tamura}}]{PiffarettiClusters}
{Piffaretti}, R., {Jetzer}, P., {Kaastra}, J.~S., \& {Tamura}, T. 2005, \aap,
  433, 101

\bibitem[{{Randall} {et~al.}(2006){Randall}, {Sarazin}, \&
  {Irwin}}]{RandallXMMNGC4649}
{Randall}, S.~W., {Sarazin}, C.~L., \& {Irwin}, J.~A. 2006, \apj, 636, 200

\bibitem[{{Snellen} {et~al.}(2003){Snellen}, {Lehnert}, {Bremer}, \&
  {Schilizzi}}]{SnellenLsigma}
{Snellen}, I.~A.~G., {Lehnert}, M.~D., {Bremer}, M.~N., \& {Schilizzi}, R.~T.
  2003, \mnras, 342, 889

\bibitem[{{Tremaine} {et~al.}(2002){Tremaine}, {Gebhardt}, {Bender}, {Bower},
  {Dressler}, {Faber}, {Filippenko}, {Green}, {Grillmair}, {Ho}, {Kormendy},
  {Lauer}, {Magorrian}, {Pinkney}, \& {Richstone}}]{TremaineBHsigma}
{Tremaine}, S., {Gebhardt}, K., {Bender}, R., {Bower}, G., {Dressler}, A.,
  {Faber}, S.~M., {Filippenko}, A.~V., {Green}, R., {Grillmair}, C., {Ho},
  L.~C., {Kormendy}, J., {Lauer}, T.~R., {Magorrian}, J., {Pinkney}, J., \&
  {Richstone}, D. 2002, \apj, 574, 740

\bibitem[{{Tully}(1988)}]{TullyRho}
{Tully}, R.~B. 1988, {Nearby galaxies catalog} (Cambridge and New York,
  Cambridge University Press, 1988, 221 p.)

\bibitem[{{Wright} {et~al.}(1996){Wright}, {Griffith}, {Burke}, \&
  {Ekers}}]{PMN6cm}
{Wright}, A.~E., {Griffith}, M.~R., {Burke}, B.~F., \& {Ekers}, R.~D. 1996,
  VizieR Online Data Catalog, 8038, 0+

\end{thebibliography}

\clearpage
\appendix
\section{Individual Temperature Profiles}

\begin{figure}[h]
\includegraphics[width=0.5\textwidth]{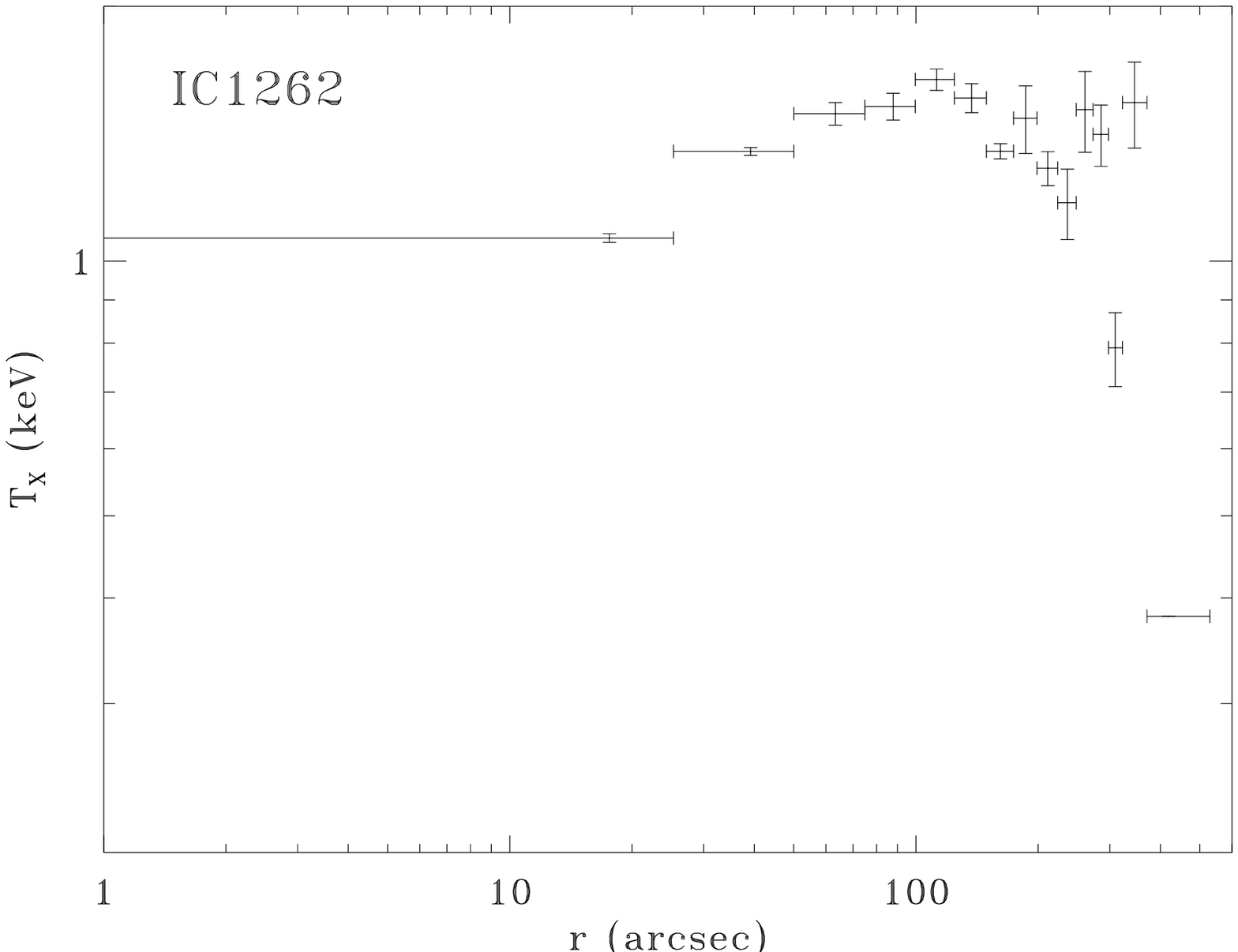}
\includegraphics[width=0.5\textwidth]{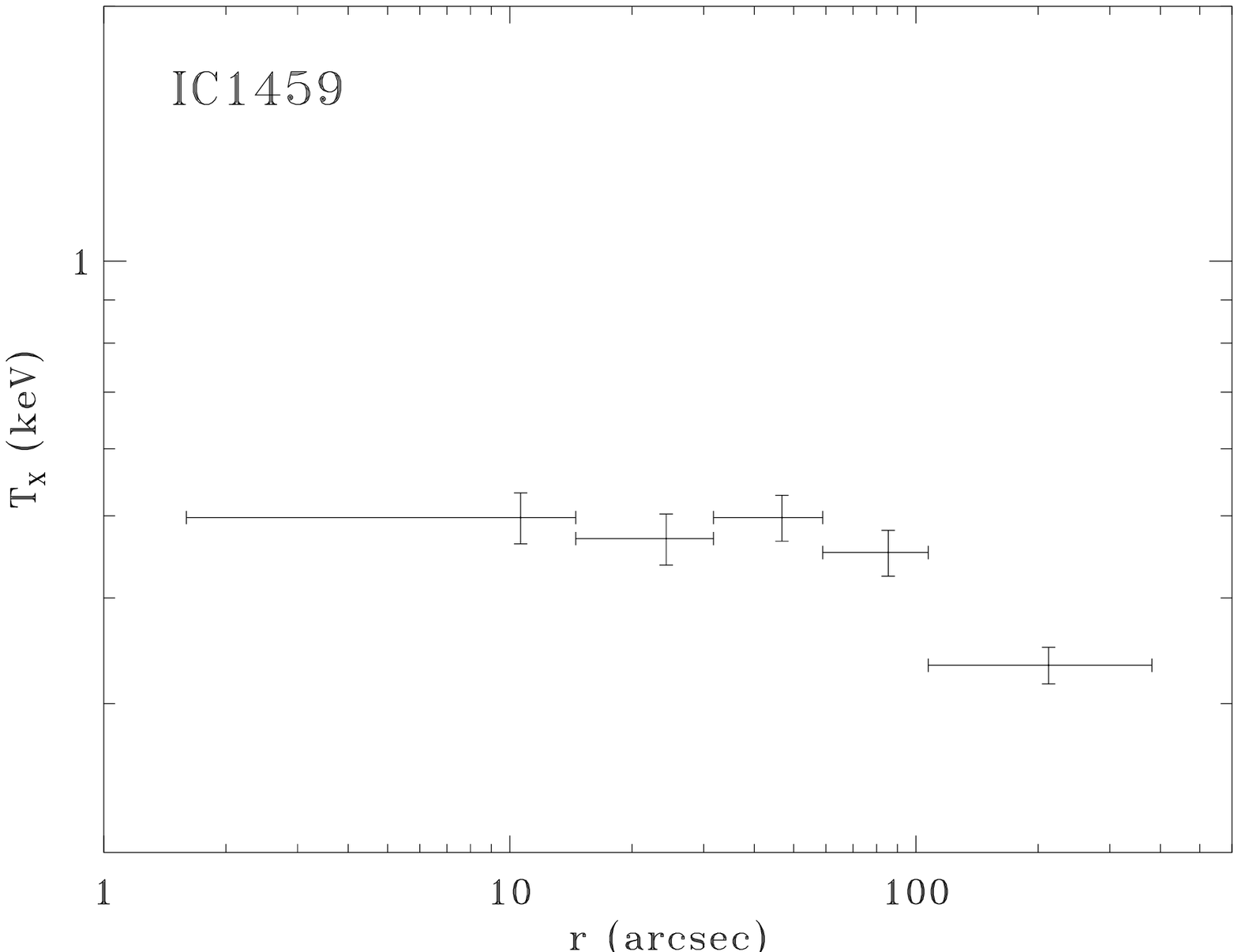}
\includegraphics[width=0.5\textwidth]{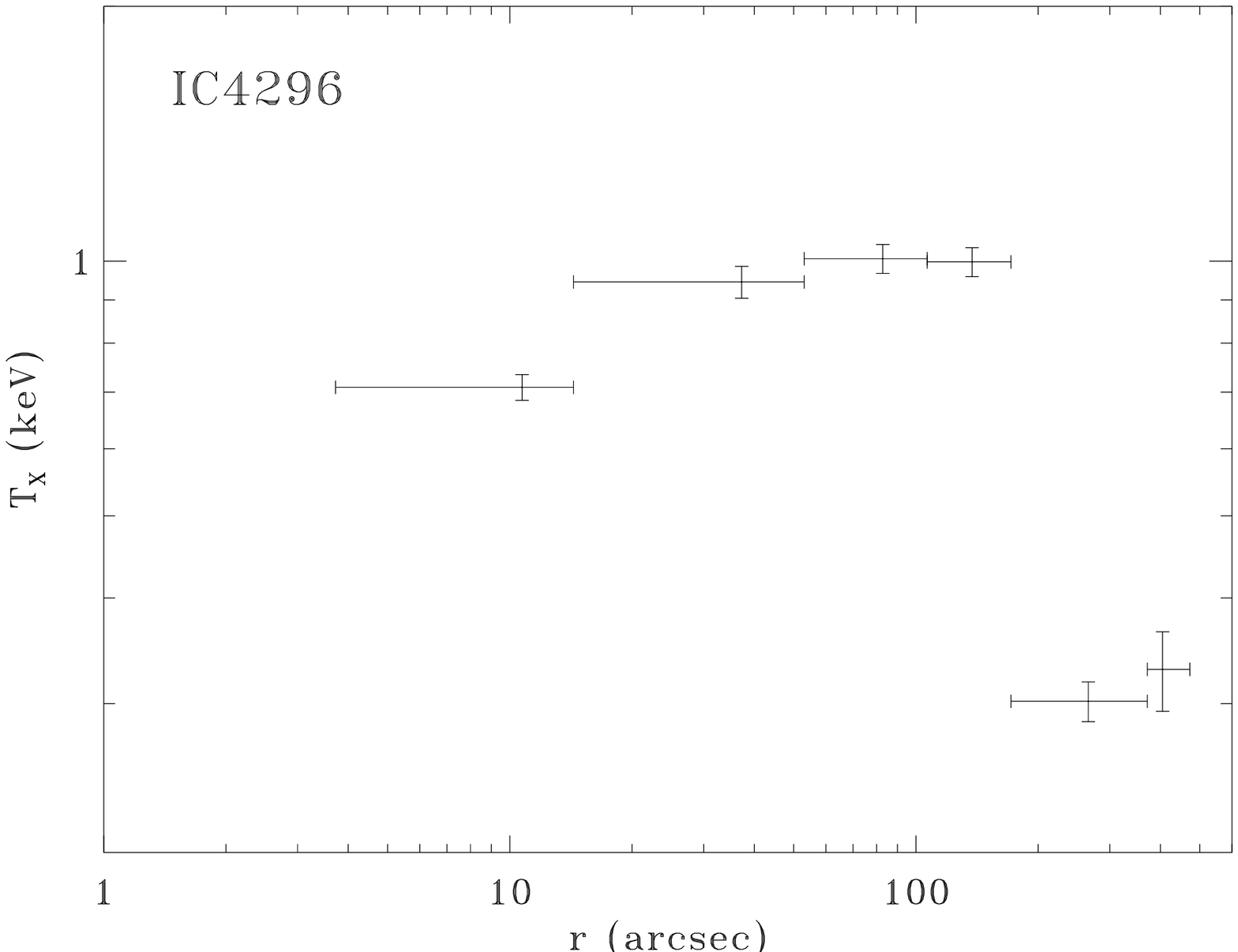}
\includegraphics[width=0.5\textwidth]{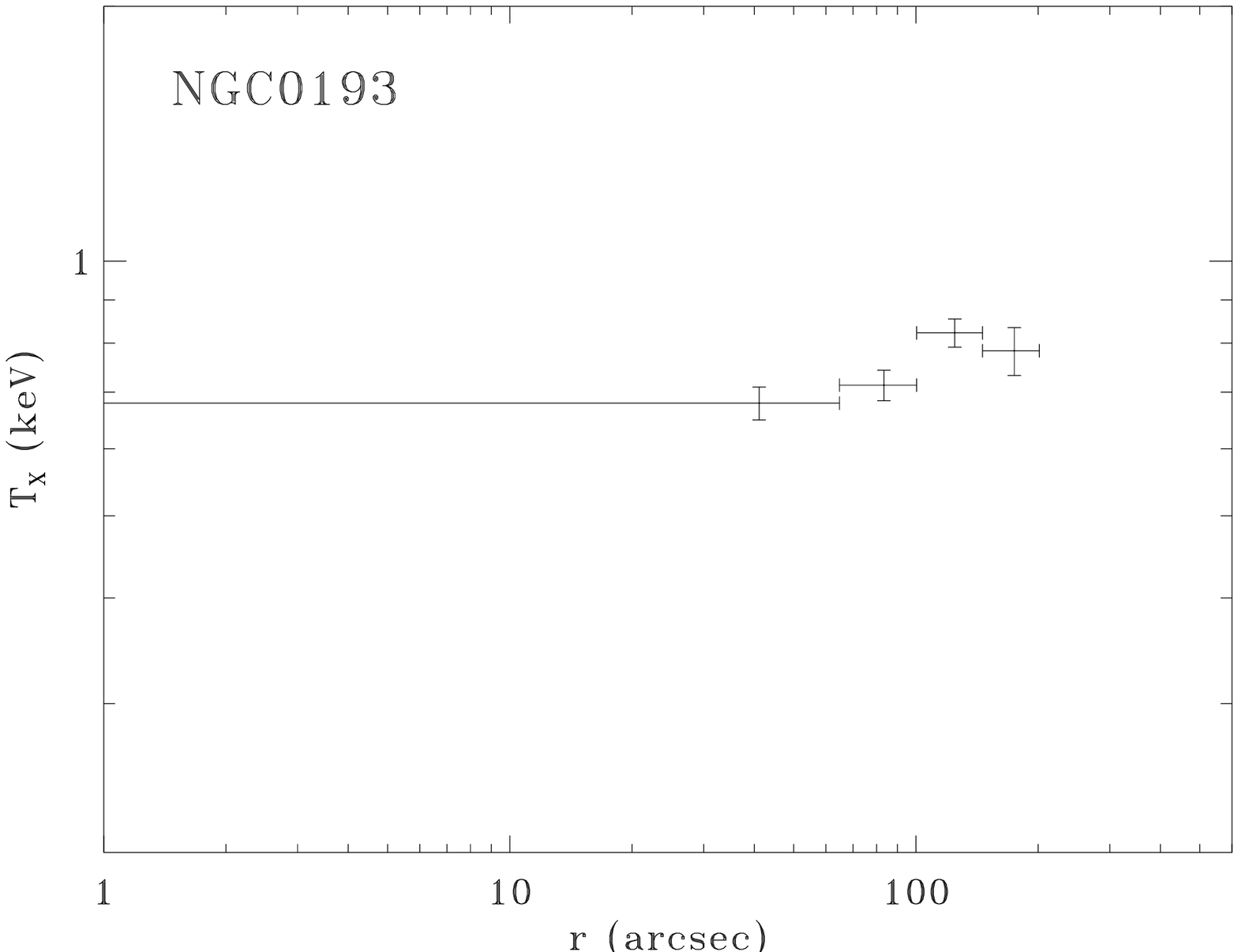}
\includegraphics[width=0.5\textwidth]{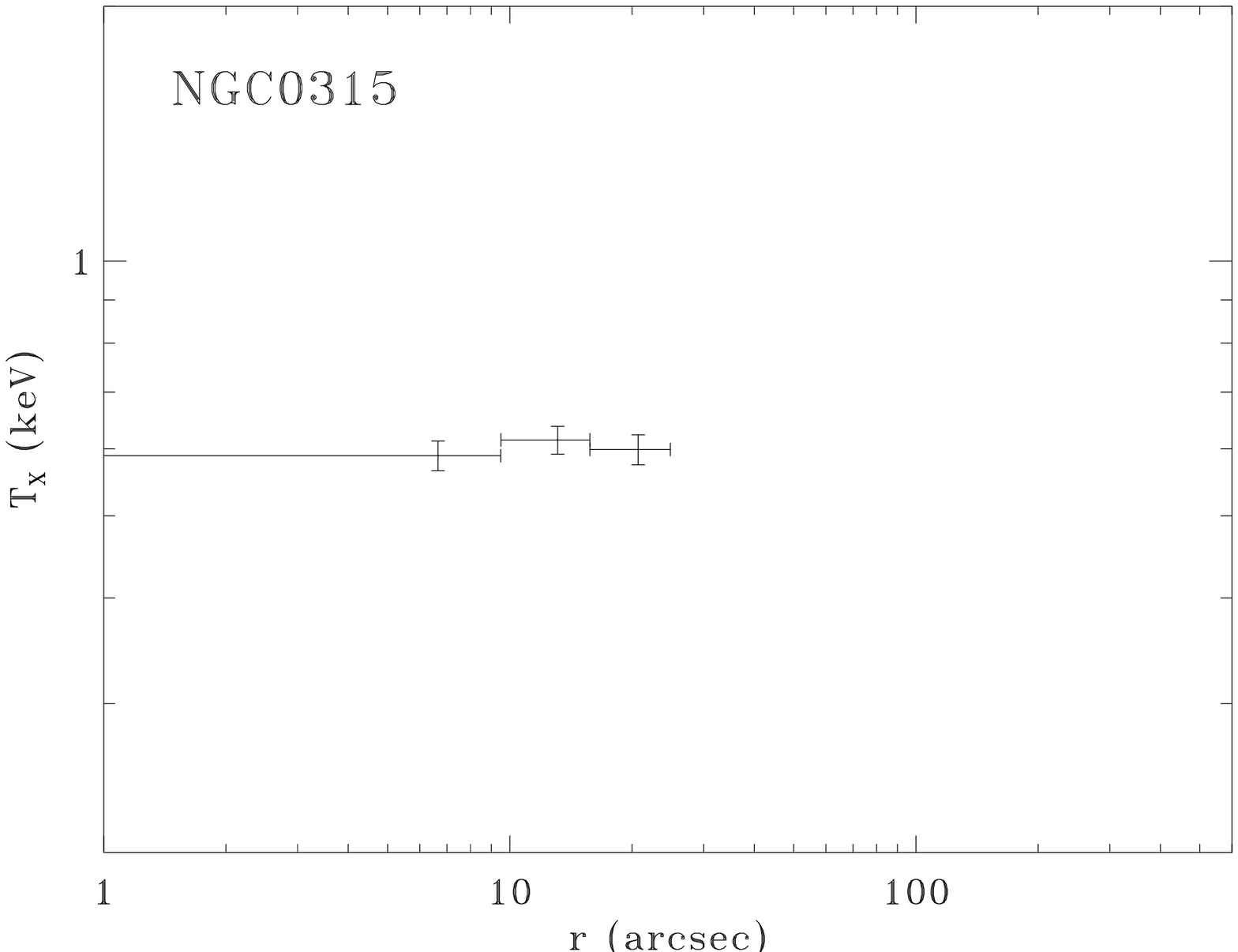}
\includegraphics[width=0.5\textwidth]{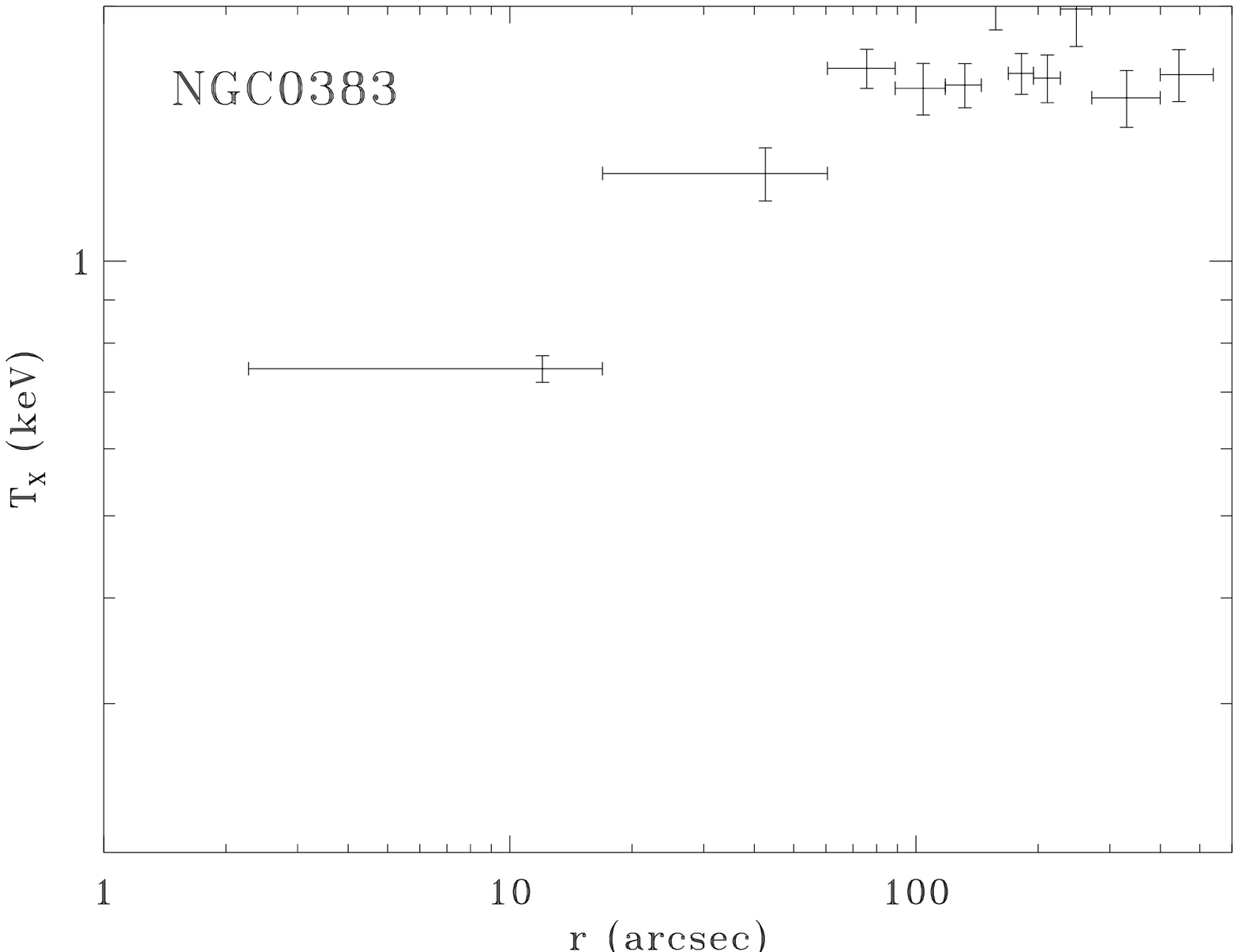}
\end{figure}
\clearpage

\begin{figure}
\includegraphics[width=0.5\textwidth]{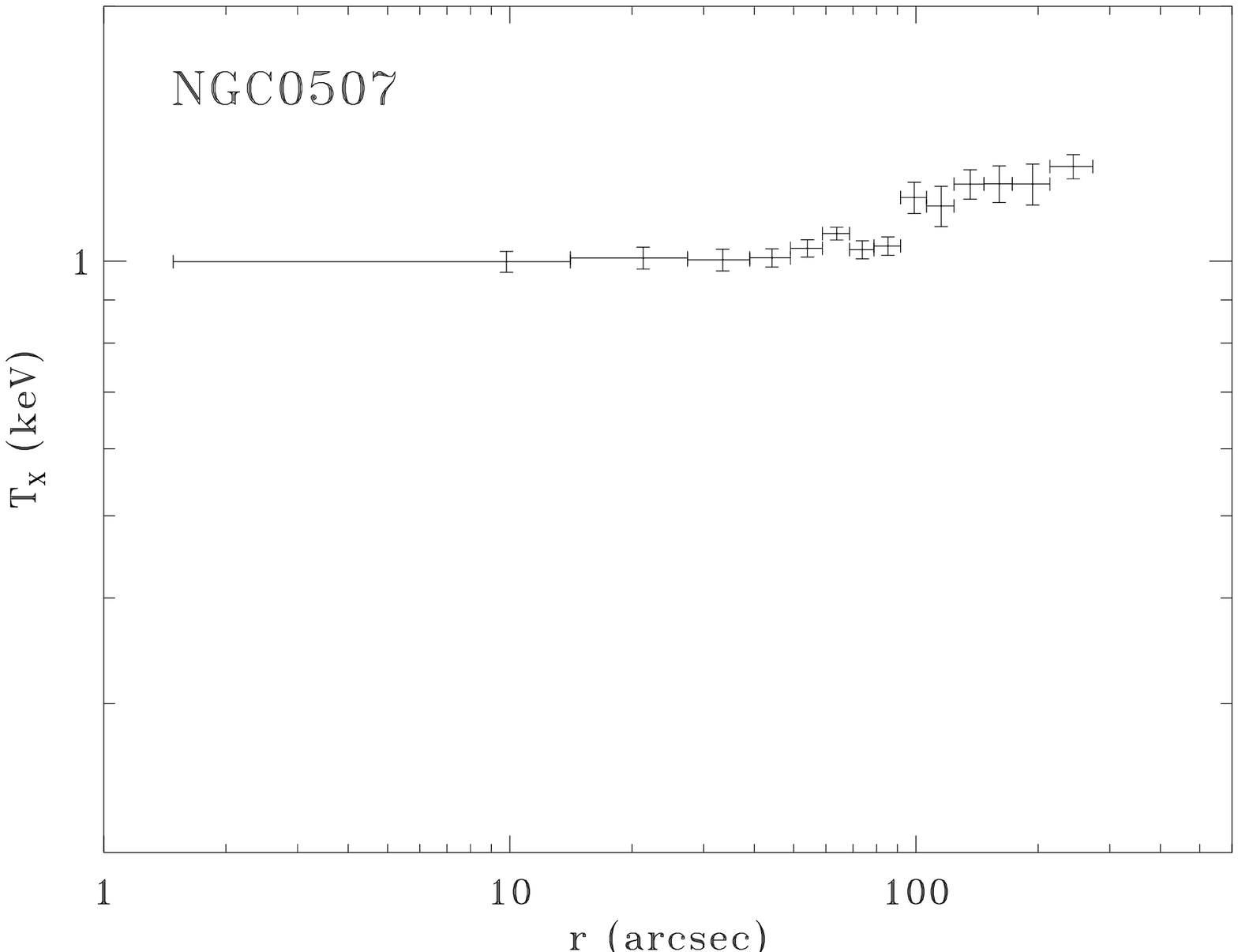}
\includegraphics[width=0.5\textwidth]{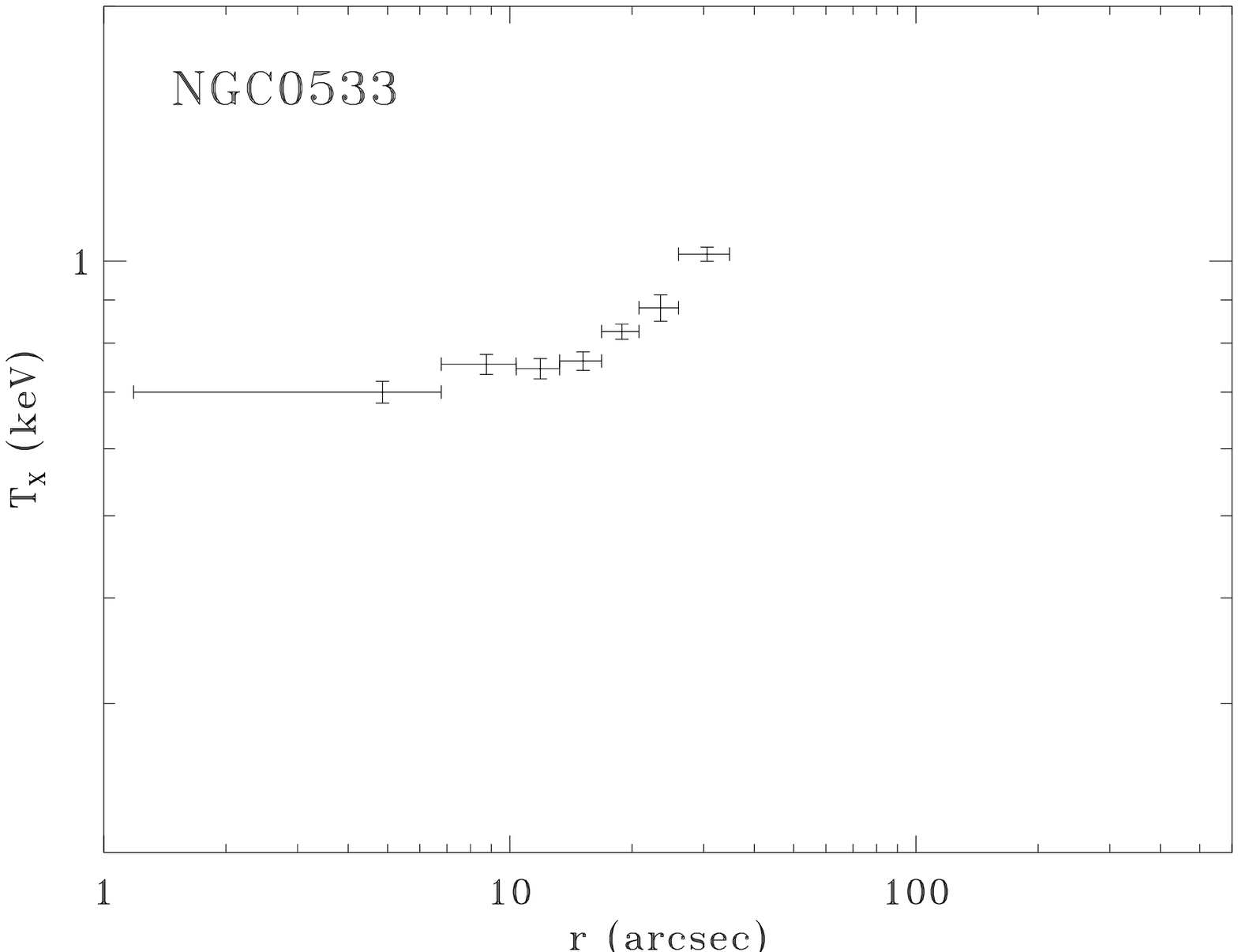}
\includegraphics[width=0.5\textwidth]{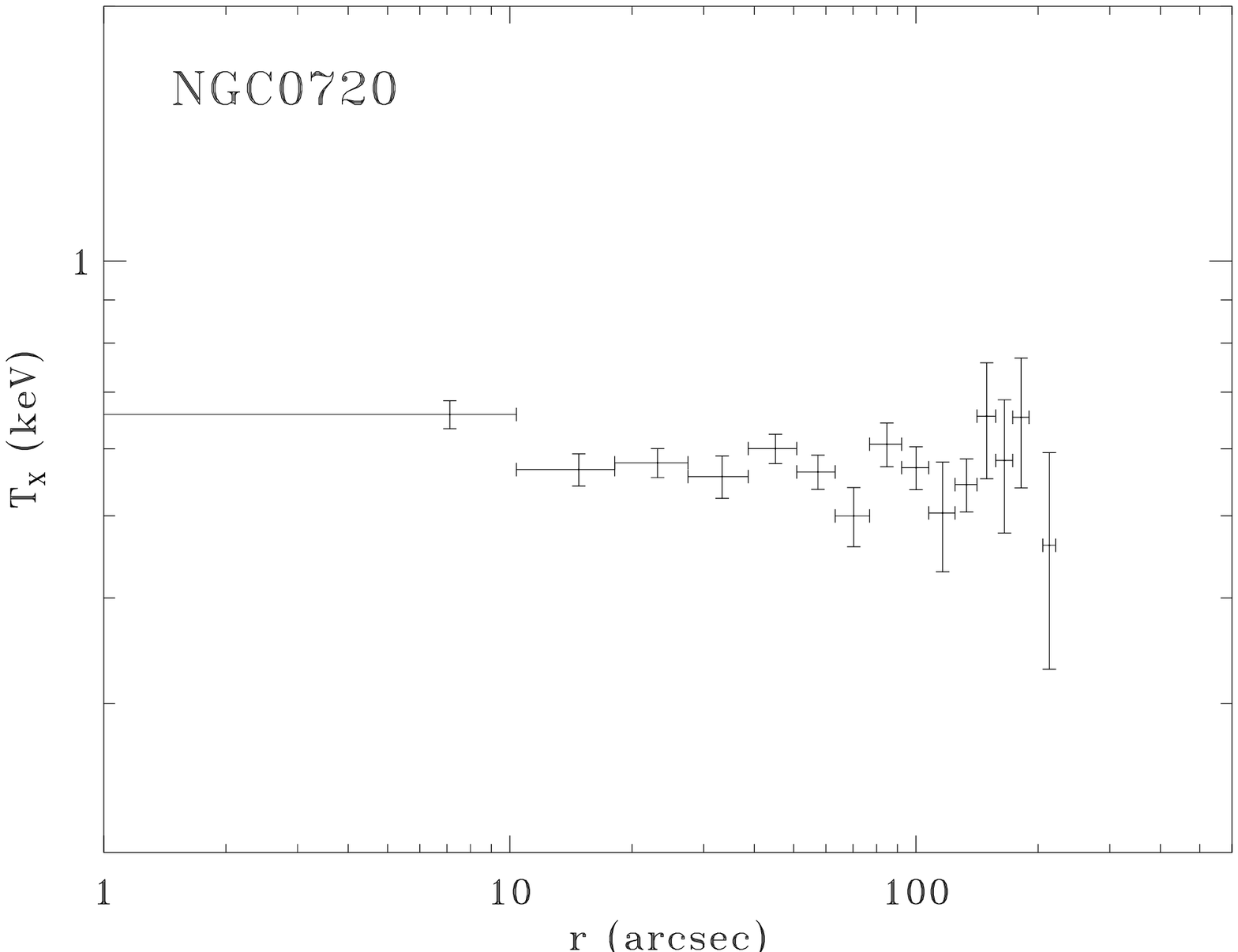}
\includegraphics[width=0.5\textwidth]{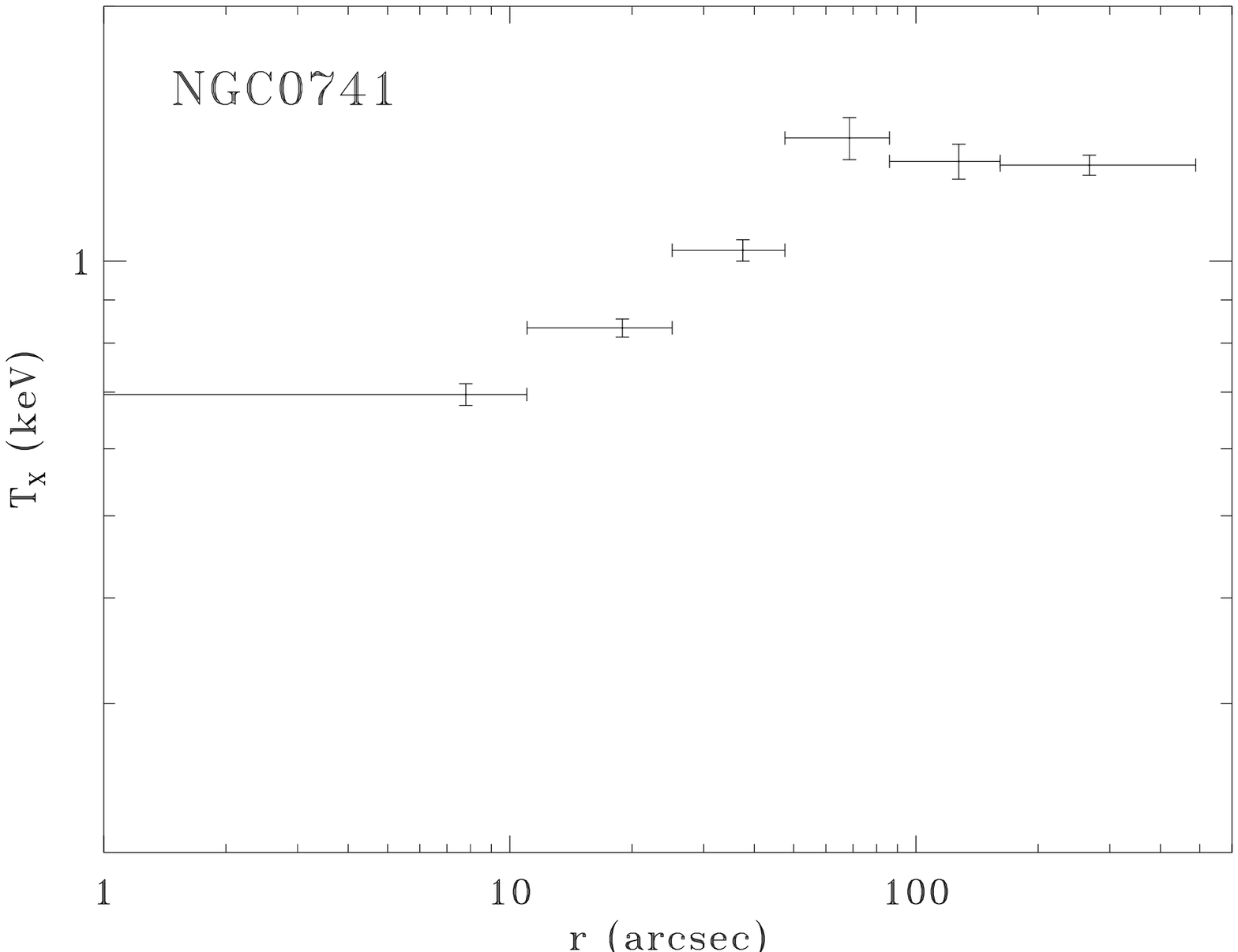}
\includegraphics[width=0.5\textwidth]{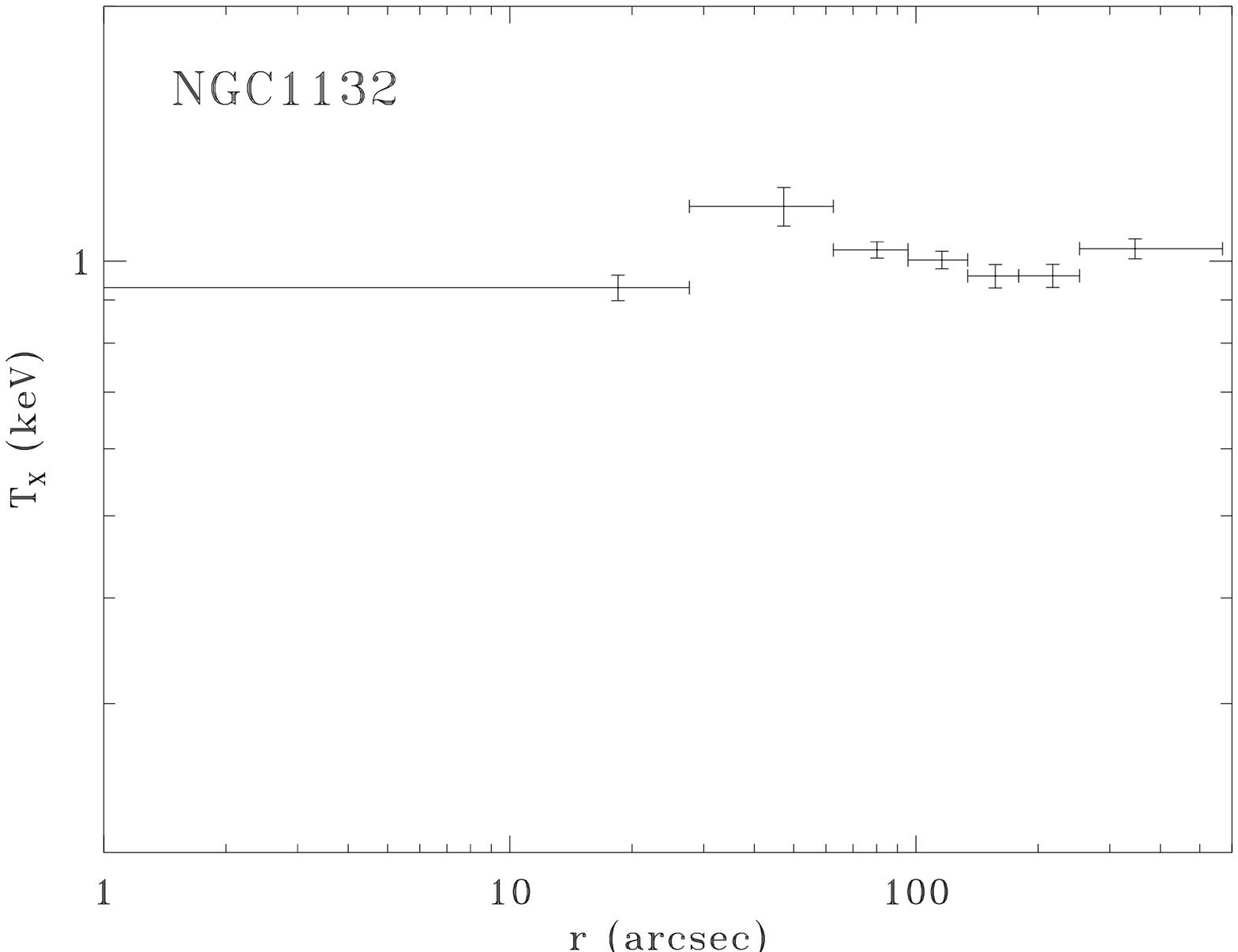}
\includegraphics[width=0.5\textwidth]{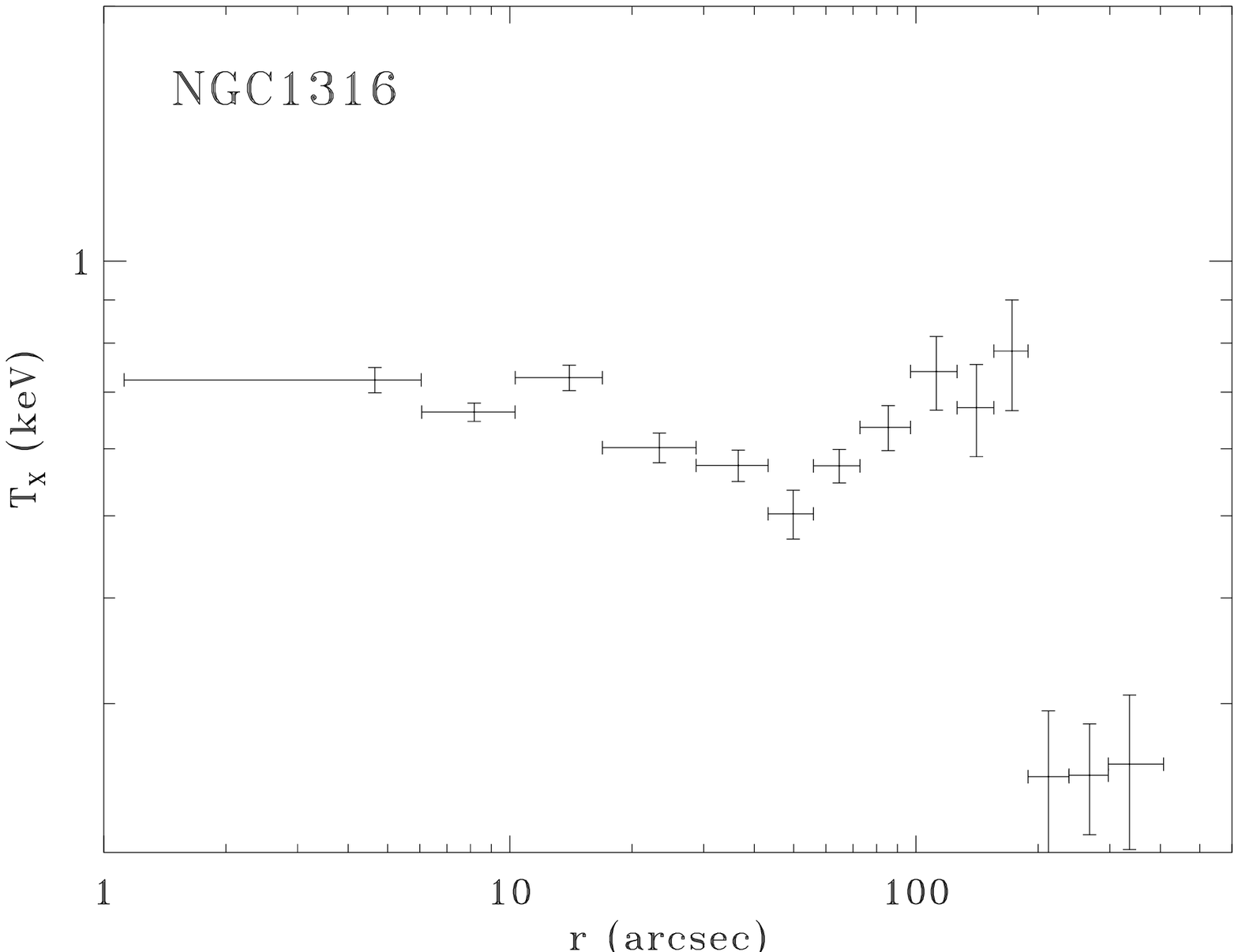}
\end{figure}
\clearpage

\begin{figure}
\includegraphics[width=0.5\textwidth]{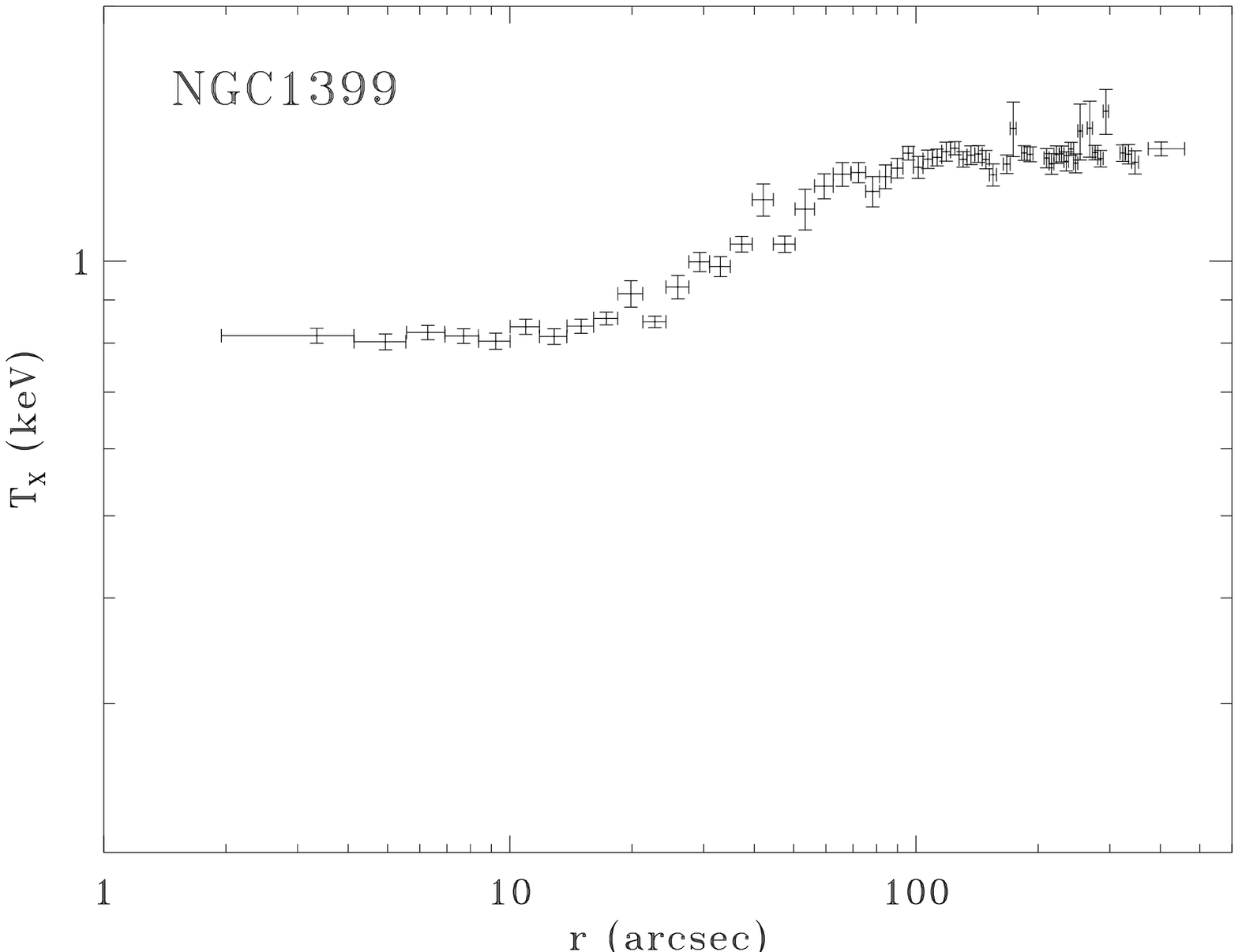}
\includegraphics[width=0.5\textwidth]{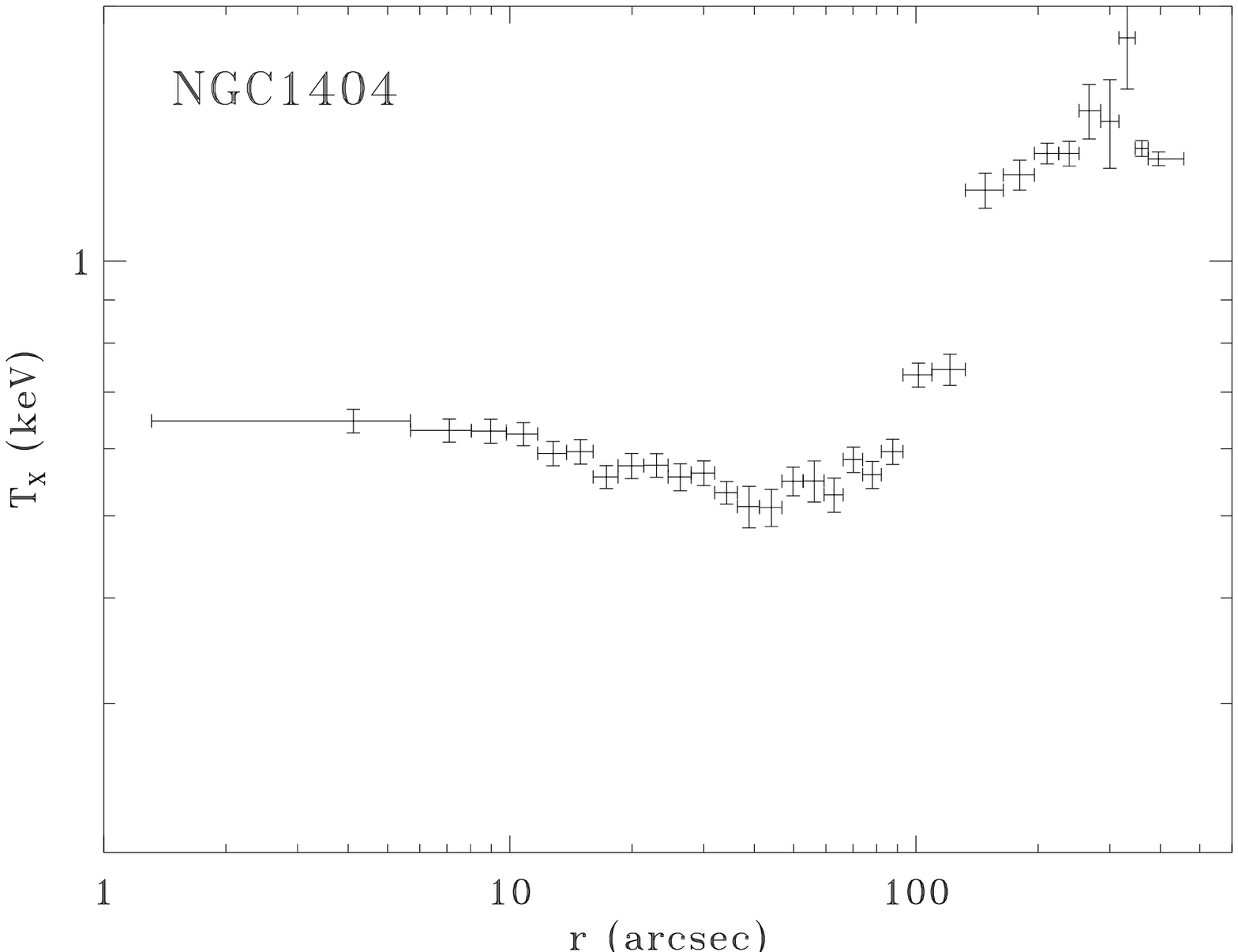}
\includegraphics[width=0.5\textwidth]{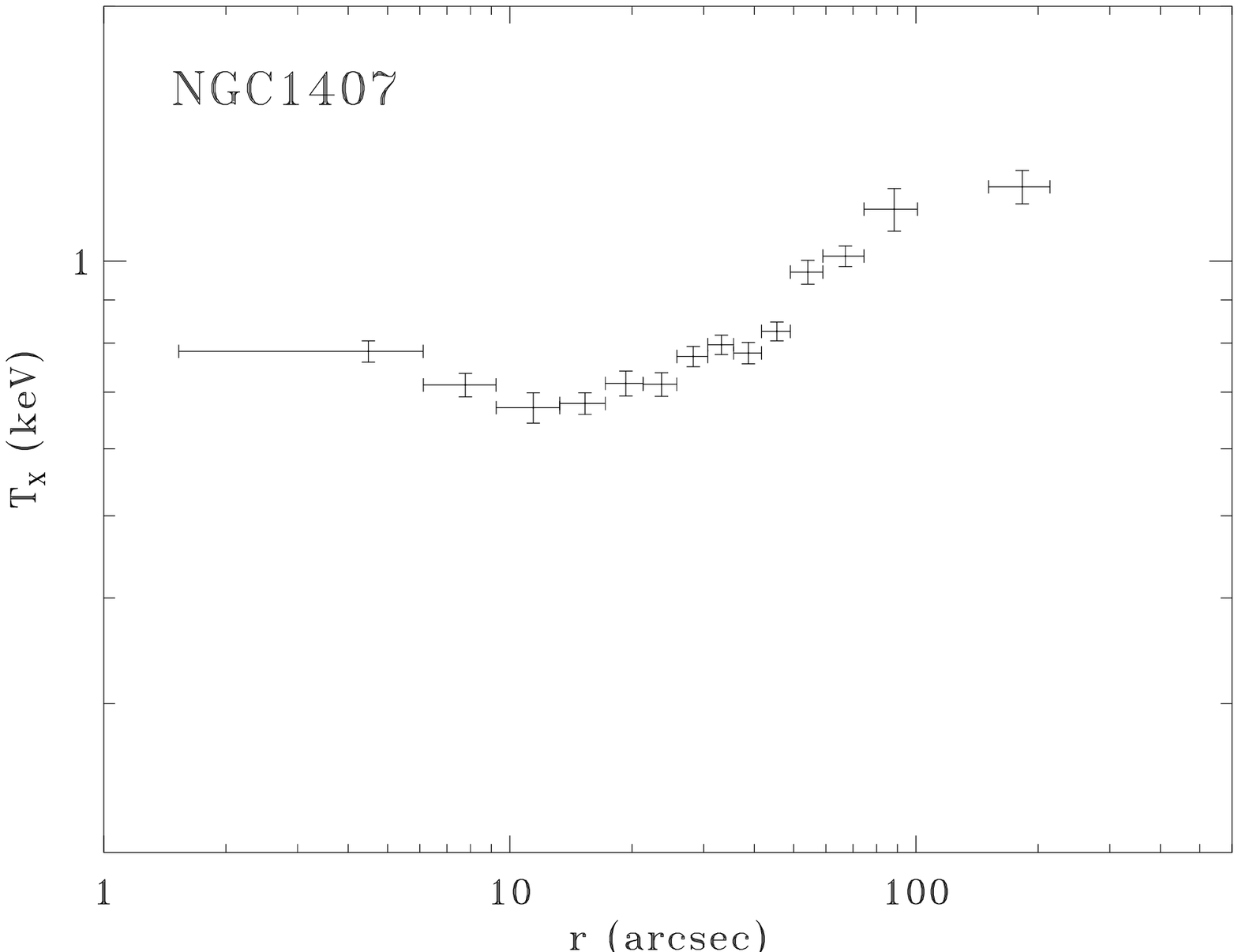}
\includegraphics[width=0.5\textwidth]{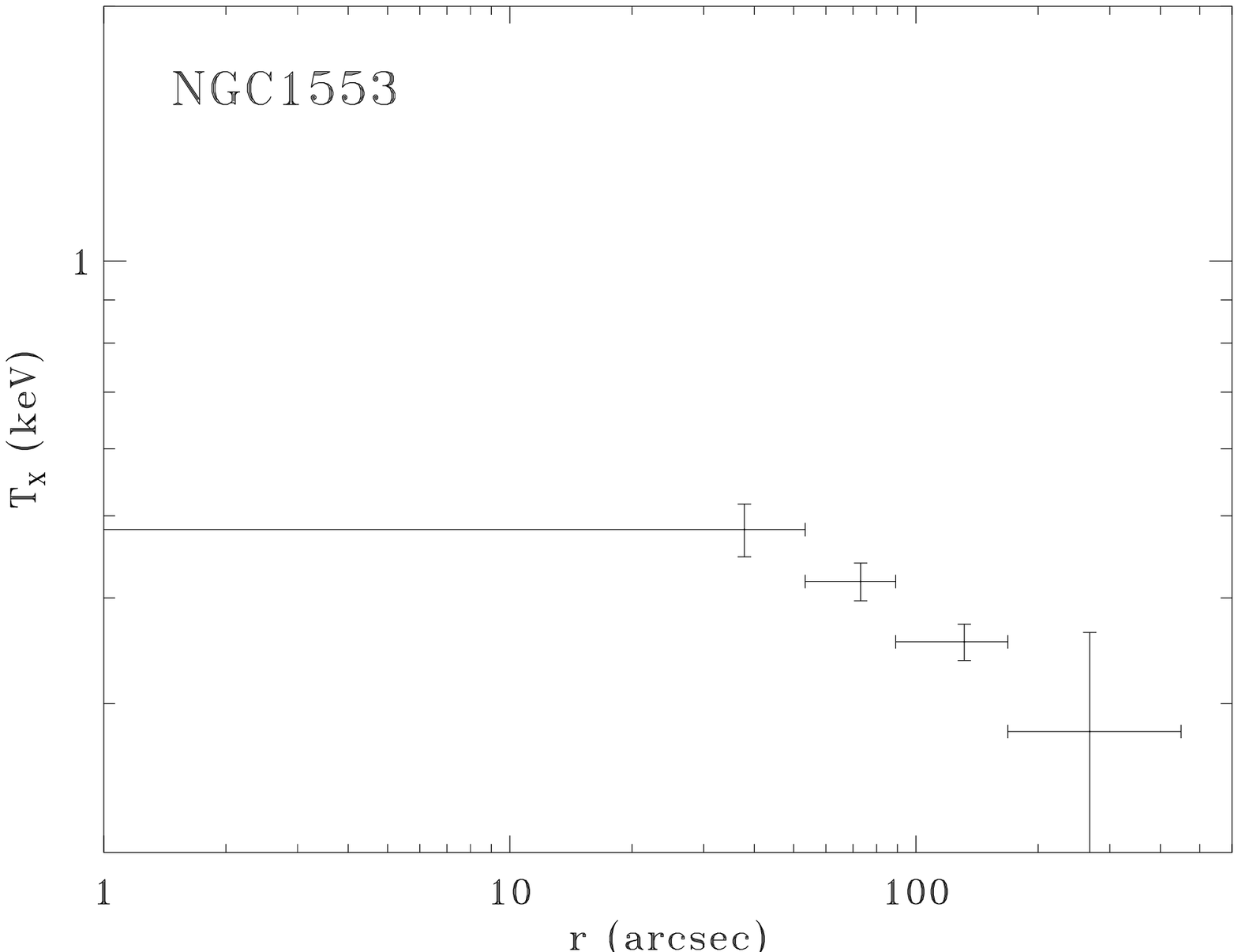}
\includegraphics[width=0.5\textwidth]{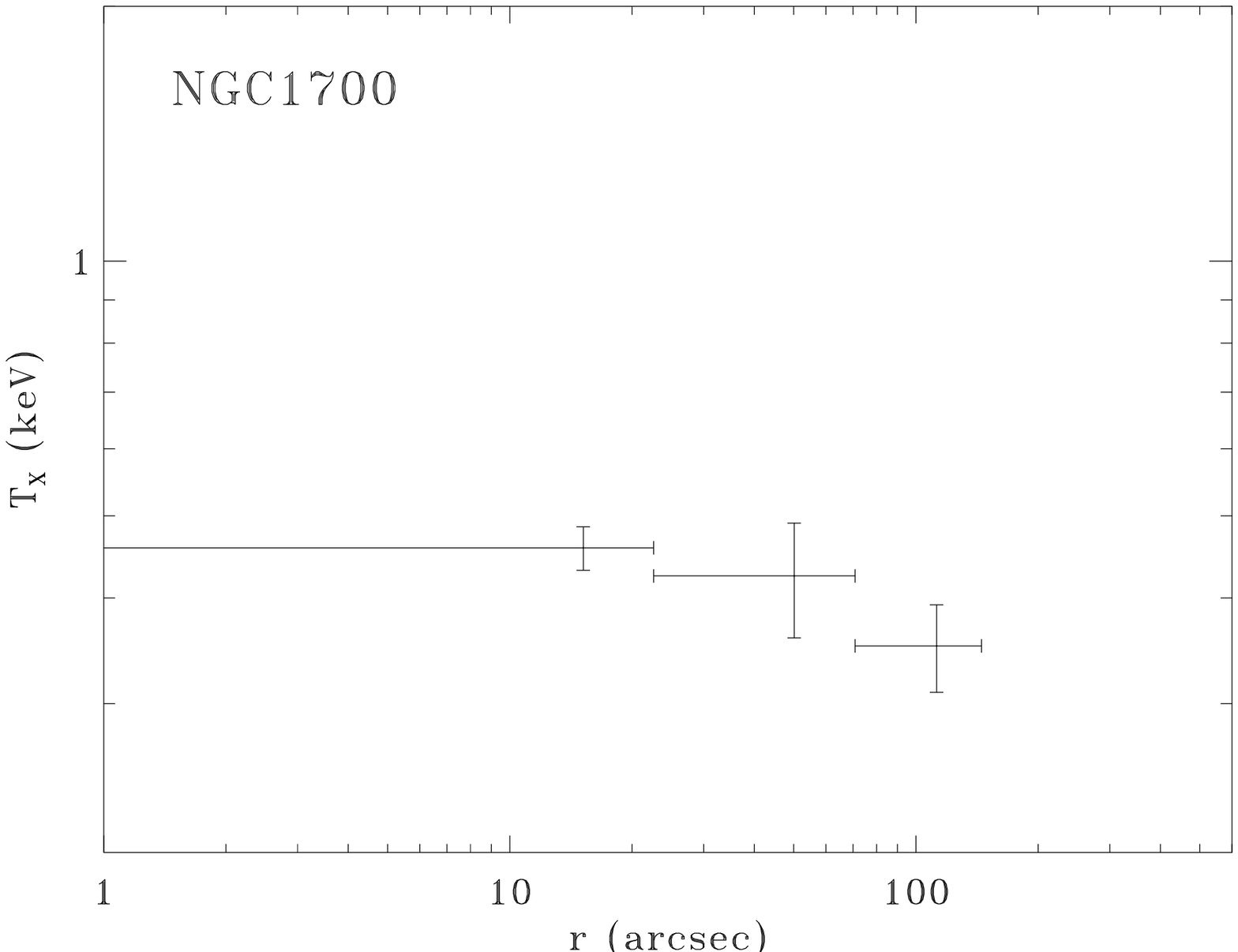}
\includegraphics[width=0.5\textwidth]{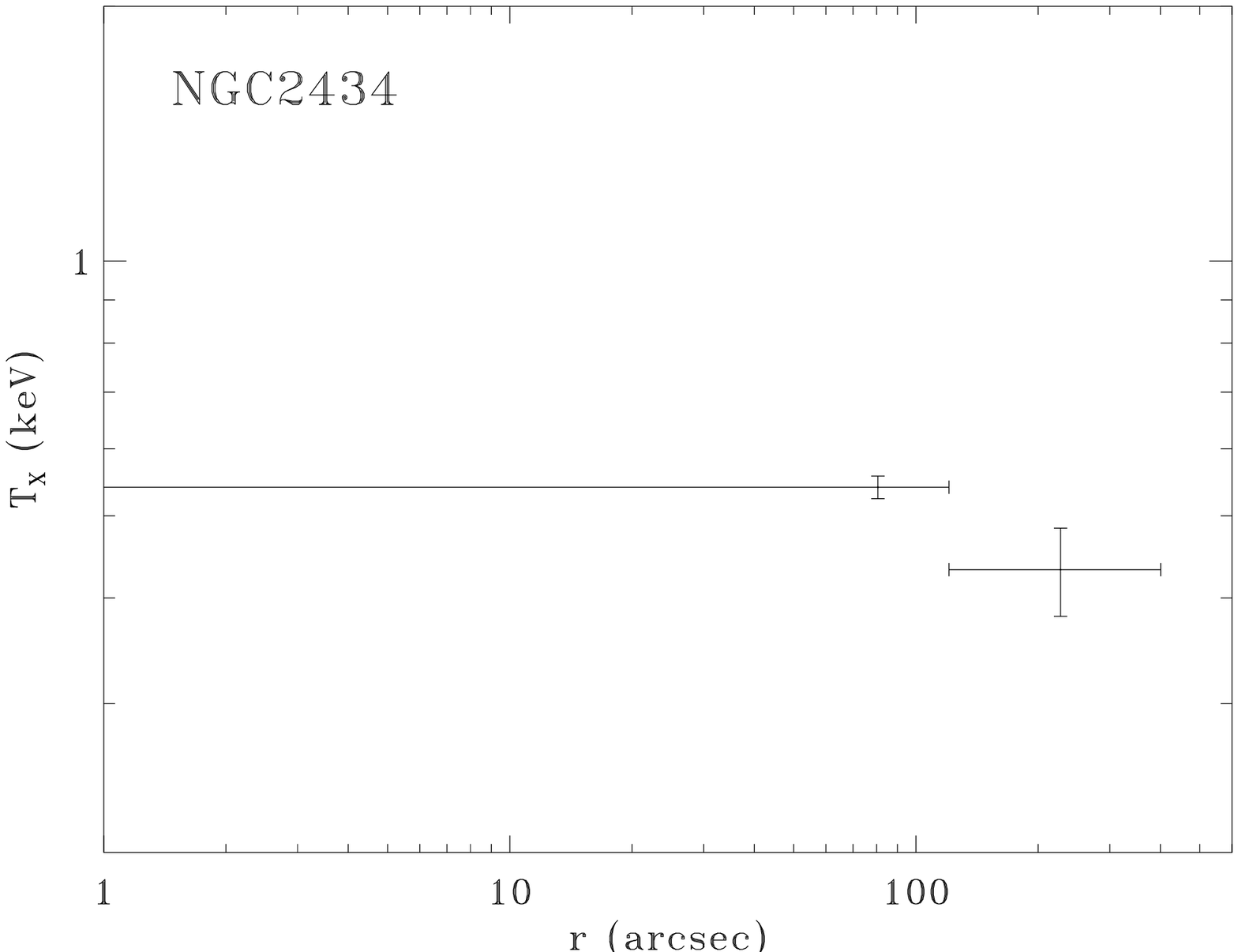}
\end{figure}
\clearpage

\begin{figure}
\includegraphics[width=0.5\textwidth]{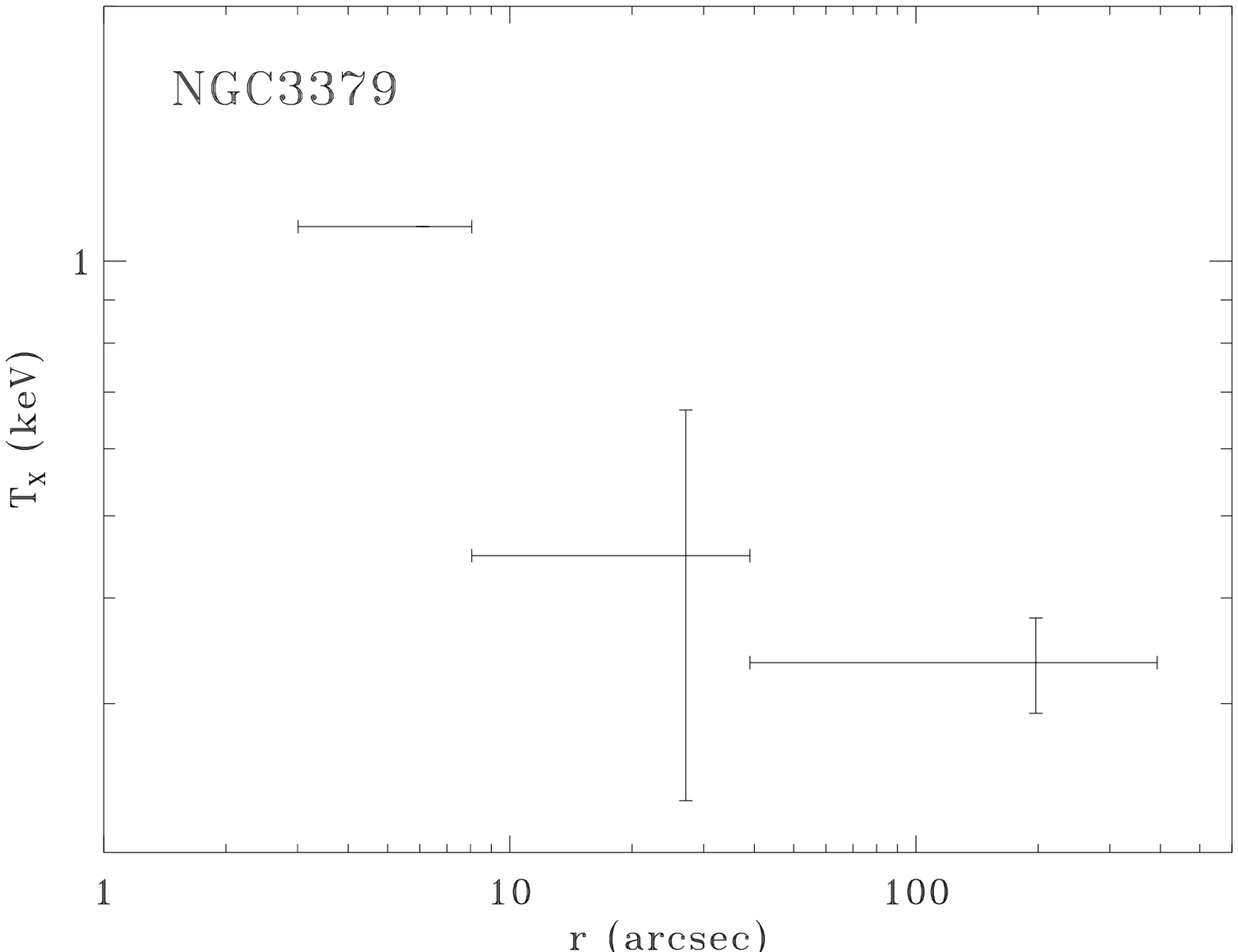}
\includegraphics[width=0.5\textwidth]{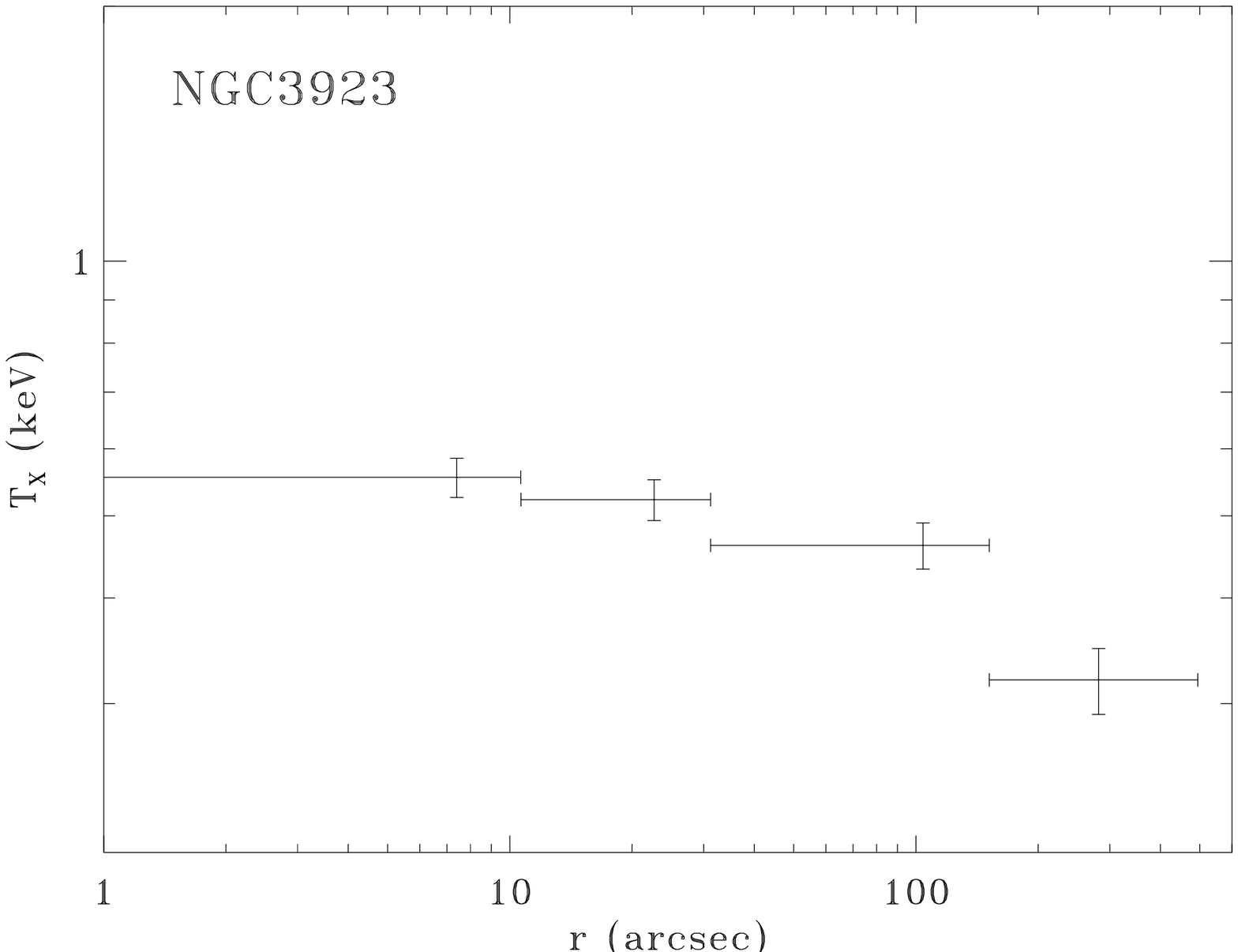}
\includegraphics[width=0.5\textwidth]{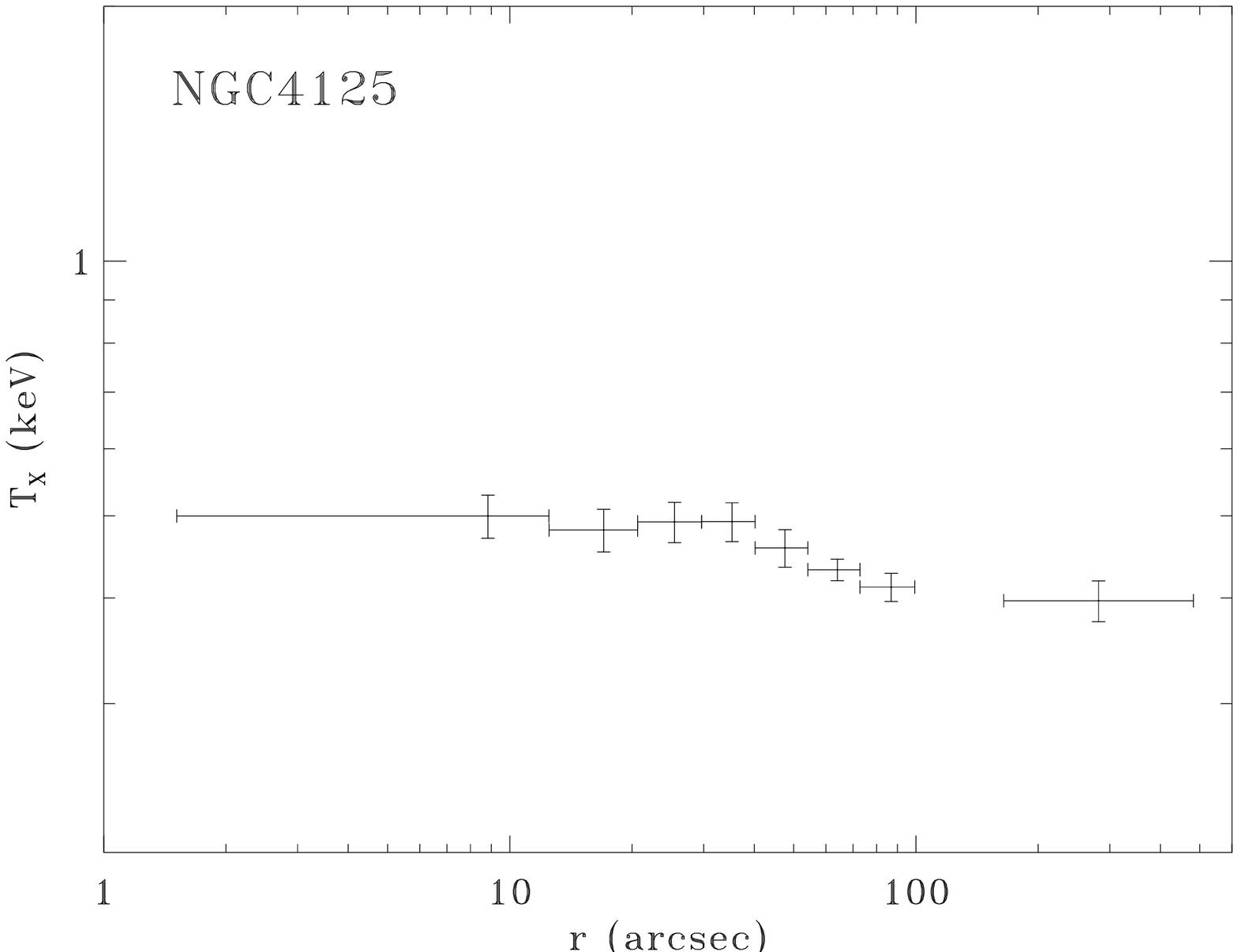}
\includegraphics[width=0.5\textwidth]{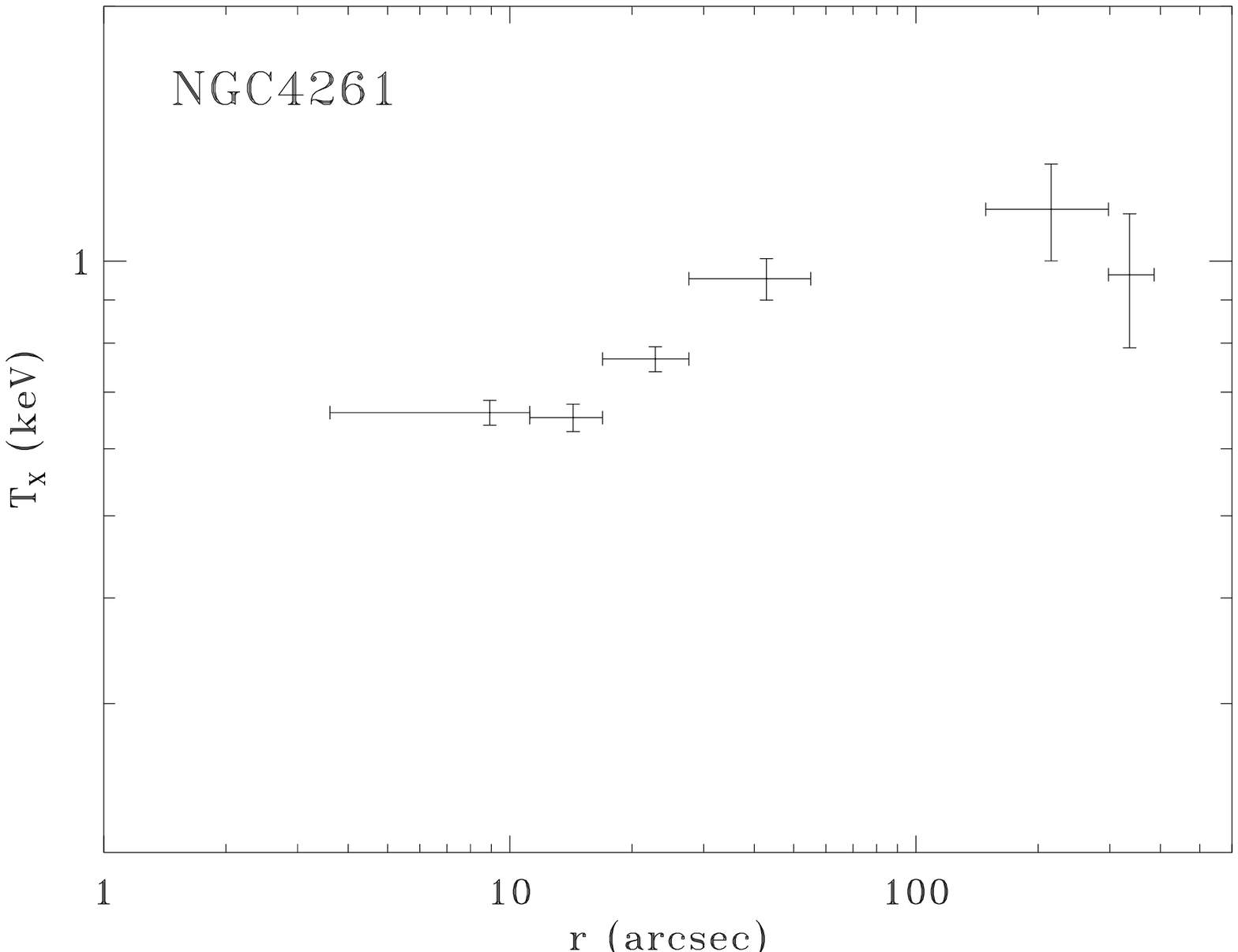}
\includegraphics[width=0.5\textwidth]{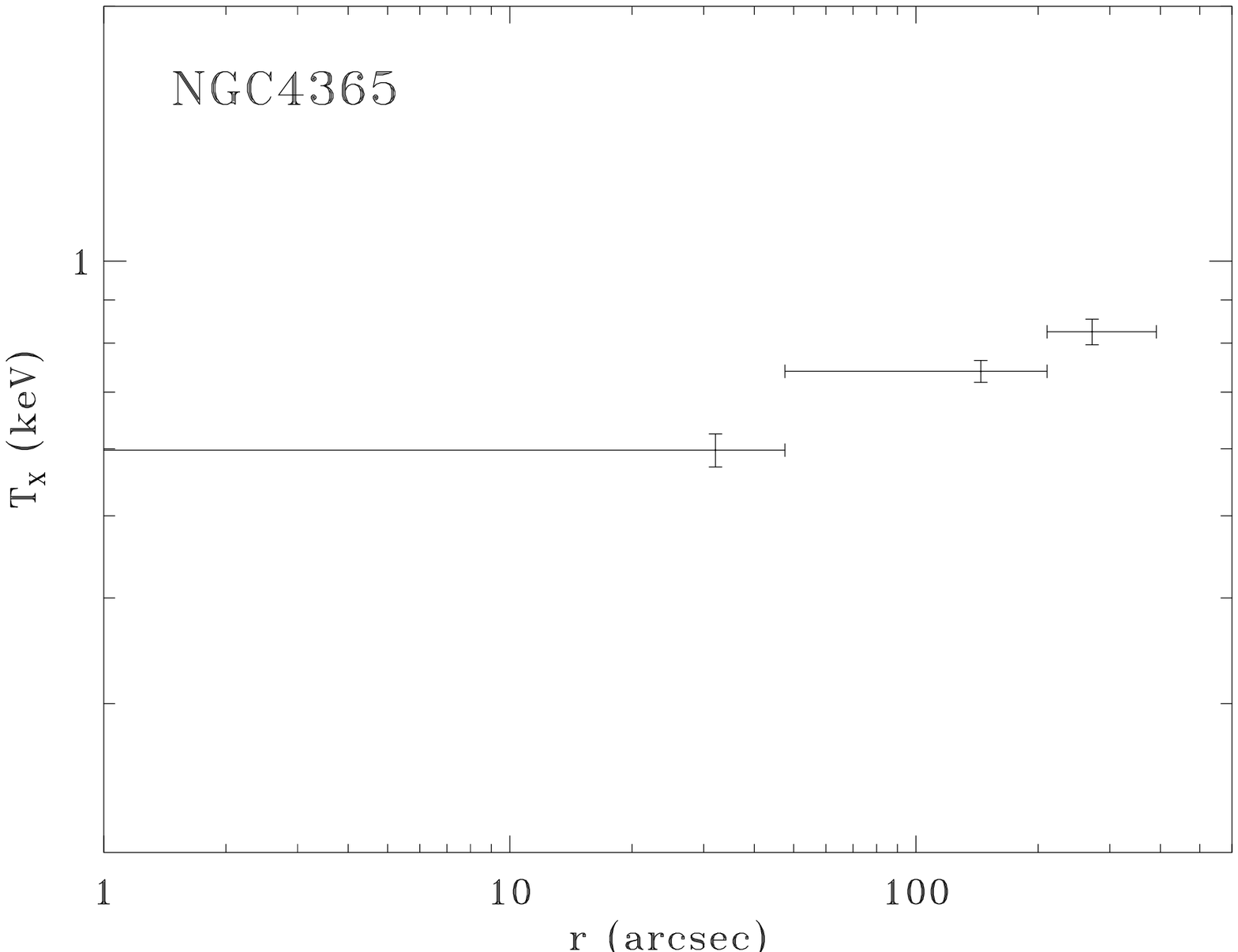}
\includegraphics[width=0.5\textwidth]{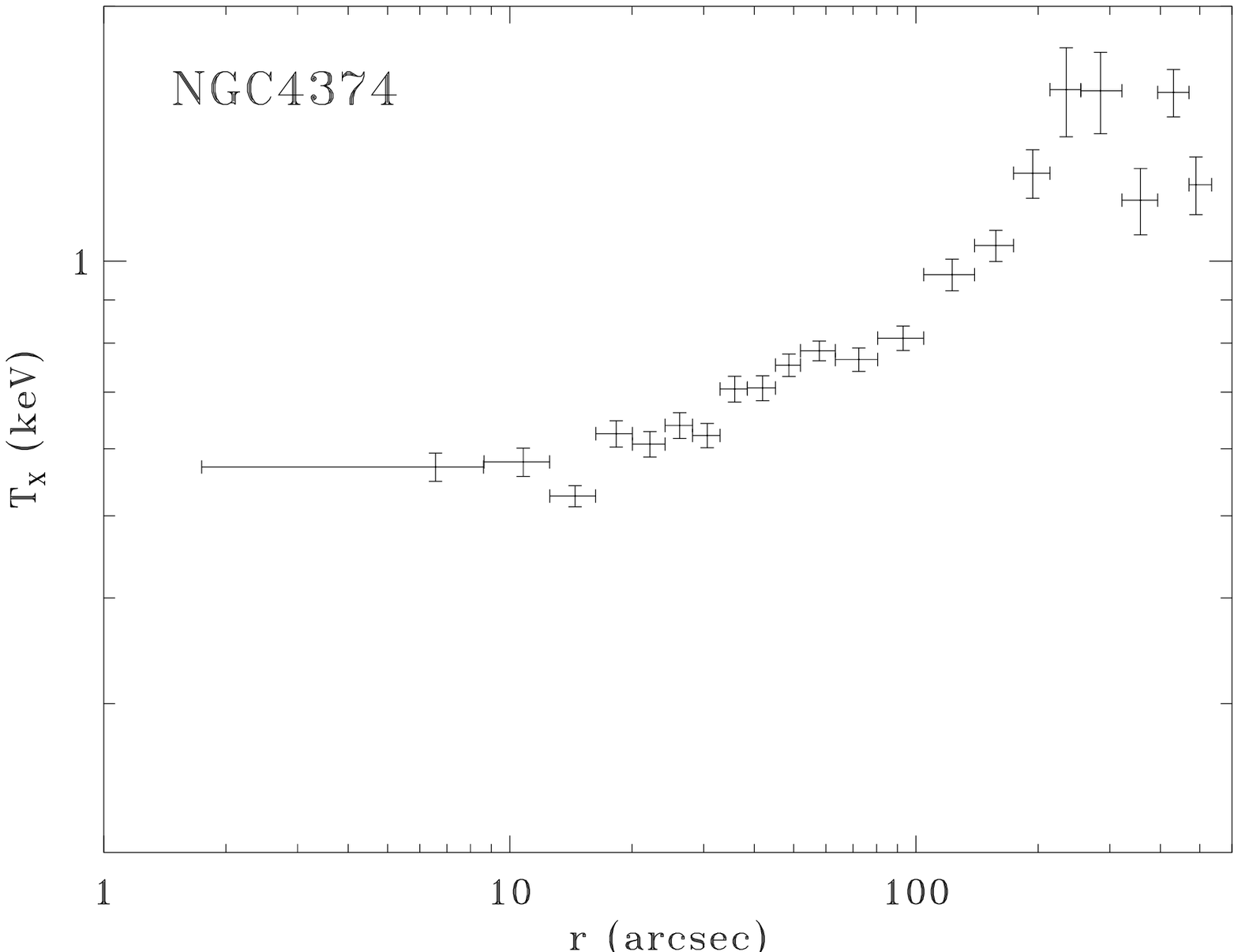}
\end{figure}
\clearpage

\begin{figure}
\includegraphics[width=0.5\textwidth]{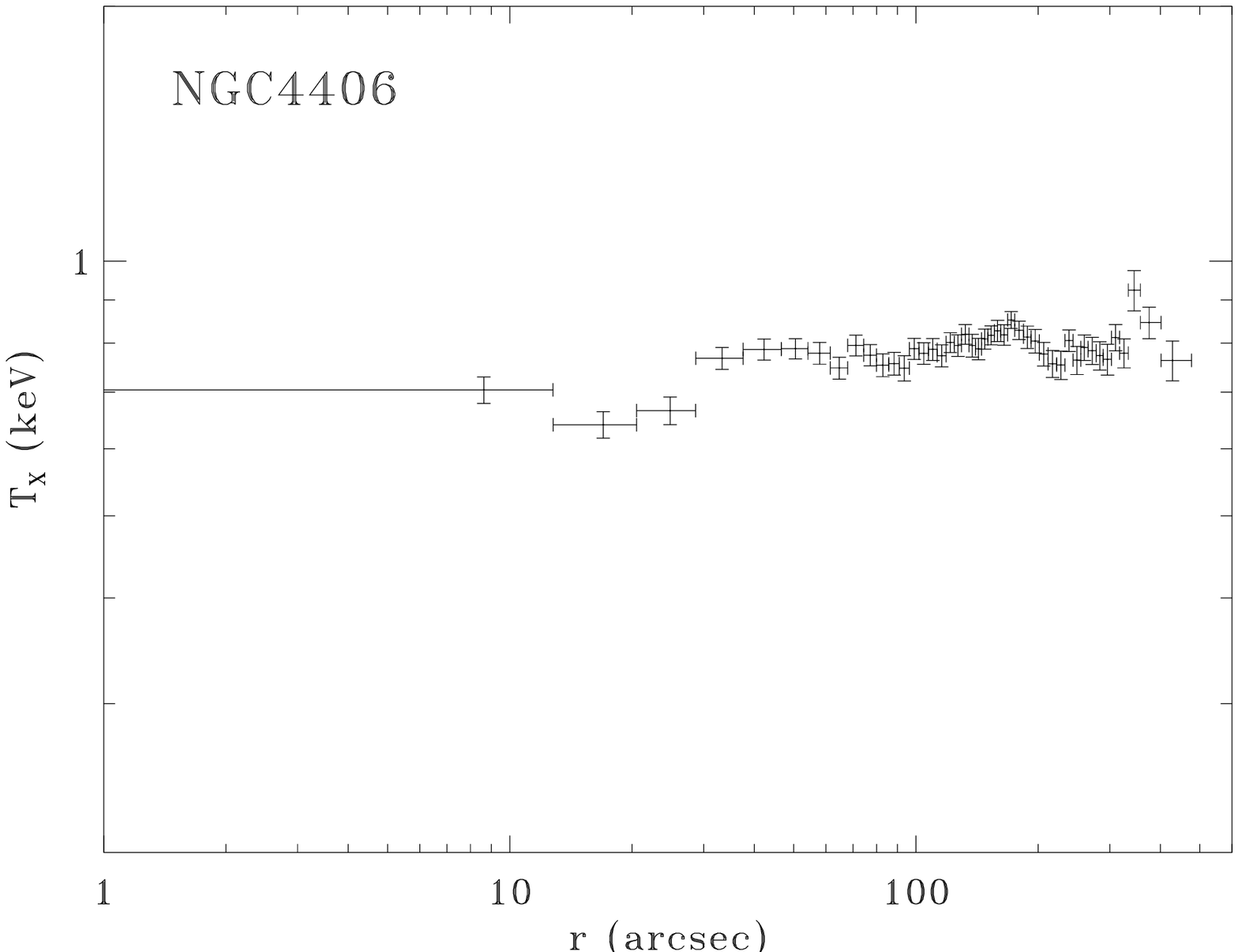}
\includegraphics[width=0.5\textwidth]{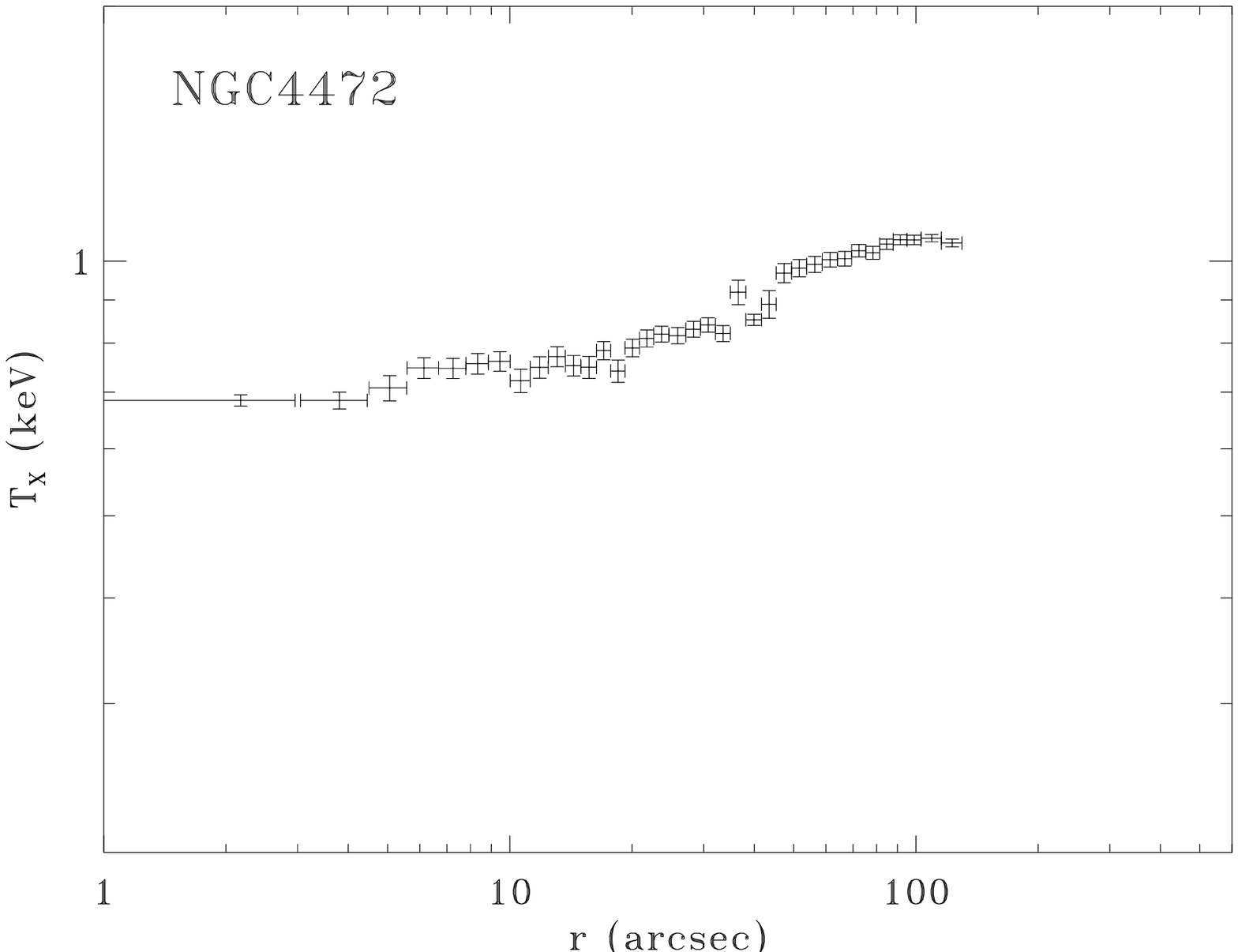}
\includegraphics[width=0.5\textwidth]{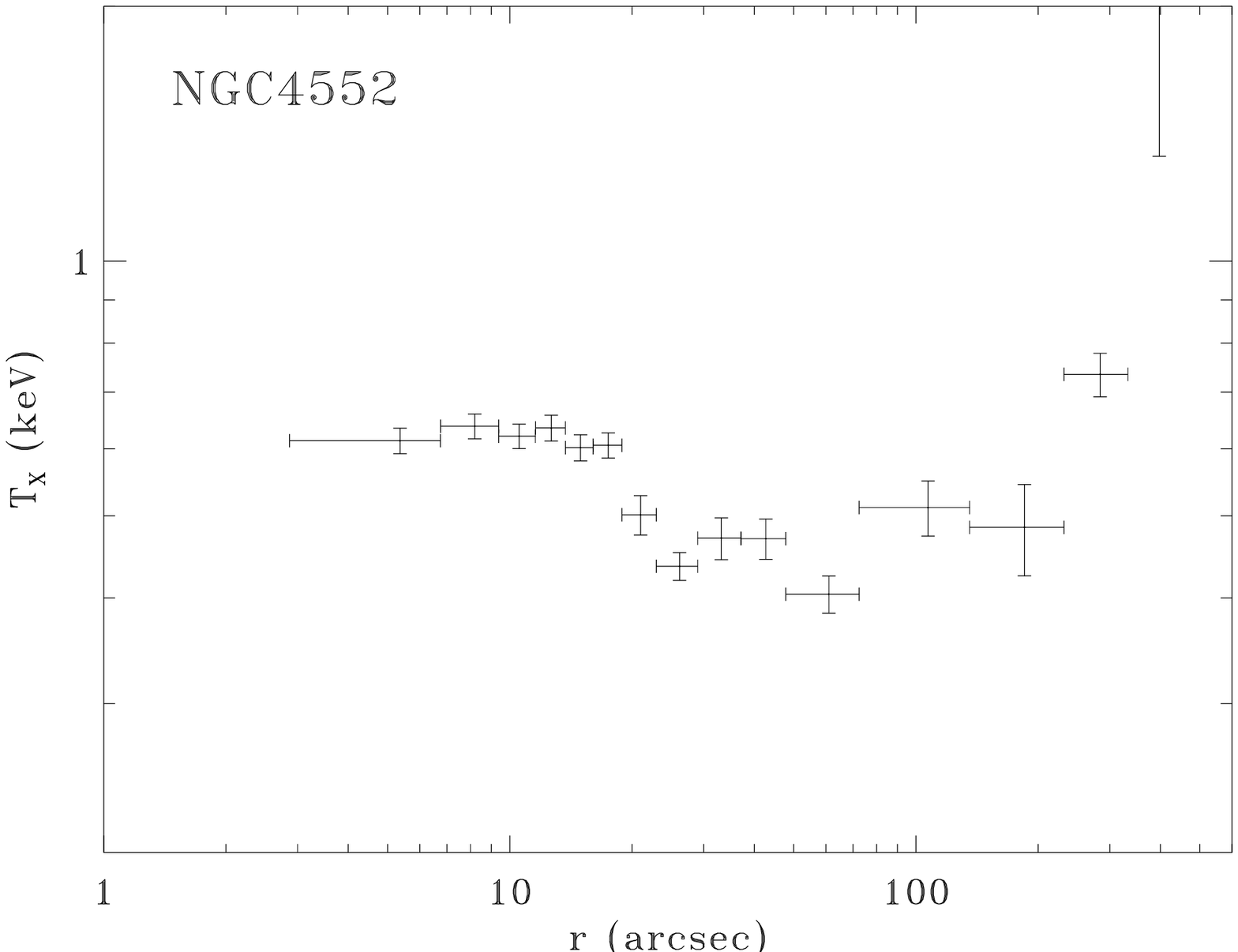}
\includegraphics[width=0.5\textwidth]{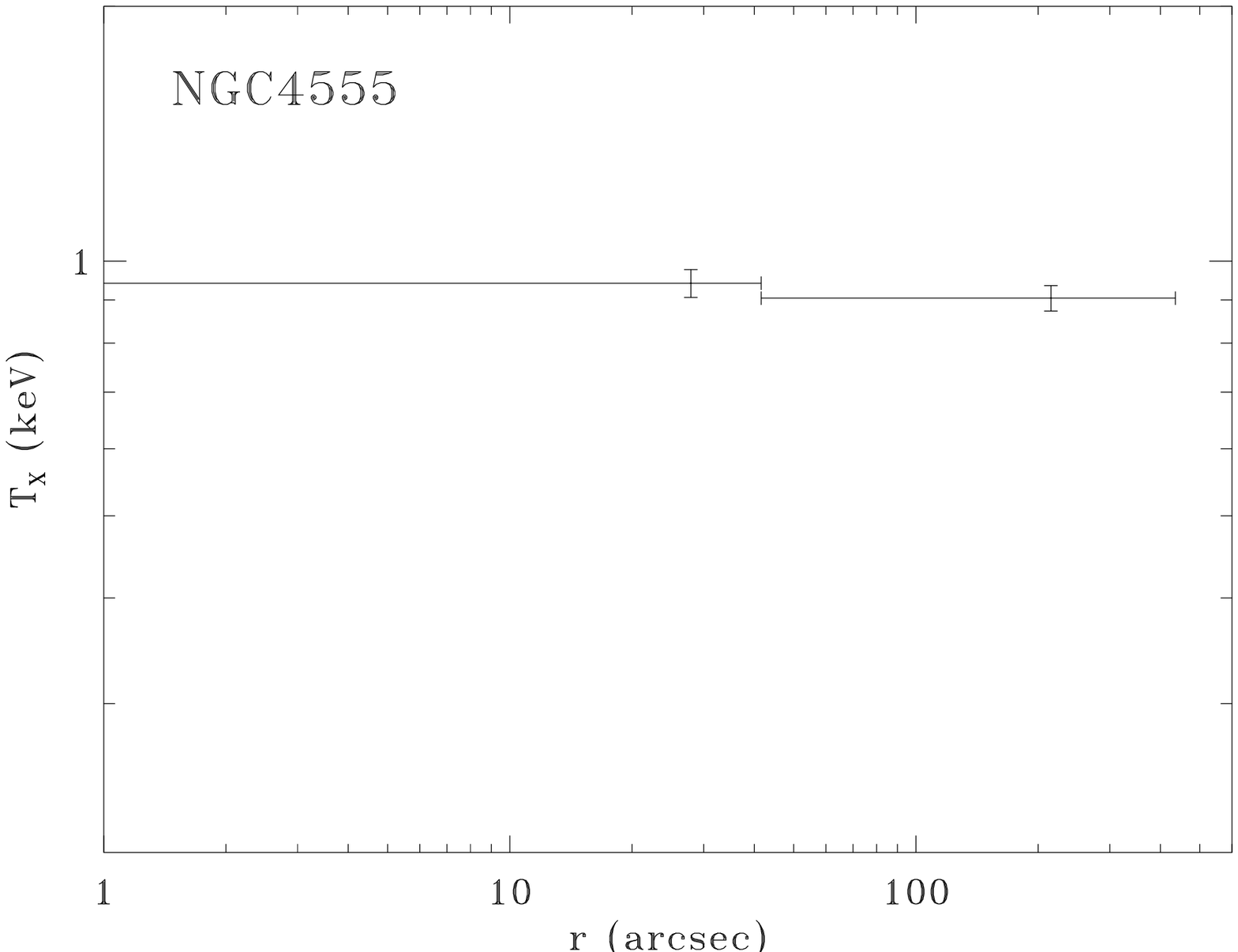}
\includegraphics[width=0.5\textwidth]{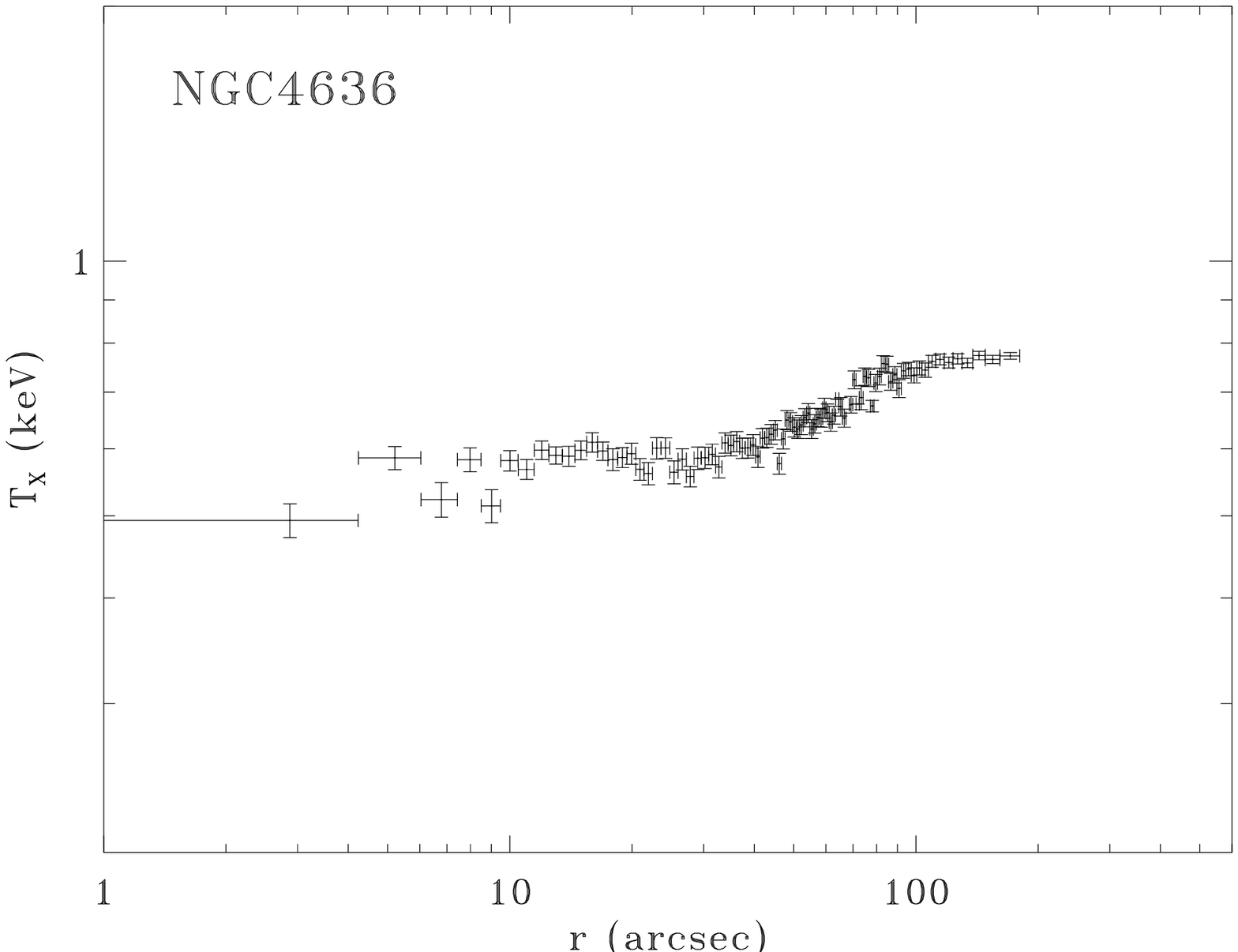}
\includegraphics[width=0.5\textwidth]{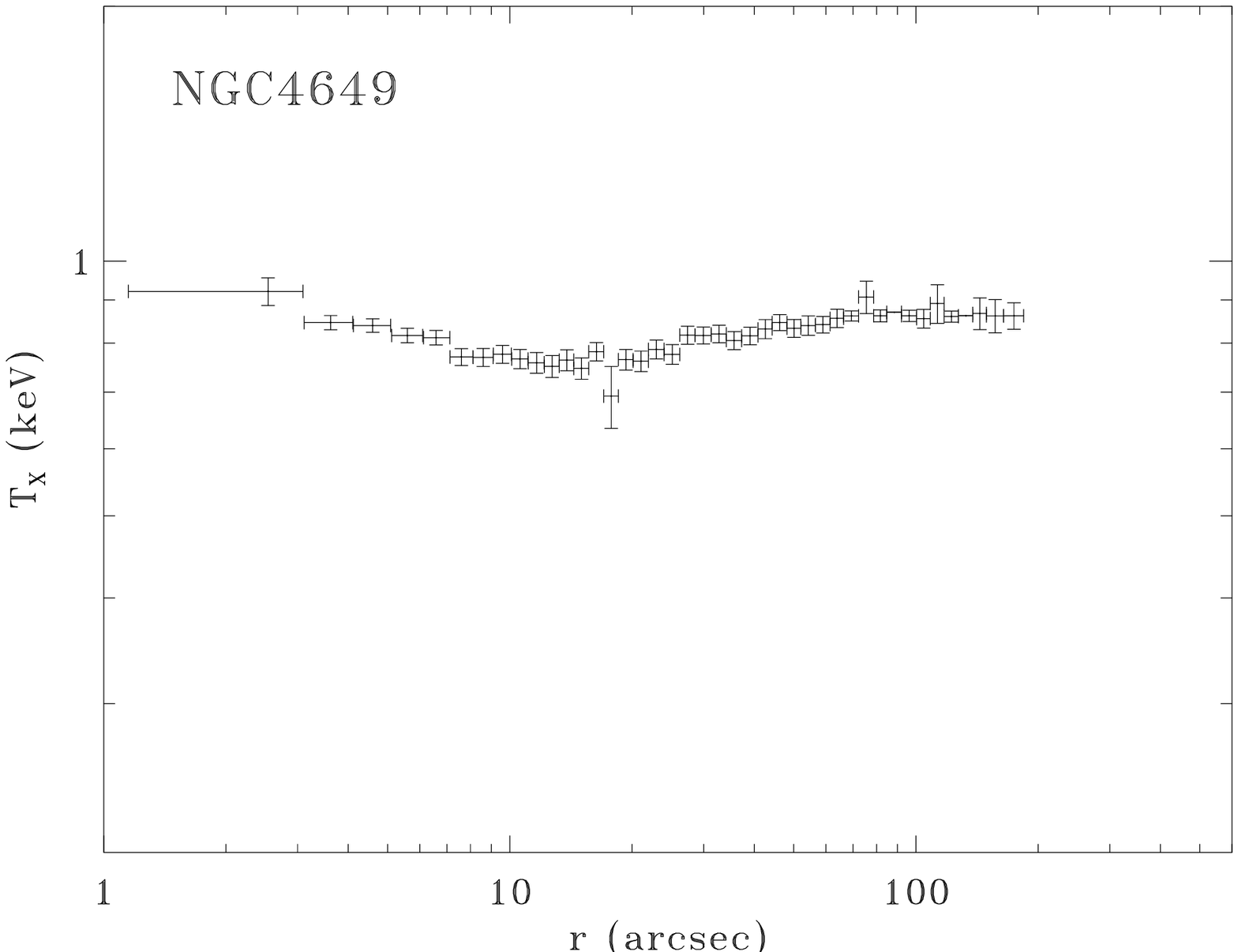}
\end{figure}
\clearpage

\begin{figure}
\includegraphics[width=0.5\textwidth]{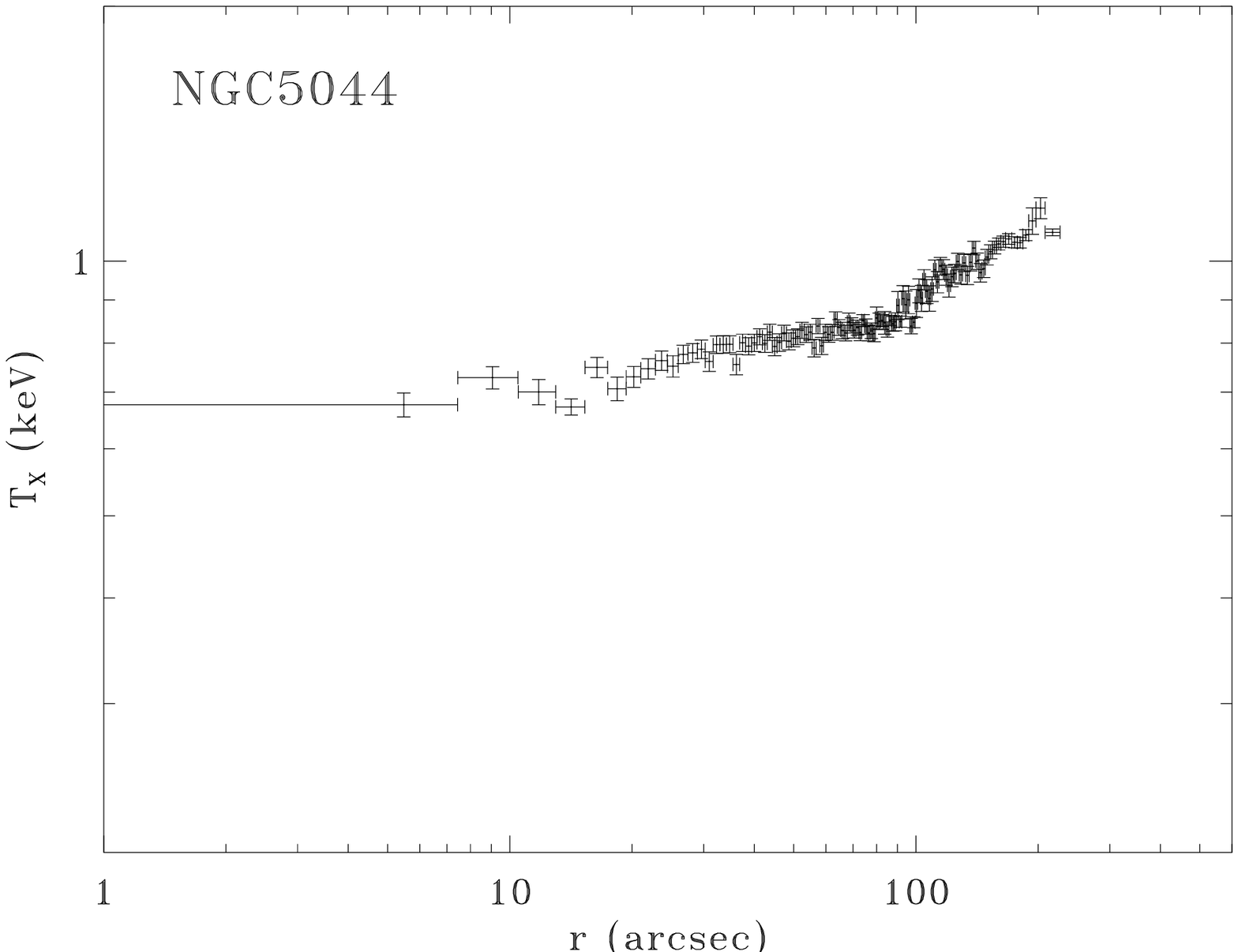}
\includegraphics[width=0.5\textwidth]{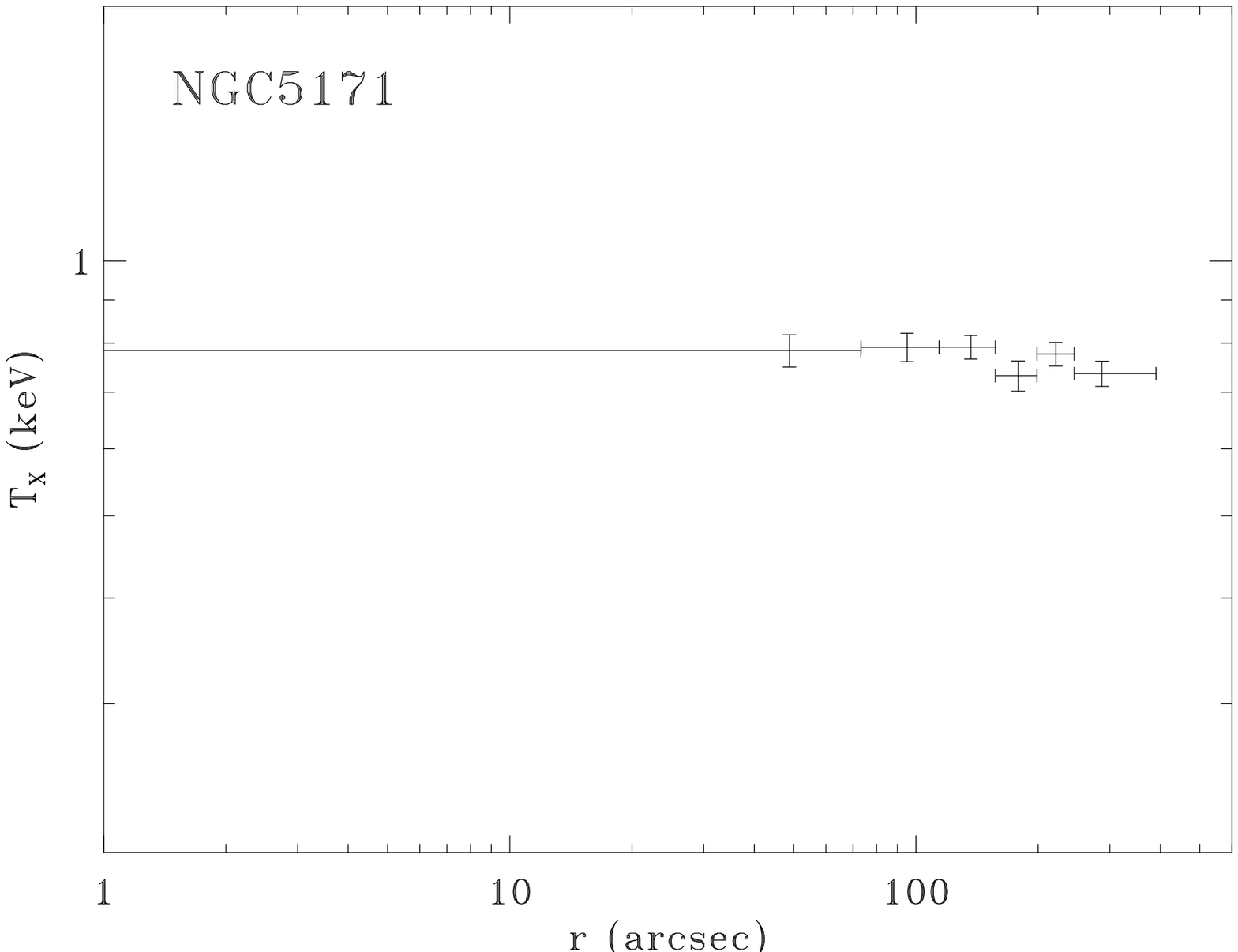}
\includegraphics[width=0.5\textwidth]{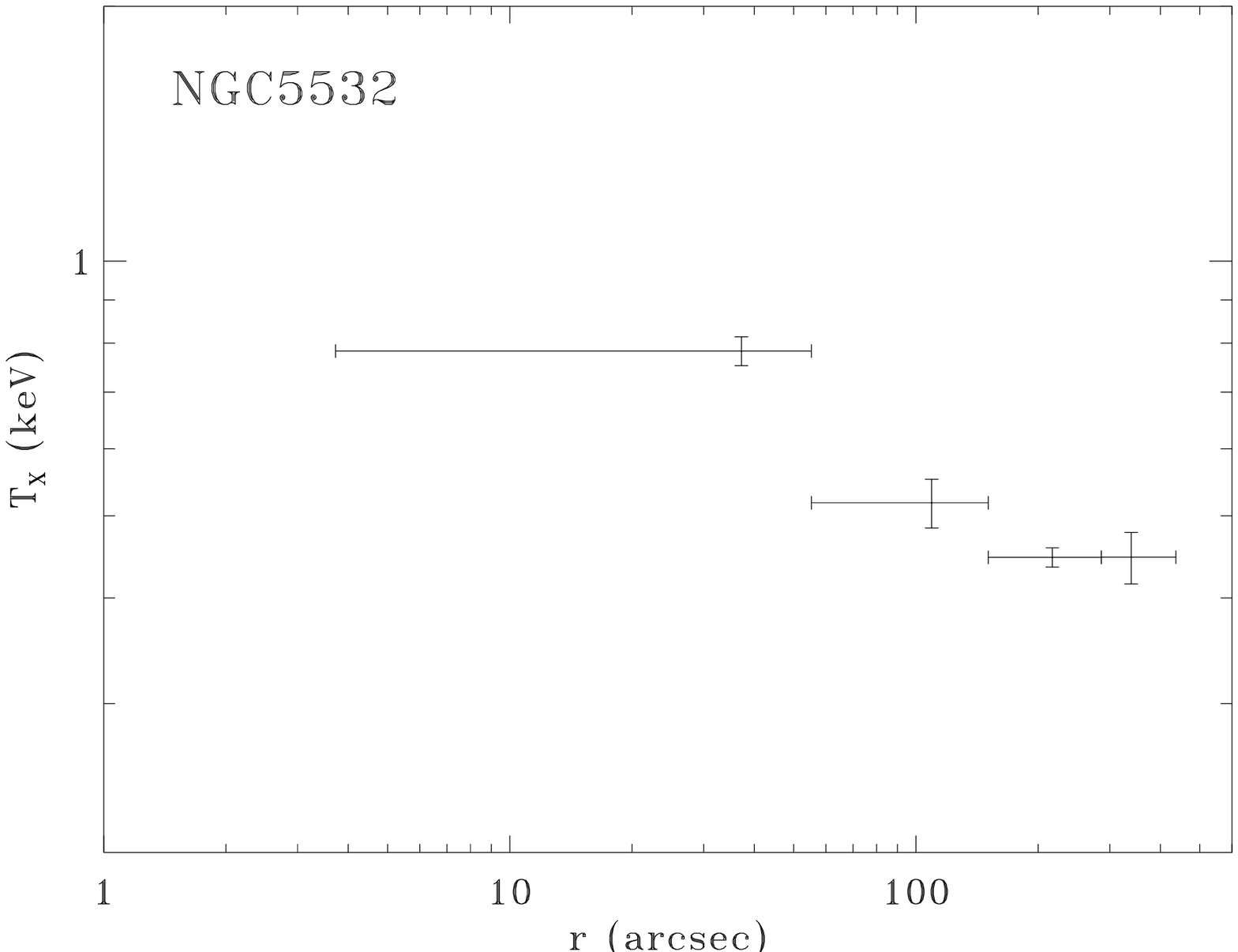}
\includegraphics[width=0.5\textwidth]{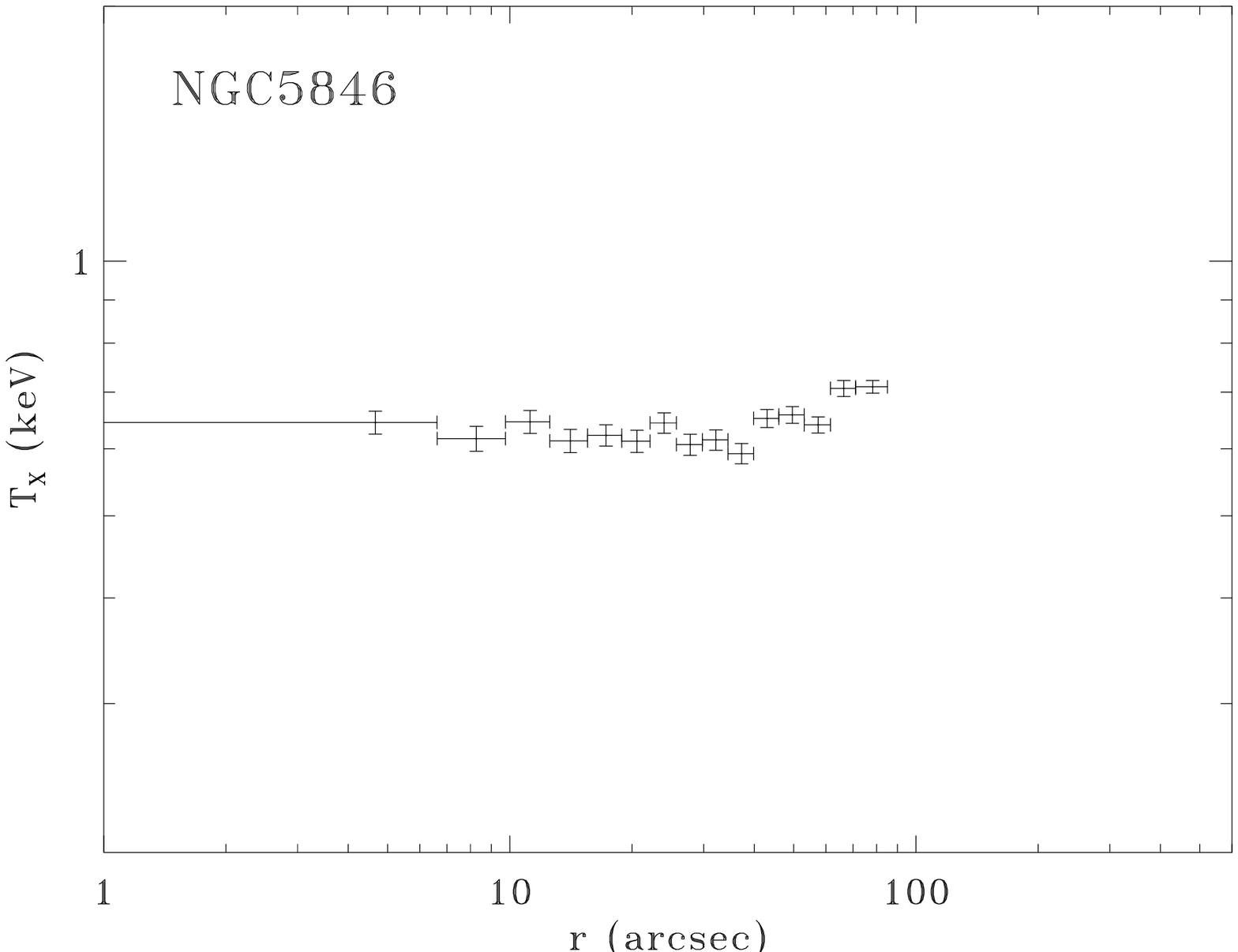}
\includegraphics[width=0.5\textwidth]{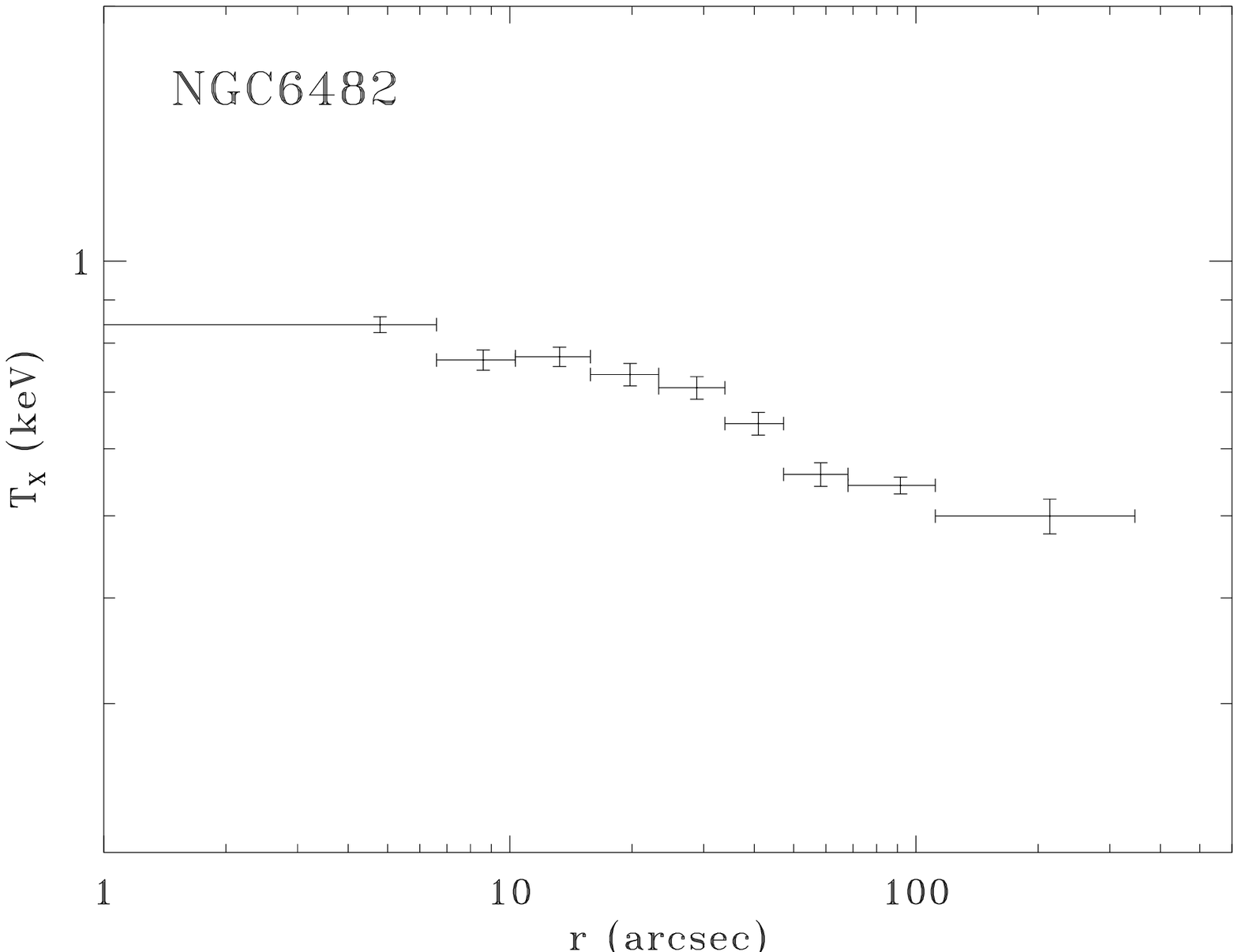}
\includegraphics[width=0.5\textwidth]{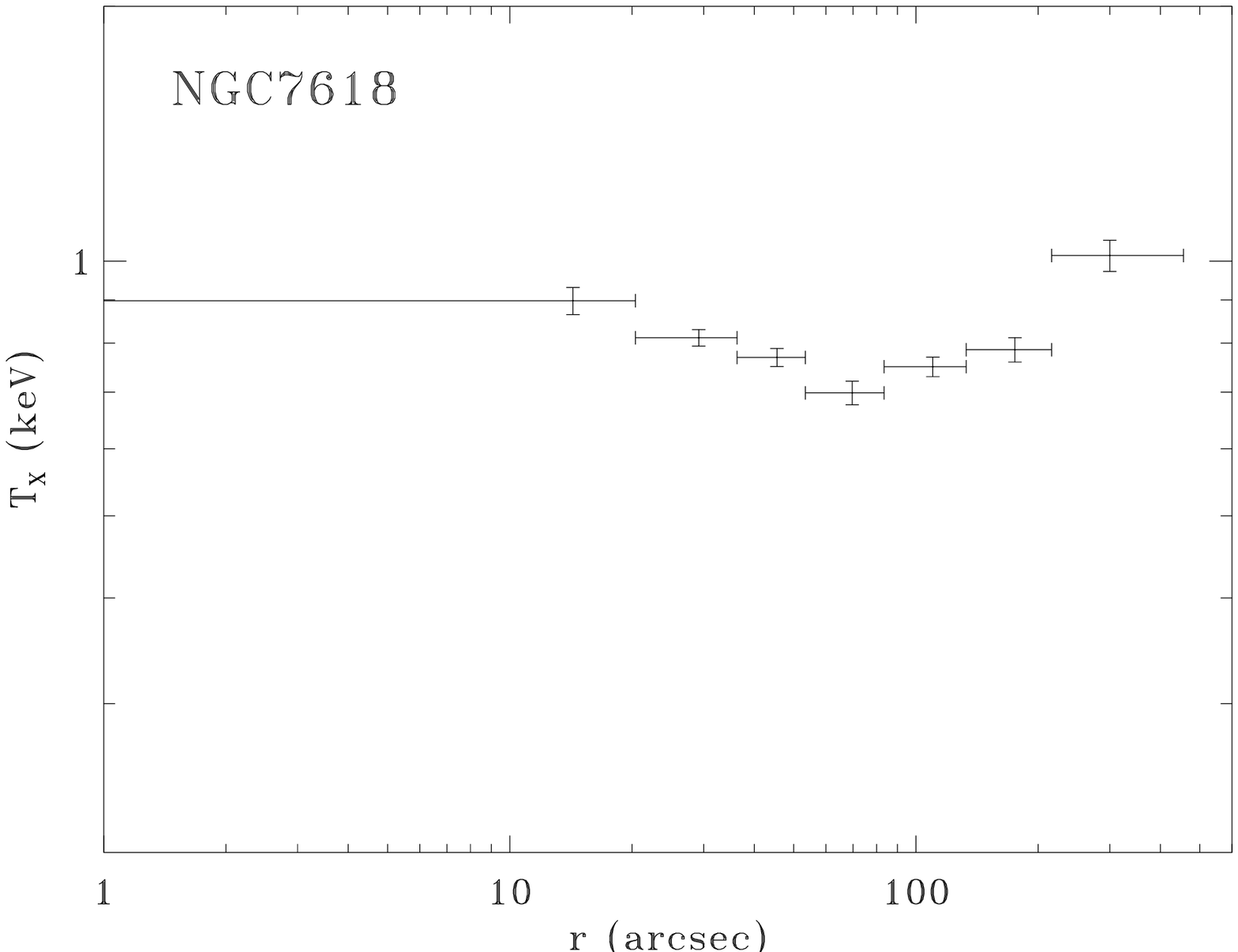}
\end{figure}

\end{document}